\documentclass[journal=jctcce,manuscript=article]{achemso}

\usepackage[version=3]{mhchem} 
\usepackage{amsfonts}
\usepackage{amssymb,amsmath}
\usepackage[linesnumbered,boxed]{algorithm2e}
\usepackage{enumerate}
\usepackage{bbm}
\usepackage{color}
\usepackage{braket}

\newcommand{\tr}{\mathrm{tr}}
\newcommand{\ii}{\mathbbm{i}}
\newcommand{\spn}{\mathrm{span}}
\newcommand{\mps}{\mathrm{MPS}}
\newcommand{\spmps}{\mathrm{SP\text{-}MPS}}

\author{Zhendong Li}\email{zhendongli2008@gmail.com}
\author{Garnet Kin-Lic Chan}\email{gkc1000@gmail.com}
\affiliation{Division of Chemistry and Chemical Engineering,
California Institute of Technology, Pasadena, CA 91125, USA}

\title[\texttt{achemso} demonstration]
{Spin-projected matrix product states (SP-MPS): a versatile tool for strongly correlated systems}

\setkeys{acs}{articletitle=true}

\begin{document}

\begin{abstract}
  We present a new wavefunction ansatz that combines the strengths of spin projection with
  the language of matrix product states (MPS) and matrix product operators
  (MPO) as used in the density matrix renormalization group (DMRG).
Specifically, spin-projected matrix product states (SP-MPS) are constructed as $|\Psi^{(N,S,M)}_{\spmps}\rangle=\mathcal{P}_S|\Psi_{\mps}^{(N,M)}\rangle$,
where $\mathcal{P}_S$ is the spin projector for total spin $S$ and $|\Psi_{\mps}^{(N,M)}\rangle$ is an MPS wavefunction with a given particle number $N$ and spin projection $M$.
This new ansatz possesses several attractive features: (1) It provides a much simpler route to achieve spin-adaptation (i.e. to create eigenfunctions of $\hat{S}^2$) compared to explicitly incorporating the non-Abelian SU(2) symmetry into the MPS. In particular, since the underlying state $|\Psi_{\mps}^{(N,M)}\rangle$ in the SP-MPS uses only Abelian symmetries, one
does not need the singlet embedding scheme for non-singlet states, as normally employed in spin-adapted DMRG, to achieve a single consistent variationally optimized state.
(2) Due to the use of $|\Psi_{\mps}^{(N,M)}\rangle$ as its underlying state,
the SP-MPS can be closely connected to broken symmetry mean-field states.
This allows to straightforwardly generate the large number of broken symmetry guesses needed to explore complex electronic landscapes
in magnetic systems. Further, this connection can be exploited in the future development of quantum embedding theories for open-shell systems. (3) The sum of MPOs representation for the Hamiltonian and spin projector $\mathcal{P}_S$ naturally leads to an embarrassingly parallel algorithm for computing expectation values and optimizing SP-MPS. (4) Optimizing SP-MPS belongs to the variation-after-projection (VAP) class of spin projected
theories. Unlike usual spin projected theories based on determinants,
the SP-MPS ansatz can be made essentially exact simply by increasing the bond dimensions in $|\Psi_{\mps}^{(N,M)}\rangle$. Computing excited states is also simple by imposing  orthogonality constraints,
which are simple to implement with MPS.

To illustrate the versatility of SP-MPS,
we formulate algorithms for the optimization of ground and excited states, develop perturbation theory
based on SP-MPS, and describe how to evaluate  spin-independent and spin-dependent properties
such as the reduced density matrices. We demonstrate the numerical performance of SP-MPS with applications
to several models typical of strong correlation, including the Hubbard model, and [2Fe-2S] and [4Fe-4S] model
complexes.
\end{abstract}

\section{Introduction}\label{introduction}
Since its invention by White\cite{white1992density,white1993density}, the density matrix
renormalization group (DMRG) has become the computational method of choice in strongly correlated
one-dimensional systems. In quantum chemistry, despite the more complex structure of the entanglement between the orbitals in
a general molecule, the DMRG has been applied successfully to a large class
of chemical systems with strong multi-configurational effects and many open shells\cite{white1999ab,daul2000full,mitrushenkov2001quantum,chan2002highly,chan2003exact,
legeza2003controlling,legeza2003optimizing,
chan2004algorithm,mitrushenkov2003quantum,chan2004state,chan2005density,
moritz2006construction,hachmann2006multireference,marti2008density,ghosh2008orbital,chan2008density,
zgid2008obtaining,marti2010density,luo2010optimizing,marti2011new,kurashige2011second,
sharma2012spin,chan2012low,wouters2012longitudinal,mizukami2012more,
kurashige2013entangled,sharma2014low,wouters2014density,wouters2014chemps2,
fertitta2014investigation,knecht2014communication,szalay2015tensor,
yanai2015density,olivares2015ab}. Part of its popularity in this context stems from its systematic improvability
by changing a single parameter (the bond dimension), its  lower computational scaling
 than other tensor
network state (TNS) methods as a function of bond dimension, as well as
various techniques that have significantly enhanced its efficiency in chemical applications, such as the use of
orbital localization, reordering, point group and spin symmetry, and parallelization\cite{chan2004algorithm}.
The identification of matrix product states (MPS) as the underlying variational
wavefunction ansatz in DMRG\cite{ostlund1995thermodynamic,rommer1997class} has further
expanded the applications of DMRG and related techniques in quantum chemistry. For example, the
similarities between MPS (DMRG) and Slater determinants (self-consistent field),
has led to the identification of the Thouless parameterization for MPS
\cite{wouters2013thouless} and the formulation of linear response
theory and tangent space methods for excited states and for time evolution\cite{dorando2009analytic,kinder2011analytic,nakatani2014linear,haegeman2011time}.
Another example is the use of MPS to develop efficient multi-reference perturbation theories\cite{sharma2014communication}, which
avoid the computation of high-order ($>2$) reduced density matrices (RDMs)
by directly minimizing the Hylleraas functional with the first-order wavefunction represented as a MPS.
Finally, MPS have recently been reformulated in Hilbert space, providing a deep connection
to configuration interaction approximations and their graphical representation~\cite{li2016hilbert}. We emphasize that
the MPS and DMRG languages are entirely complementary. The MPS language provides a precise mathematical description of the wavefunctions
used in DMRG algorithms as a product of site tensors $A[k]$. The DMRG language, on the other hand, is the appropriate way to describe
many of the efficient algorithms to evaluate and manipulate MPS.


In this paper, using MPS and related matrix product operator (MPO) techniques~\cite{verstraete2004matrix,mcculloch2007density,verstraete2008matrix,pirvu2010matrix,chan2016mpo}, we revisit the problem of constructing spin
eigenfunctions from MPS, i.e. spin adaptation.
Previously, spin adapted MPS were formulated by imposing non-Abelian SU(2) symmetry structure
on the site tensors\cite{sierra1997density,mcculloch2000density,mcculloch2001total,mcculloch2002non,sharma2012spin,wouters2014chemps2,keller2016spin}. This amounts to
generating spin-adapted reduced bases from direct products of two spin-adapted reduced bases during the DMRG blocking procedure.
However, although it provides a practical advantage in terms of reducing the computational bond dimension by approximately half\cite{sharma2012spin},
there are some formal and practical drawbacks to the explicitly spin-adapted MPS (SA-MPS) approach.
First, for non-singlet states, the one-site DMRG optimization leads to
different MPS wavefunctions, even at the same site of the sweep, when moving in the
left and right directions.
A simple example
 illustrates this\cite{sharma2012spin}: consider a reduced
triplet wavefunction written as $\|\Psi_{S=1}\rangle
=\frac{1}{\sqrt{2}}\|S_L=1\rangle\times(\|S_R=0\rangle
+\|S'_R=2\rangle)$, where $\|S\rangle$ represents
a reduced spin state with total spin quantum number $S$. In this case, the left reduced density matrix
is $\rho^L=\left[1\right]$, while the right reduced density matrix is
$\rho^R=\frac{1}{2}\left[\begin{array}{cc}
1 & 1 \\
1 & 1 \\
\end{array}\right]$.
The density matrix $\rho^L$ is of rank one, which yields a discarded weight of zero in the DMRG optimization
if the bond dimension $D$ (the number of states to be kept) is $\geq 1$. However, although the density matrix $\rho^R$ is also of rank one, in order to ensure that the renormalized states have well-defined spin quantum numbers, only the block diagonal
part of $\rho^R$ (the so-called quasi-density matrix\cite{mcculloch2000density,mcculloch2001total,mcculloch2002non,zgid2008spin})
enters the spin-adapted renormalization procedure, generating
 {\it two} renormalized states.
 This leads to a truncation of the wavefunction if $D=1$ and thus, even at the same site,
 the DMRG wavefunctions generated by a sweep to the left or to the right are different
 and yield different energies.
 While this is not a severe problem when  $D$ is large as the difference is very small,
 the DMRG sweep no longer strictly corresponds to an energy minimization. This complicates
 the computation of properties, such
as the nuclear gradients, since
$\partial E/\partial A[k]$ may not be exactly zero.
To avoid the above problem,  one usually uses the singlet embedding scheme\cite{tatsuaki2000interaction,sharma2012spin,wouters2014chemps2}
to create a singlet total wavefunction by coupling the non-singlet physical state to a set of noninteracting fictitious spins.
Within this singlet total state, the spin quantum numbers and dimensions of the left and right renormalized Hilbert spaces
must always match, ensuring that the one-site DMRG optimization algorithm converges to a single consistent MPS.
However, the singlet embedding scheme introduces its own complications, for example
in the evaluation
of transition density matrices between two states with different spins, as
required in the state interaction treatment of spin-orbit coupling with SA-MPS\cite{sayfutyarova2016state}.


The second, and perhaps more important in practice, deficiency of the spin-adapted MPS formulation
is that it does not provide a simple connection
to the intuitive spin structures and charge configurations of broken symmetry Slater determinants. 
This connection is important to retain for two reasons.
First, it helps in interpreting the wavefunction when understanding the electronic
structure of systems with many open shells, as is found in
transition metal clusters\cite{noodleman1988broken}. Second, the connection to  broken
symmetry determinants can be used to easily prepare
many different types of  initial guess for the DMRG optimization. In complex systems where there
are many competing low-energy spin states, the ability to systematically prepare many such initial guesses ensures that the subsequent optimization avoids
physically irrelevant local minima. 

To address these issues, in this work we propose
an alternative way to construct spin eigenfunctions from MPS using spin projection.
Spin projection techniques have a long history in quantum
chemistry, dating  back to the work by L{\"o}wdin\cite{lowdin1955quantum}
using a spin projector with broken symmetry Slater determinants. The general idea has more
recently been revived\cite{scuseria2011projected,jimenez2012projected,jimenez2013multi,jimenez2013excited,tsuchimochi2016communication,tsuchimochi2016black}
using group theoretical projectors rather than L{\"o}wdin's original projector.
Here, we will use the wavefunction ansatz $|\Psi^{(N,S,M)}_{\spmps}\rangle=\mathcal{P}_S|\Psi_{\mps}^{(N,M)}\rangle$
to obtain spin eigenstates, where $\mathcal{P}_S$ is the spin projector for total spin $S$ and
$|\Psi_{\mps}^{(N,M)}\rangle$ is an MPS wavefunction with given particle number
$N$ and spin projection $M$.
Such spin-projected matrix product states (SP-MPS) display the following interesting features:
\begin{itemize}
\item They allow for a consistent energy minimization procedure for non-singlet states without using the singlet embedding scheme
  as only Abelian symmetries are used in the underlying state $|\Psi_{\mps}^{(N,M)}\rangle$. As mentioned above, this is mainly a formal advantage over SA-MPS.

\item The underlying MPS can reduce at $D=1$ to a ``broken symmetry'' determinant.
Although in principle, different spatial orbitals for different spins (DODS) can be used for
the underlying MPS,  throughout this paper the same spatial orbitals for different spins (SODS) will be used, to simplify the representation of the spin projector $\mathcal{P}_S$ (see
Sec. \ref{theory:spmps}).
Thus, the term ``broken symmetry'' used here refers to the breaking of spin symmetry at the level
of configurations instead of orbitals. Compared with SA-MPS, this connection to a
``broken symmetry'' determinant helps in the interpretation of complex electronic wavefunctions,
as well as in generating initial guesses for $|\Psi_{\mps}^{(N,M)}\rangle$, to systematically explore electronic landscapes with many competing states. Further, this connection may be used to extend the density matrix embedding theory (DMET)\cite{knizia2012density,knizia2013density} to overall open-shell systems, using a mean-field state and an SP-MPS state for the low- and high-level descriptions of the open-shell system, respectively.

\item It is a variation-after-projection (VAP) ansatz, where the underlying broken-symmetry MPS can be efficiently optimized with a DMRG sweep algorithm. Unlike other spin projected theories in quantum chemistry, it is easy to make the variational state more accurate, as one can simply increase the bond dimension in $|\Psi_{\mps}^{(N,M)}\rangle$. The resulting spin-projected MPS retains all the advantages of conventional MPS. For example, computing excited states is straightforward,
as imposing  orthogonality constraints is simple with MPS.
\item A final advantage is that the computational implementation of SP-MPS algorithms is in many respects simpler than the
  implementation of algorithms involving  spin-adapted MPS. This simplicity is potentially even more advantageous
  when working with more complex tensor network states, such as projected entangled pair states (PEPS)\cite{verstraete2004renormalization}.
\end{itemize}

The remainder of the paper is organized as follows. In Sec.
\ref{theory:mpsmpo}, we recapitulate the MPS and MPO representations
for many-body states and operators, which are essential for
describing how to work with the SP-MPS ansatz. In particular, we provide an explicit recursive algorithm
to derive an  MPO representation for the second quantized \emph{ab initio} Hamiltonian.
The wavefunction ansatz
for SP-MPS is then introduced and its properties discussed
in Sec. \ref{theory:spmps}. The DMRG-like sweep algorithm for
optimizing the SP-MPS representation of ground and excited states, as well as a parallelization scheme
based on the sum of MPOs representation for the Hamiltonian
and spin projector $\mathcal{P}_S$, are also formulated in Sec. \ref{theory:algorithm}.
The evaluation of properties such as spin-free and spin-dependent reduced density matrices is
considered in Sec. \ref{theory:properties}. Pilot applications to
some prototypical open shell strongly correlated systems are described in Sec. \ref{results}. Conclusions
and outlines for future directions are presented in Sec. \ref{conclusion}.

\section{Theory}\label{theory}

\subsection{Matrix product states (MPS) and matrix product operators (MPO)}\label{theory:mpsmpo}
In this section, we recapitulate  essential aspects of the MPS and MPO representations
for many-body states and operators. This will be necessary to formulate the theory of SP-MPS.
A more detailed description of this language can be found in Refs. \cite{schollwock2011density,chan2016mpo,keller2015efficient,keller2016spin}.
As an important example of how to work with the MPO representation, we will give an explicit algorithm
to derive the MPO representation for the second quantized \emph{ab initio} Hamiltonian.

A simple way to introduce matrix product states is from the perspective of a repeated singular value decompositions (SVDs)
of a general wavefunction defined in Fock space,
\begin{align}
  \ket{\Psi} = \sum_{\{n_k\}} \Psi^{n_1 n_2 \cdots n_K} \ket{ n_1 n_2 \cdots n_K},\label{FSwf}
\end{align}
where $\ket{n_1 n_2 \cdots n_K}$ ($n_k\in\{0,1\}$) is an occupancy basis state with $K$ spin-orbitals.
A successive SVD for the coefficient tensor $\Psi^{n_1 n_2 \cdots n_K}$ leads to an MPS representation
with open boundary conditions (OBC)\cite{schollwock2011density},
\begin{eqnarray}
\Psi^{n_1n_2\cdots n_K}
=\sum_{\{\alpha_k\}} A^{n_1}_{\alpha_1}[1]A^{n_2}_{\alpha_1\alpha_2}[2]\cdots A^{n_K}_{\alpha_{K-1}}[K],\label{FSMPS}
\end{eqnarray}
where the site tensor $A[k]$ away from the two boundaries  is a three-way tensor of dimension $2\times D\times D$,
and the two tensors at the boundary are matrices of dimension $2\times D$ or $D\times 2$.
The so-called bond dimension $D$ controls the number of retained (renormalized) states,
and therefore controls the accuracy of the representation when used
as a variational ansatz. Similarly to Eq. \eqref{FSwf}, within the occupation number representation, with basis vectors $\ket{n_1 n_2 \cdots n_K}$, an operator $\hat{O}$ is represented by its coefficient tensors $O^{n_1 n_2 \cdots n_K}_{n_1' n_2' \cdots n_K'}$, viz.,
\begin{align}
\hat{O} = \sum_{\{n_k\}\{n_k'\} } O^{n_1 n_2 \cdots n_K}_{n_1' n_2' \cdots n_K'} |n_1 n_2 \cdots n_K\rangle \langle n_1' n_2' \cdots n_K'|,\quad
O^{n_1 n_2 \cdots n_K}_{n_1' n_2' \cdots n_K'} \triangleq\langle n_1 n_2 \cdots n_K|O|n_1' n_2' \cdots n_K'\rangle,\label{FSop}
\end{align}
and similarly to Eq. \eqref{FSMPS}, an MPO representation for the coefficient can be obtained through successive SVDs as
\begin{align}
O^{n_1 n_2 \cdots n_K}_{n_1' n_2' \cdots n_K'}  = \sum_{\{ \beta_k\}} W^{n_1 n_1'}_{\beta_1 }[1] W^{n_2n_2'}_{\beta_1 \beta_2}[2] \cdots W^{n_Kn_K'}_{\beta_{K-1}}[K].\label{FSMPO}
\end{align}

\begin{figure}
    \begin{tabular}{cc}
    {\resizebox{0.4\textwidth}{!}{\includegraphics{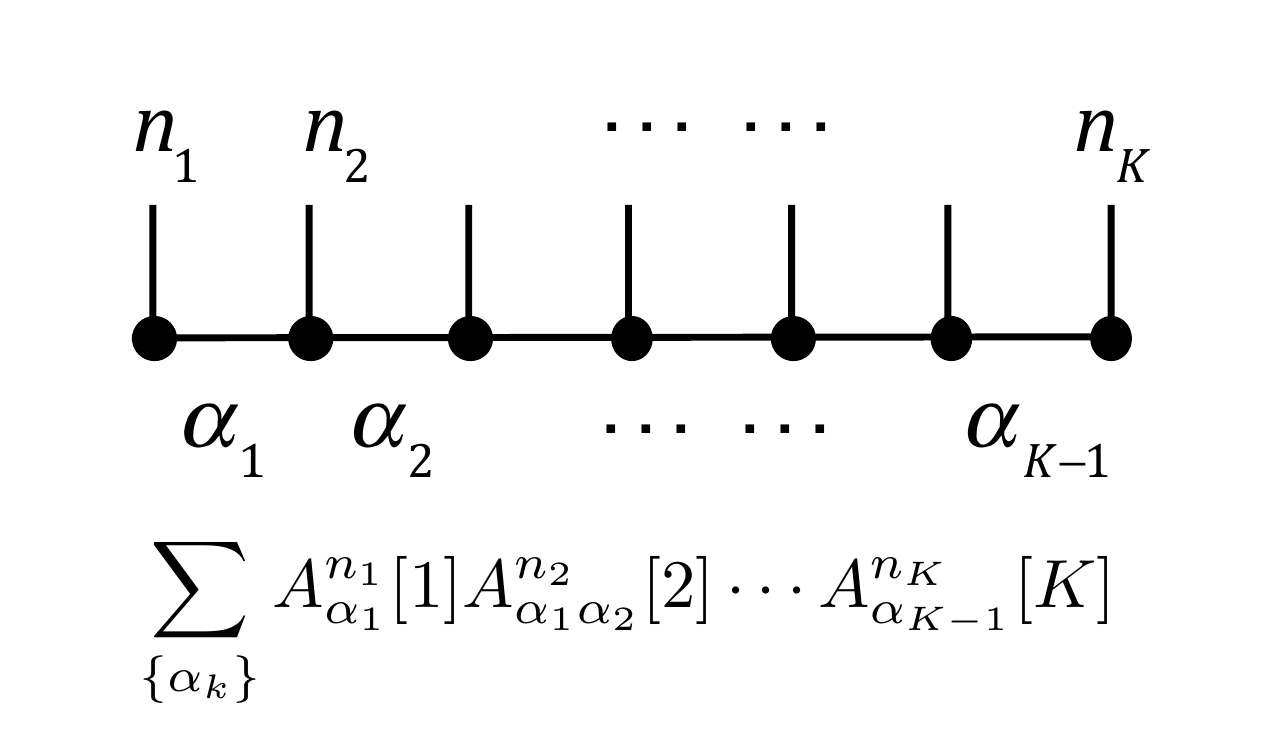}}} &
    {\resizebox{0.4\textwidth}{!}{\includegraphics{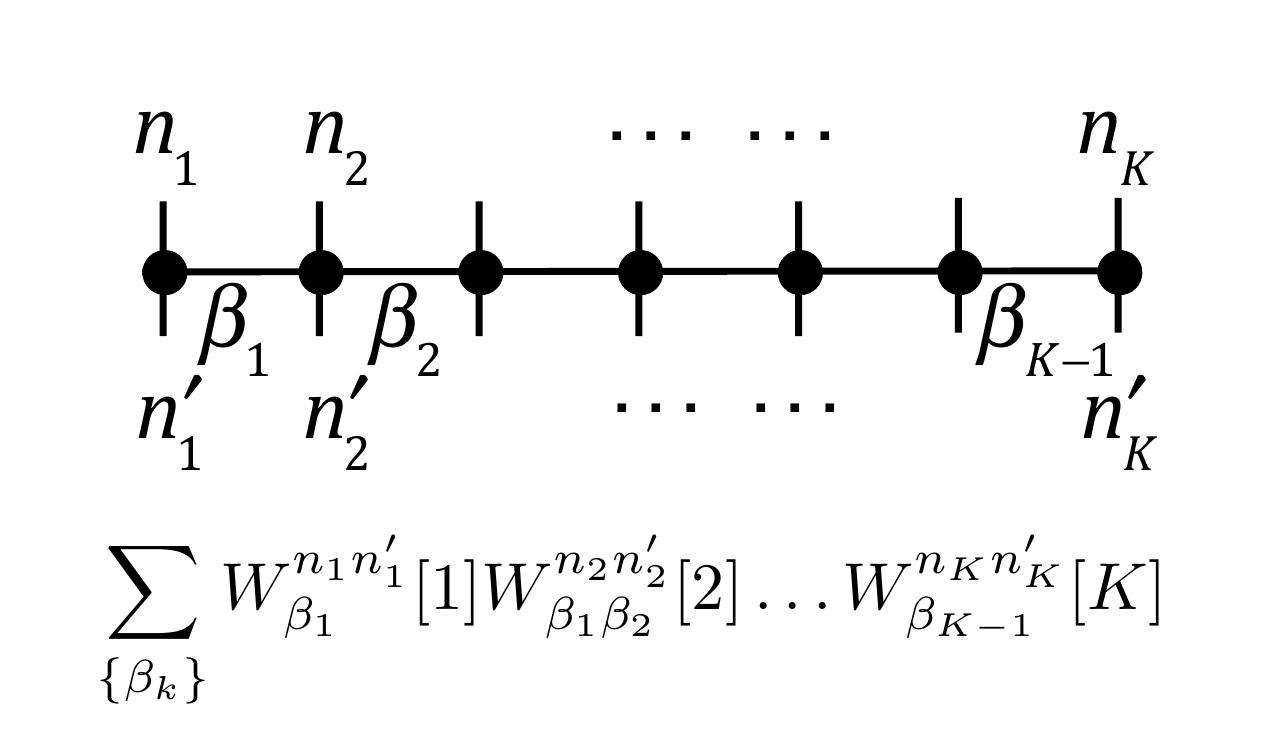}}} \\
    (a) MPS & (b) MPO \\
    \end{tabular}
\caption{Graphical representations for MPS \eqref{FSMPS} and MPO \eqref{FSMPO}.}\label{fig:mpsmpo}
\end{figure}

To simplify the discussion of MPS and MPO, it is convenient to use a graphical representation. In Figure \ref{fig:mpsmpo}, the graphical representations for a MPS \eqref{FSMPS} and MPO \eqref{FSMPO} are shown. A key simplification comes
from the convention that the sum over auxiliary (dummy) indices is replaced by a bond between two tensors graphically represented
by dots. This notation eliminates the need
to write out nested algebraic summations to express contractions between MPS and MPO.

An important point that deserves emphasis is that in both the MPS and MPO representations,
the site tensors such as $A[k]$ in Eq. \eqref{FSMPS} and $W[k]$ in Eq. \eqref{FSMPO} are not uniquely determined.
Given a set of occupancies, Eq. \eqref{FSMPS} tells us that the coefficient is expressed as a product of matrices with
vectors at the boundaries. There is thus a gauge degree of freedom arising from the trivial fact that the product of two matrices $AB$ is not affected
by inserting a pair of matrices $G G^{-1}$, i.e. $AB=(AG)(G^{-1}B)$, if $G$ is nonsingular. In practice, we
 use certain types of gauges to simplify computations and improve numerically stability.

As an important example of the MPO formalism in practice,  we now derive an explicit MPO
representation for the generic second quantized Hamiltonian written as
\begin{eqnarray}
\hat{H}=\sum_{pq}h_{pq} a_p^\dagger a_q + \frac{1}{2}\sum_{pqrs}g_{pq,sr}a_p^\dagger a_q^\dagger a_r a_s,\quad
g_{pq,sr}=\langle pq|sr\rangle.\label{sqH}
\end{eqnarray}
The MPO representation we obtain does not depend on the symmetry of the integrals $h_{pq}$ and $g_{pq,rs}$, and therefore also applies to
other two-body operators, such as the doubles excitation operator $\hat{T}_2$ used in correlation methods.
To begin, following Ref. \cite{chan2016mpo}, we rewrite $\hat{H}$ in Eq. \eqref{sqH} as a sum of $K$ operators,
\begin{eqnarray}
\hat{H}=\sum_{p}\hat{H}_p,\quad \hat{H}_p=a_p^\dagger\left(\sum_{q}h_{pq}a_q +\sum_{q,r<s}g^A_{pqrs} a_q^\dagger a_r a_s\right),
\quad g^A_{pqrs}\triangleq \frac{1}{2}(g_{pq,sr}-g_{pq,rs}).\label{sumH}
\end{eqnarray}
such that it suffices to examine the construction of the MPO for an operator $\hat{V}$ of the following form,
\begin{eqnarray}
\hat{V}=\sum_{q}h_{q}a_q+\sum_{qrs}V_{qrs} a_q^\dagger a_r a_s,\quad V_{q,r\ge s}=0.\label{sqV}
\end{eqnarray}
Once the MPO form for $\hat{V}$ is obtained, we can substitute it in the bracket
in Eq. \eqref{sumH}, giving $\hat{H}$  as a sum of $K$ MPOs. As will be shown
below, the bond dimension for representing $\hat{V}$ exactly as an MPO is of $O(K)$,
and since $a_p^\dagger$ is a simple MPO with bond dimension 1, each $\hat{H}_p$ will also be an
MPO with bond dimension of $O(K)$. Consequently, adding together the $\hat{H}_p$, we will have an exact MPO
representation for $\hat{H}$ with bond dimension $O(K^2)$, which is the correct scaling
of the bond dimension for generic two-body operators\cite{chan2016mpo}.

Assuming  one spin orbital per site, there are only four basis operators $\{I_k,a^\dagger_k,a_k,n_k=a_k^\dagger a_k\}$
per site. We will use the notation $\hat{V}^{[k,K]}$ to denote
the operator $\hat{V}$ defined with spin orbitals only in the range from $k$ to $K$.
To derive the formula for the MPO representations of $\hat{V}=\hat{V}^{[1,k]}$ \eqref{sqV}, we start by
examining the recurrence relation for $\hat{V}^{[k,K]}$,
\begin{eqnarray}
\hat{V}^{[k,K]}&\triangleq&\sum_{k\le q\le K}h_{q}a_q+\sum_{k\le qrs\le K}V_{qrs} a_q^\dagger a_r a_s\nonumber\\
&=&
h_{k}a_k\otimes I^{[k+1,K]}+\widetilde{I}_k\otimes \hat{V}^{[k+1,K]}\nonumber\\
&+&a_k^\dagger\otimes \hat{P}_{k}^{[k+1,K]}
-a_k\otimes \hat{Q}_{k}^{[k+1,K]}
+\sum_{r}(V_{kkr}\widetilde{n}_k)\otimes (a_r)^{[k+1,K]},\label{eq:VkK}
\end{eqnarray}
where the symbol $\otimes$ represents the Kronecker product, the tilde in $\widetilde{I}_k$
is used to indicate the presence of the fermionic sign factor that needs to be
taken into account when going to the matrix representation of the operator (vide post), and the intermediates (complementary operators) $\hat{P}$ and $\hat{Q}$ are defined by
\begin{eqnarray}
\hat{P}^{[k,K]}_{l} &\triangleq& (\sum_{rs}V_{lrs}a_r a_s)^{[k,K]}\nonumber\\
&=& I_k\otimes \hat{P}^{[k+1,K]}_l +
\sum_{r}(V_{lkr}\widetilde{a}_k)\otimes (a_r)^{[k+1,K]},\quad l\in[1,k-1],\label{eq:PkK}\\
\hat{Q}^{[k,K]}_{l} &\triangleq& (\sum_{qs}V_{qls}a_q^\dagger a_s)^{[k,K]}\nonumber\\
&=& V_{klk}n_k\otimes I^{[k+1,K]}+I_k\otimes \hat{Q}^{[k+1,K]}_l \nonumber\\
&+&\sum_{r}(-V_{rlk}\widetilde{a}_k)\otimes (a_r^\dagger)^{[k+1,K]}
+\sum_{r}(V_{klr}\widetilde{a}_k^\dagger)\otimes (a_r)^{[k+1,K]},\quad l\in [1,k-1].\label{eq:QkK}
\end{eqnarray}
Putting Eqs. \eqref{eq:VkK}, \eqref{eq:PkK}, and \eqref{eq:QkK} together, we recast
these recurrence relations in a compact matrix-vector product form,
\begin{eqnarray}
\left[\begin{array}{c}
\hat{V}^{[k,K]} \\
\hline\hline
(a^{\dagger}_{[k]})^{[k,K]} \\
(a^{\dagger}_{[k+1,K]})^{[k,K]} \\
\hline
(a_{[k]})^{[k,K]} \\
(a_{[k+1,K]})^{[k,K]} \\
\hline
\hat{Q}^{[k,K]}_{[1,k-1]}\\
\hline
\hat{P}^{[k,K]}_{[1,k-1]}\\
\hline\hline
I^{[k,K]} \\
\end{array}\right]
=
\left[\begin{array}{c||c|c|c|c|c|c||c}
\widetilde{I}_k & 0 & V_{kkr}\widetilde{n}_k & 0 & -a_k & 0 & a_k^\dagger & h_{k}a_k \\
\hline\hline
0 & 0 & 0 & 0 & 0 & 0 & 0 & a^\dagger_{k} \\
0 & \widetilde{I}_k & 0 & 0 & 0 & 0 & 0 & 0 \\
\hline
 0 & 0 & 0 & 0 & 0 & 0 & 0 & a_k \\
0 & 0 & \widetilde{I}_k & 0 & 0 & 0 & 0 & 0 \\
\hline
0 & -V_{rlk}\widetilde{a}_{k} & V_{klr}\widetilde{a}^\dagger_{k} & I_k & 0 & 0& 0 & V_{klk}n_k \\
\hline
0 & 0 & V_{lkr}\widetilde{a}_k & 0 &0 & I_k & 0 & 0 \\
\hline\hline
 0 & 0 & 0 & 0 & 0 & 0 & 0& I_k \\
\end{array}\right]
\left[\begin{array}{c}
\hat{V}^{[k+1,K]} \\
\hline\hline
(a^{\dagger}_{[k+1,K]})^{[k+1,K]} \\
\hline
(a_{[k+1,K]})^{[k+1,K]} \\
\hline
\hat{Q}^{[k+1,K]}_{[1,k-1]}\\
\hat{Q}^{[k+1,K]}_{[k]}\\
\hline
\hat{P}^{[k+1,K]}_{[1,k-1]} \\
\hat{P}^{[k+1,K]}_{[k]} \\
\hline\hline
I^{[k+1,K]} \\
\end{array}\right],\label{Wfac}
\end{eqnarray}
where the subscript in operators such as $\hat{Q}^{[k+1,K]}_{[1,k-1]}$ indicates
the range for the index $l$ in Eq. \eqref{eq:QkK}.
Since $\hat{V}^{[1,K]}$ is the operator to be written as an MPO, one can immediately
recognize that the $2(K+1)$-by-$2(K+1)$ coefficient matrix in Eq. \eqref{Wfac} with operator entries is just the operator counterpart
of the site tensor $W[k]$ \eqref{FSMPS}, while at the left and right boundaries,
the first row and the last column in Eq. \eqref{Wfac} can be read off for $W[1]$ and $W[K]$,
respectively. To obtain the site tensor $W[k]$,
we use the matrix representations for operators such as $a_k$ in the space of $\{|0\rangle,|1\rangle\}$.
The necessary matrix representations for all the operators in Eq. \eqref{Wfac} are as follows:
\begin{eqnarray}
\protect[I]&=&I=
\left[\begin{array}{cc}
1 & 0 \\
0 & 1 \\
\end{array}\right],
\quad
\protect[\widetilde{I}]=\sigma_z=
\left[\begin{array}{cc}
1 & 0 \\
0 & -1 \\
\end{array}\right],
\nonumber\\
\protect[a^\dagger]&=&\sigma_-=
\left[\begin{array}{cc}
0 & 0 \\
1 & 0 \\
\end{array}\right],
\quad
\protect[\widetilde{a}^\dagger]=\sigma_-\sigma_z=
\left[\begin{array}{cc}
0 & 0 \\
1 & 0 \\
\end{array}\right],
\nonumber\\
\protect[a]&=&\sigma_{+}=
\left[\begin{array}{cc}
0 & 1 \\
0 & 0 \\
\end{array}\right],
\quad
\protect[\widetilde{a}]=\sigma_+\sigma_z=
\left[\begin{array}{cc}
0 & -1 \\
0 & 0 \\
\end{array}\right],
\nonumber\\
\protect[n]&=&\sigma_-\sigma_+=
\left[\begin{array}{cc}
0 & 0 \\
0 & 1 \\
\end{array}\right],
\quad
\protect[\widetilde{n}]=\sigma_-\sigma_+\sigma_z=
\left[\begin{array}{cc}
0 & 0 \\
0 & -1 \\
\end{array}\right].\label{JWtrans}
\end{eqnarray}
The Pauli matrices in Eq. \eqref{JWtrans} remind us that
the above matrix representations are simply an expression of the
Jordan-Wigner transformation\cite{jordan1928pauli} that maps fermions
to spins and vice versa. A final remark is that
to obtain the MPO in the spatial orbital basis, in which
the local dimension is 4 instead of 2, one
need only to merge two adjacent MPO site tensors defined in Eq. \eqref{Wfac},
 placing the $\alpha$ and $\beta$ spin orbitals together. The resulting
spatial MPO is the one we will use in the energy minimization algorithm with SP-MPS. In the
 following, we will assume the use of spatial orbitals as sites, the index $k$
for spatial orbitals, and $K$ for the total number of spatial orbitals.

\subsection{Spin-projected matrix product states (SP-MPS)}\label{theory:spmps}
\begin{figure}
    \begin{tabular}{cc}
    {\resizebox{0.4\textwidth}{!}{\includegraphics{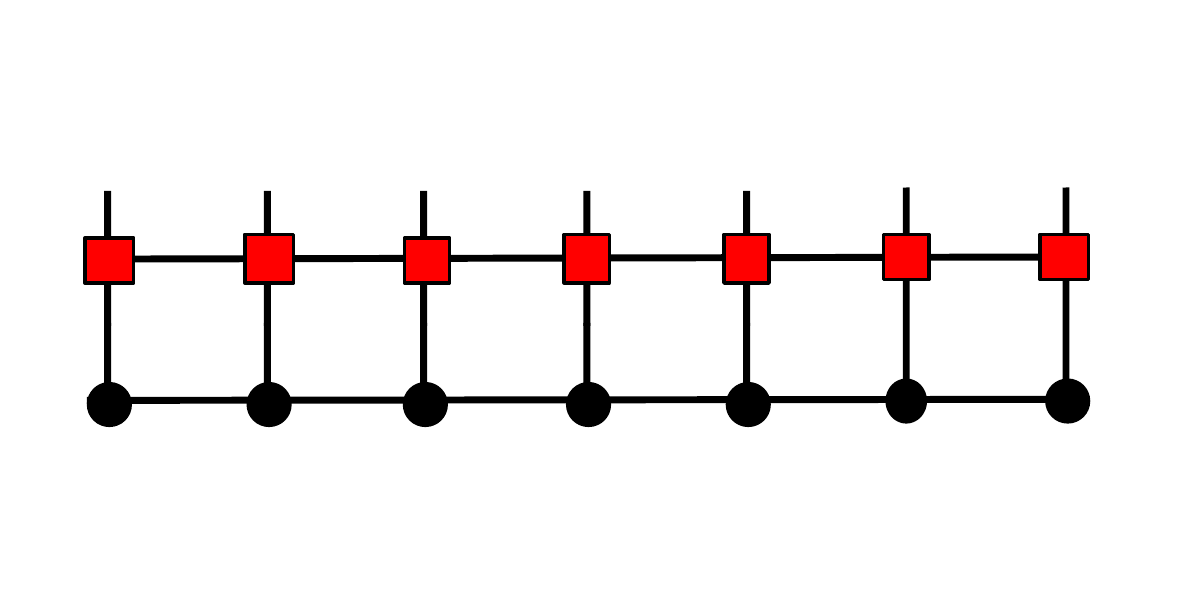}}} &
    {\resizebox{0.45\textwidth}{!}{\includegraphics{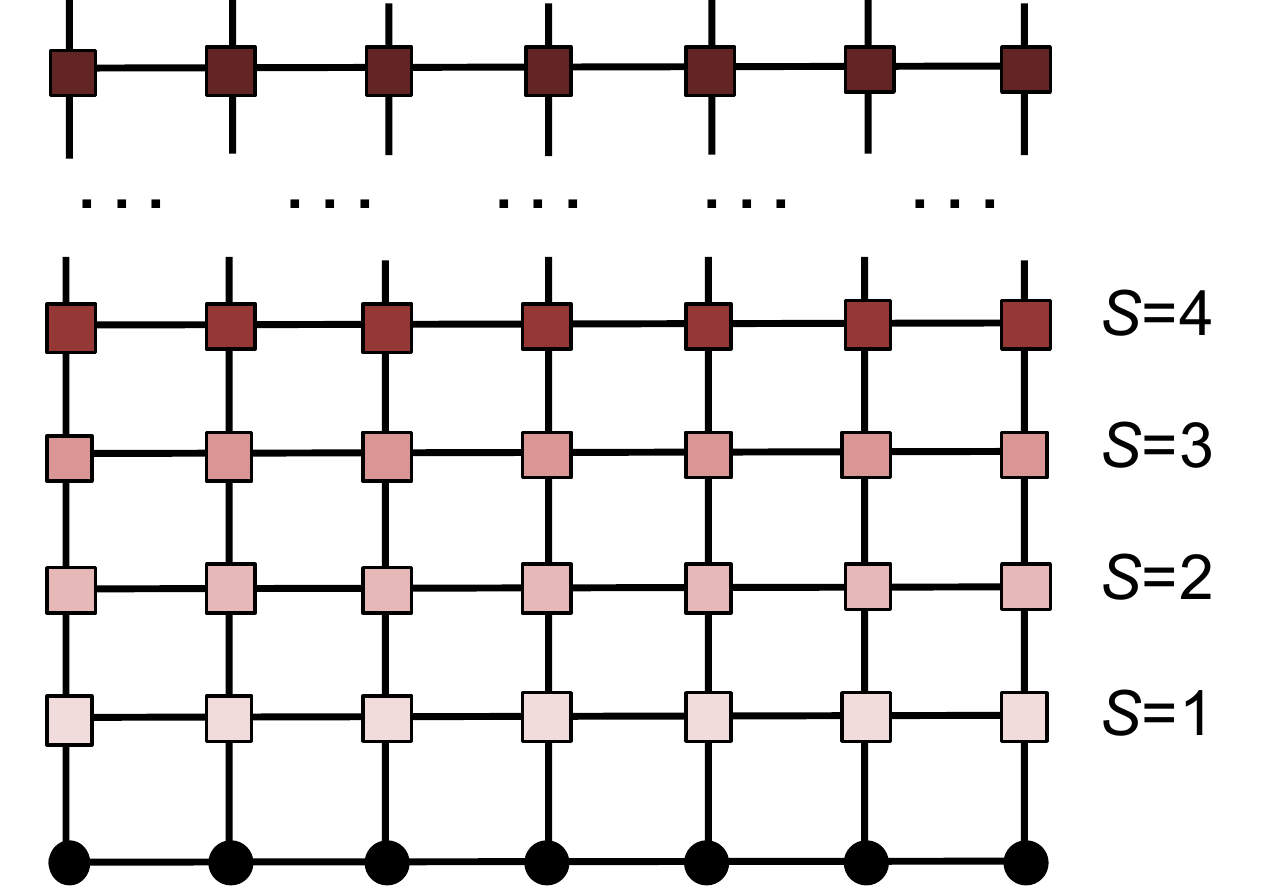}}}\\
    (a) Generic form of SP-MPS  &
    (b) SP-MPS with the L{\"o}wdin's projector $\mathcal{P}_S$ \\
    \end{tabular}
\caption{Graphical representations for SP-MPS: (a) a generic form of SP-MPS
where the MPO in red represents the projector $\mathcal{P}_S$,
(b) an example of SP-MPS with the L{\"o}wdin projector \eqref{lowdin}
for singlet states $S=0$, where each layer of MPOs represent one projector $\hat{P}_{S'}$ in
Eq. \eqref{lowdin}.}\label{fig:spmps}
\end{figure}

Based on the above MPS and MPO formalism, we can now introduce the SP-MPS wavefunction $|\Psi^{(N,S,M)}_{\spmps}\rangle$
as  resulting from acting a spin projector $\mathcal{P}_S$
on an MPS $|\Psi_{\mps}^{(N,M)}\rangle$ with given particle number
$N$ and spin projection $M$, viz.,
\begin{eqnarray}
|\Psi^{(N,S,M)}_{\spmps}\rangle=\mathcal{P}_S|\Psi_{\mps}^{(N,M)}\rangle,\label{SPMPS}
\end{eqnarray}
where its graphical representation is shown in Figure \ref{fig:spmps}(a).
It is possible to develop an even more general family of MPS via additional symmetry projections, such
as  particle number projection, which may allow to use smaller bond dimensions
in the underlying MPS to achieve the same accuracy. However,  since our purpose here is
mainly to circumvent the problems arising from SA-MPS as mentioned in Sec. \ref{introduction},
we will only consider the simplest case, where the underlying MPS
uses the physical Abelian symmetries of particle number and spin projection. We will also assume
the SODS scheme which helps to significantly reduce the bond dimension to represent the projector $\mathcal{P}_S$.
Physically, this means that the task of describing fluctuations around a classical broken symmetry determinant
is largely performed by the underlying MPS, similarly to as in the normal DMRG case. This is conceptually rather different
from other spin projected methods\cite{scuseria2011projected,jimenez2012projected,jimenez2013multi,jimenez2013excited,tsuchimochi2016communication,tsuchimochi2016black},
where the spin projection itself is essential for restoring fluctuations and correlation via
a  deliberate symmetry breaking and restoration mechanism.

There are various choices for the spin projector $\mathcal{P}_S$ in Eq. \eqref{SPMPS}.
L\"{o}wdin's spin projector\cite{lowdin1955quantum} takes the  form,
\begin{eqnarray}
\mathcal{P}_S=\prod_{S'\ne S}\hat{P}_{S'},\quad
\hat{P}_{S'}=\frac{\hat{S}^2-S'(S'+1)}{S(S+1)-S'(S'+1)},\label{lowdin}
\end{eqnarray}
where for given numbers of $\alpha$ and $\beta$ electrons,
assuming  $N_\alpha\ge N_\beta$, the allowed values for $S'$
range from $\frac{N_\alpha-N_\beta}{2}$ to $\frac{N_\alpha+N_\beta}{2}$, such that
the projector in Eq. \eqref{lowdin} is a product of $N_\beta$ terms.
The use of this operator is a formidable task in conjunction
with standard quantum chemistry methods due to its complicated operator form.
However, within the MPS and MPO formalism,
as long as the operators involved are local, the computation
is tractable regardless of the rank of the operators.
This is the case for the operator $\hat{S}^2$ in the projector \eqref{lowdin}.
Using $\hat{S}^2=\frac{1}{2}(\hat{S}_{+}\hat{S}_{-}+\hat{S}_{-}\hat{S}_+)+\hat{S}_z^2$ and following
the same recursive method for $\hat{H}$, one can find a compact
MPO representation for $\hat{S}^2$ with a bond dimension of only 5, viz.,
\begin{eqnarray}
\left[\begin{array}{c}
(\hat{S}^2)^{[k,K]} \\
\hline\hline
(\hat{S}_+)^{[k,K]} \\
\hline
(\hat{S}_-)^{[k,K]} \\
\hline
(\hat{S}_z)^{[k,K]} \\
\hline\hline
I^{[k,K]} \\
\end{array}\right]
=
\left[\begin{array}{c||c|c|c||c}
I_k & \hat{S}_{-,k} & \hat{S}_{+,k} & 2\hat{S}_{z,k} & \hat{S}^2_k \\
\hline\hline
0 & I_k & 0 & 0 & \hat{S}_{+,k} \\
\hline
0 & 0 & I_k & 0 & \hat{S}_{-,k} \\
\hline
0 & 0 & 0 & I_k & \hat{S}_{z,k} \\
\hline\hline
 0 & 0 & 0 & 0& I_k \\
\end{array}\right]
\left[\begin{array}{c}
(\hat{S}^2)^{[k+1,K]} \\
\hline\hline
(\hat{S}_+)^{[k+1,K]} \\
\hline
(\hat{S}_-)^{[k+1,K]} \\
\hline
(\hat{S}_z)^{[k+1,K]} \\
\hline\hline
I^{[k+1,K]} \\
\end{array}\right],\label{Sfac}
\end{eqnarray}
where both $k$ and $K$ represent indices for spatial orbitals.
It must be emphasized that this simplicity is associated
with the SODS scheme, while using the DODS scheme would lead to a more complicated
representation for $\hat{S}^2$. The matrix representations for the local operators  appearing
in Eq. \eqref{Sfac} in the space
$\spn\{|0\rangle,|k_{\alpha}\rangle\}
\otimes\spn\{|0\rangle,|k_{\beta}\rangle\}
=
\spn\{|0\rangle,|k_{\beta}\rangle,
|k_{\alpha}\rangle,|k_{\alpha}k_{\beta}\rangle\}$ are the same for all $k$, and are
\begin{eqnarray}
\protect[I] &=& \left[\begin{array}{cccc}
1 & 0 & 0 & 0 \\
0 & 1 & 0 & 0 \\
0 & 0 & 1 & 0 \\
0 & 0 & 0 & 1 \\
\end{array}\right],\quad
\protect[\hat{S}_{+}] = \left[\begin{array}{cccc}
0 & 0 & 0 & 0 \\
0 & 0 & 0 & 0 \\
0 & 1 & 0 & 0 \\
0 & 0 & 0 & 0 \\
\end{array}\right],\quad
\protect[\hat{S}_{-}] = \left[\begin{array}{cccc}
0 & 0 & 0 & 0 \\
0 & 0 & 1 & 0 \\
0 & 0 & 0 & 0 \\
0 & 0 & 0 & 0 \\
\end{array}\right], \nonumber\\
\protect[\hat{S}_{z}] &=& \left[\begin{array}{cccc}
0 & 0 & 0 & 0 \\
0 & -1/2 & 0 & 0 \\
0 & 0 & 1/2 & 0 \\
0 & 0 & 0 & 0 \\
\end{array}\right],\quad
\protect[\hat{S}^2] = \left[\begin{array}{cccc}
0 & 0 & 0 & 0 \\
0 & 3/4 & 0 & 0 \\
0 & 0 & 3/4 & 0 \\
0 & 0 & 0 & 0 \\
\end{array}\right].
\end{eqnarray}
Since adding a constant and multiplying by a factor to go from $\hat{S}^2$ to $\hat{P}_{S'}$ in Eq. \eqref{lowdin}
does not change the MPO bond dimension in Eq. \eqref{Sfac}, we can conclude that the bond dimension for the L{\"o}wdin projector is at most $5^{N_\beta}$, with its graphical representation shown in Figure \ref{fig:spmps}(b).
Although formally the bond dimension is exponential in $N_\beta$, in practice one can
expect that the wavefunction $\mathcal{P}_S|\Psi_{\mps}^{(N,M)}\rangle$
will be highly compressible at least for ground states,
in the sense that when one projector $\hat{P}_{S'}$ is applied
to $|\Psi_{\mps}^{(N,M)}\rangle$ with bond dimension $D$, the resulting wavefunction should be compressible
back to an MPS without increasing the bond dimension too much in order to
achieve a good accuracy, such as ca. 1mH in the ground state energies.
Thus, unlike with other quantum chemistry methods, the combination of the L{\"o}wdin projector with MPS is
in principle possible. However, this ansatz does not naturally fit into the
DMRG sweep optimization algorithm, and hence needs to be optimized by other techniques.

In this paper, we consider another form of projector that is more compatible with DMRG sweep optimization,
viz., the group theoretical projector\cite{percus1962exact},
\begin{eqnarray}
\mathcal{P}^S_{M,M'}=\frac{2S+1}{8\pi^2}\int d\Omega D_{M,M'}^{S*}(\Omega)\hat{R}(\Omega),\quad
\hat{R}(\Omega)=e^{-\ii\alpha \hat{S}_z}e^{-\ii \beta \hat{S}_y}e^{-\ii \gamma \hat{S}_z},\label{group}
\end{eqnarray}
where $\Omega=(\alpha,\beta,\gamma)$ are the Euler angles,
$\hat{R}(\Omega)$ is the rotation operator, $D_{M,M'}^{S}(\Omega)=\langle SM|\hat{R}(\Omega)|SM'\rangle
=e^{-\ii M\alpha}d_{M,M'}^{S}(\beta)e^{-\ii M'\gamma}$ is the
Wigner $D$-matrix, and $d_{M,M'}^{S}(\beta)$ is an element of Wigner's
small $d$-matrix, which can be chosen to be real for simplicity. The projector
$\mathcal{P}^S_{M,M'}$ \eqref{group} can then be rewritten as
\begin{eqnarray}
\mathcal{P}^S_{M,M'}&=&\mathcal{P}_{M}\hat{P}^S_{M,M'}\mathcal{P}_{M'},\\
\mathcal{P}_{M}&=& \frac{1}{2\pi}\int_{0}^{2\pi}d\gamma \;e^{\ii M S_z} e^{-\ii\gamma \hat{S}_z}, \\
\hat{P}^{S}_{M,M'}&=&\frac{2S+1}{2}\int_0^\pi d\beta\sin\beta d_{M,M'}^{S}(\beta)e^{-\ii\beta \hat{S}_y}.\label{PSMM}
\end{eqnarray}
The calligraphic symbols $\mathcal{P}^S_{M,M'}$ and $\mathcal{P}_{M}$ are used to indicate that
these operators are Hermitian idempotent projectors (if $M=M'$) in the $N$-electron Hilbert space, whereas $\hat{P}^S_{M,M'}$,
which induces local mixtures of $\alpha$ and $\beta$ orbitals, is not a projector.
To construct the SP-MPS using $\mathcal{P}^S_{M,M'}$ \eqref{group},
we require $M'=M$ in $|\Psi_{\mps}^{(N,M)}\rangle$ such that Eq. \eqref{SPMPS}
becomes,
\begin{eqnarray}
|\Psi^{(N,S,M)}_{\spmps}\rangle=
\mathcal{P}^S_{M,M}|\Psi_{\mps}^{(N,M)}\rangle=
\mathcal{P}_{M}\hat{P}^{S}_{M,M}|\Psi_{\mps}^{(N,M)}\rangle.\label{SPMPS2}
\end{eqnarray}
The energy to be variationally optimized can then be considered as a functional of the underlying MPS $|\Psi_{\mps}^{(N,M)}\rangle$, and its explicit functional form reads as,
\begin{eqnarray}
E[|\Psi_{\spmps}^{(N,S,M)}\rangle]&\equiv&E[|\Psi_{\mps}^{(N,M)}\rangle]\nonumber\\
&=&\frac{\langle\Psi^{(N,S,M)}_{\spmps}|\hat{H}|\Psi^{(N,S,M)}_{\spmps}\rangle}{\langle\Psi^{(N,S,M)}_{\spmps}|\Psi^{(N,S,M)}_{\spmps}\rangle}\nonumber\\
&=&\frac{\langle\Psi^{(N,M)}_{\mps}|\mathcal{P}^S_{M,M}
\hat{H}\mathcal{P}^S_{M,M}|\Psi^{(N,M)}_{\mps}\rangle}{\langle\Psi^{(N,M)}_{\mps}|\mathcal{P}^S_{M,M}|
\Psi^{(N,M)}_{\mps}\rangle}\nonumber\\
&=&\frac{\langle\Psi_{\mps}^{(N,M)}|\hat{H}\hat{P}^S_{M,M}|\Psi_{\mps}^{(N,M)}\rangle}{\langle\Psi_{\mps}^{(N,M)}|
\hat{P}^S_{M,M}|
\Psi_{\mps}^{(N,M)}\rangle},\label{efunctional}
\end{eqnarray}
where the fact that $\hat{H}$ is spin-free ($[\mathcal{P}^S_{M,M},\hat{H}]=0$) has been used,
and $\mathcal{P}_{M}$ has been dropped in the last two identities due to the left projection
to the bra state with a good quantum number $M$.
Before proceeding to the sweep algorithm
for optimizing $E[|\Psi_{\mps}^{(N,M)}\rangle]$, we should point out that
similar to other spin projected implementations\cite{scuseria2011projected,jimenez2012projected,jimenez2013multi,jimenez2013excited,tsuchimochi2016communication,tsuchimochi2016black},
 the integration in $\hat{P}^{S}_{M,M}$ is carried out in practice by numerical quadrature.
Specifically, we employ  Gauss-Legendre quadrature via the transformation $x=\cos\beta\in[-1,1]$,
\begin{eqnarray}
\hat{P}^S_{M,M}=\sum_{g=1}^{N_g}w_g d^S_{M,M}(\beta_g) e^{-\ii\beta_g \hat{S}_y},\label{pmpo}
\end{eqnarray}
where $N_g$ is the number of quadrature points. Eq. \eqref{pmpo} in the MPO language
becomes a sum of $N_g$ simple MPOs with bond dimension 1, since the exponential $e^{-\ii\beta \hat{S}_y}$
of a sum of local operators $\hat{S}_{y,k}$ is just a product of local operators (as they commute with each other)
$e^{-\ii\beta \hat{S}_y}= e^{-\ii\beta \sum_{k=1} \hat{S}_{y,k}}
=\prod_{k=1} e^{-\ii\beta \hat{S}_{y,k}}$, where the matrix representation
of $e^{-\ii\beta \hat{S}_{y,k}}$ is
\begin{eqnarray}
\protect[e^{-\ii\beta \hat{S}_{y,k}}]=
\left[\begin{array}{cccc}
1 & 0 & 0 & 0 \\
0 & c & s & 0 \\
0 & -s & c & 0\\
0 & 0 & 0 & 1 \\
\end{array}\right],\quad c=\cos(\beta/2),\quad s=\sin(\beta/2).\label{expSy}
\end{eqnarray}
From Eq. \eqref{expSy}, we note that only real algebra is needed
to implement Eq. \eqref{pmpo}, even though
the imaginary unit appears in the formulation.
For a system of $N$ electrons in $K$ spatial orbitals,
the quadrature \eqref{pmpo} is exact if $N_g$
is chosen to be at least
\begin{eqnarray}
N_g &=& \lceil(\Omega_{max}/2+S+1)/2\rceil,\nonumber\\
\Omega_{max}&=&\min(N_\alpha,K-N_\beta)+\min(N_\beta,K-N_\alpha)=
\left\{\begin{array}{cc}
N, & (N\le K) \\
2K-N, & (N>K)
\end{array}\right.,\label{exactQuadrature}
\end{eqnarray}
where $\Omega_{max}$ is the maximal seniority number (number of singly occupied orbitals) and $\lceil x\rceil$
is the ceiling function. Eq. \eqref{exactQuadrature} follows from
the Gauss quadrature rule, which is constructed
to be exact for polynomials of degree $2n-1$ or lower\cite{press2007numerical}
for an $n$-point quadrature, and
the observation that the integrand
in either $\langle\Psi_{\mps}^{(N,M)}|\hat{H}\hat{P}^S_{M,M}|\Psi_{\mps}^{(N,M)}\rangle$
or $\langle\Psi_{\mps}^{(N,M)}|\hat{P}^S_{M,M}|\Psi_{\mps}^{(N,M)}\rangle$ is a polynomial in $x=\cos\beta$, see Eq. \eqref{expSy}, whose maximal degree is $\Omega_{max}/2+S$ determined by the maximally singly occupied configurations.

Combined with the sum of MPOs representation for $\hat{H}$ \eqref{sumH},
the operator $\hat{H}\hat{P}^{S}_{M,M}$ in the numerator of
Eq. \eqref{efunctional} becomes a sum of MPOs with bond dimension
$2KN_g$. Distributing groups of MPOs to different processors leads to an embarrassingly parallel
scheme to compute expectation values over $|\Psi_{\mps}^{(N,M)}\rangle$,
which will be exploited in the following DMRG sweep optimizations.
Based on these MPO representations for both $\hat{H}$ and $\hat{P}^S_{M,M}$,
the energy functional \eqref{efunctional} possesses a very nice
graphical representation, as shown in Figure \ref{fig:spmpsOpt}(a), as a quotient of two
fully contracted tensor networks.

\subsection{Sweep algorithms for ground and excited states}\label{theory:algorithm}

\subsubsection{Ground-state DMRG optimization}
\begin{figure}
    \begin{tabular}{ccc}
    {\resizebox{0.3\textwidth}{!}{\includegraphics{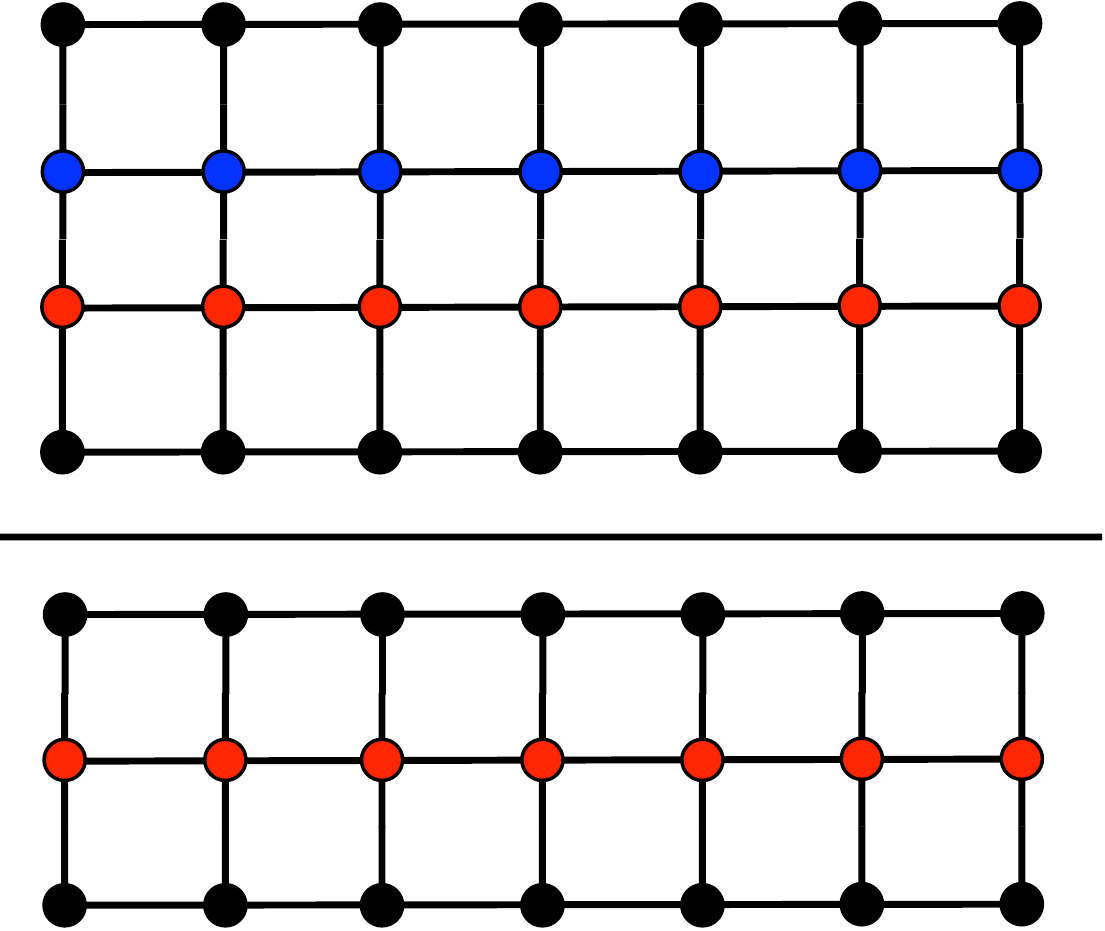}}} &
    {\resizebox{0.3\textwidth}{!}{\includegraphics{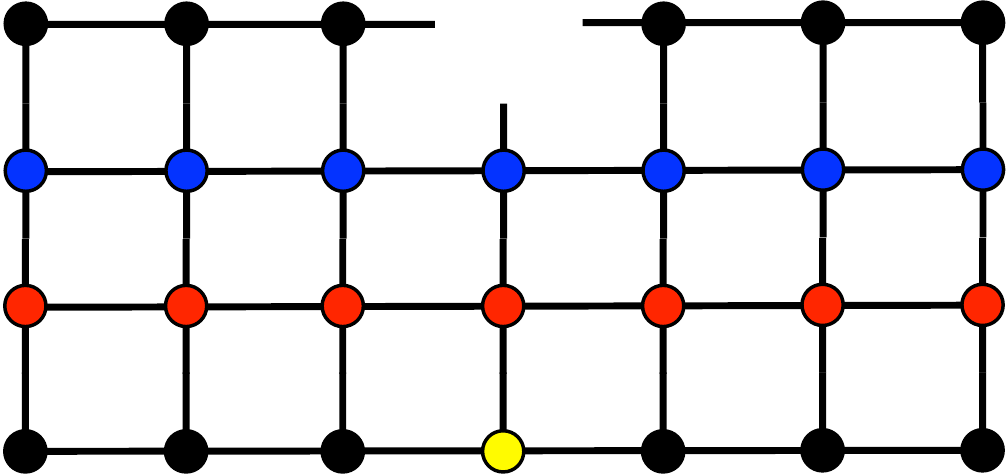}}} &
    {\resizebox{0.3\textwidth}{!}{\includegraphics{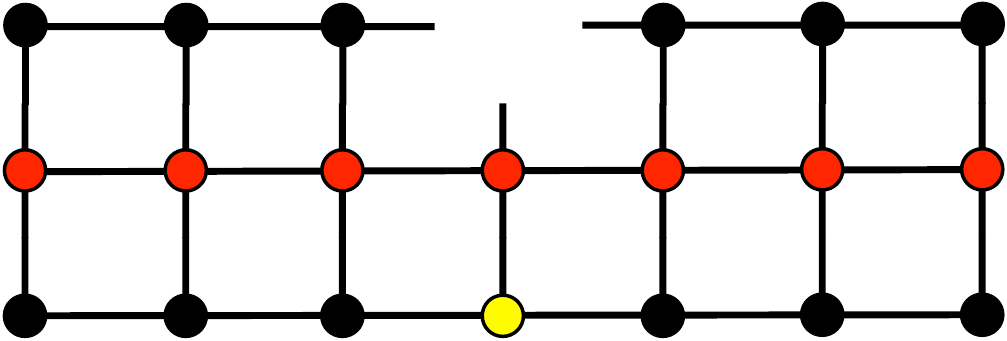}}}\\
    (a) $E[|\Psi_{\spmps}^{(N,S,M)}\rangle]=E[|\Psi_{\mps}^{(N,M)}\rangle]$ &
    (b) $\frac{\partial\langle\Psi_{\mps}^{(N,M)}|\hat{H}\hat{P}^{S}_{M,M}|\Psi_{\mps}^{(N,M)}\rangle}{\partial A^*[k]}=H_{\mathrm{eff}}A[k]$ &
    (c) $\frac{\partial\langle\Psi_{\mps}^{(N,M)}|\hat{P}^{S}_{M,M}|\Psi_{\mps}^{(N,M)}\rangle}{\partial A^*[k]}=N_{\mathrm{eff}}A[k]$ \\
    \end{tabular}
\caption{The ground state SP-MPS optimization problem: (a) the energy functional for SP-MPS, (b) and (c) for the left and right
 hand sides of Eq. \eqref{spmpsopt}, respectively.}\label{fig:spmpsOpt}
\end{figure}

The DMRG sweep algorithm can be used to minimize the
spin-projected energy functional \eqref{efunctional}. Specifically, assuming the one-site formalism for simplicity in this
discussion,
the optimization of the set of site tensors $A[k]$ is carried out one at a time,
and at site $k$, the local minimization problem corresponds to solving the
following stationary condition,
\begin{eqnarray}
\frac{\partial\langle\Psi_{\mps}^{(N,M)}|\hat{H}\hat{P}^{S}_{M,M}|\Psi_{\mps}^{(N,M)}\rangle}{\partial A^*[k]}
=\;E\;\frac{\partial\langle\Psi_{\mps}^{(N,M)}|\hat{P}^{S}_{M,M}|\Psi_{\mps}^{(N,M)}\rangle}{\partial A^*[k]},\label{spmpsopt}
\end{eqnarray}
which, when $A^{n_k}_{l_{k-1}r_{k}}[k]$ is viewed as a vector,
leads to a generalized eigenvalue problem,
\begin{eqnarray}
H_{\mathrm{eff}}A[k]=N_{\mathrm{eff}}A[k]E.\label{geproblem}
\end{eqnarray}
While the explicit algebraic form of the effective Hamiltonian $H_{\mathrm{eff}}$ and the metric $N_{\mathrm{eff}}$, involving numerous
sums of products, is very lengthy, the graphical representation introduced earlier
enables a very compact expression. This is depicted in
Figure \ref{fig:spmpsOpt}, where the yellow dot represents $A[k]$,
and deleting it from the Figures \ref{fig:spmpsOpt}(b) and \ref{fig:spmpsOpt}(c) leads to $H_{\mathrm{eff}}$ and
$N_{\mathrm{eff}}$, respectively.

We mention here that the metric $N_{\mathrm{eff}}$ does not arise from the choice
of gauge for the MPS, as it comes from the introduction of the projector,
as clearly shown by Figure \ref{fig:spmpsOpt}(c). This means that during the
optimization sweeps, we are free to use the mixed canonical form for the underlying MPS as usual,
which ensures that the renormalized configuration basis is orthonormal, leading to more
stable numerical algorithms.
This differs from the situation arising in the optimization of MPS with periodic boundary conditions (PBC)\cite{verstraete2004density,pippan2010efficient}, where the metric arises from the impossibility of choosing an orthonormal gauge due to the
cyclic structure of the MPS with PBC. Thus, the attractive feature of
SP-MPS is that almost all the usual DMRG machinery can be reused without any modification.
In particular, after solving the generalized eigenvalue problem \eqref{geproblem},
the site tensor can be chosen in left or right canonical form by using a SVD or density matrix renormalization
in exactly the same way as  in the usual DMRG. More importantly, since
the underlying MPS $|\Psi^{(N,M)}_{\mps}\rangle$ possesses Abelian symmetries only,
there is no need to use the singlet embedding scheme for non-singlet states,
and the one-dot algorithm naturally leads to a consistent MPS at convergence that fully minimizes
the energy functional with respect to all site tensors\eqref{efunctional}.

The computational cost for optimizing the SP-MPS scales as $O(N_g(D^3K^3+D^2K^4))$, which
is a factor of $N_{g}$ higher than the usual DMRG. By using the sum of MPO representations for
$\hat{H}$ \eqref{sumH} and $\hat{P}^S_{M,M}$ \eqref{pmpo},
the calculations can be  parallelized easily over up to $2KN_{p}$ processors,
where $K$ represents the number of spatial orbitals here. In the present pilot implementation,
the actual cost is higher than $N_g$ times that of a normal DMRG calculation. This is because
due to the presence of $\exp(-\ii \beta \hat{S}_y)$, the trial vector
for $|\Psi^{(N,M)}_{\mps}\rangle$ at a given site needs to first be subducted
to a lower symmetry with particle number symmetry only, and only after the application of
$\hat{H}\exp(-\ii \beta \hat{S}_y)$ to the trial vector, is the resulting vector projected back to
the space with symmetry $(N,M)$. Therefore, in the matrix-vector product
step, there is less symmetry to use compared with a normal DMRG calculation
with symmetries $(N,M)$. However, this can in principle be alleviated
by exploiting block data sparsity, since it is reasonable to imagine that as the bond dimension
becomes large in a finite system, the underlying state $|\Psi^{(N,M)}_{\mps}\rangle$ should be close
to the target state, such that the action of $\exp(-\ii \beta \hat{S}_y)$ will not
introduce too many states in the other $M$ sectors. This strategy for reducing
the computational cost will be explored in our future studies.

\subsubsection{State-specific excited-state optimizations}
\begin{figure}
    \begin{tabular}{cc}
    {\resizebox{0.3\textwidth}{!}{\includegraphics{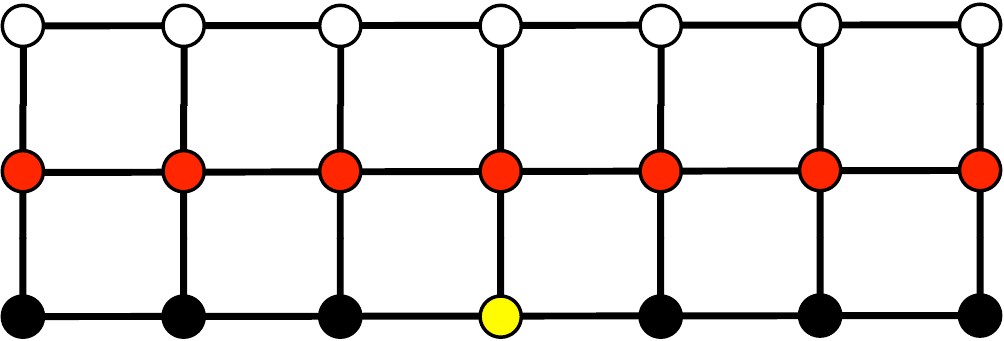}}} &
    {\resizebox{0.1\textwidth}{!}{\includegraphics{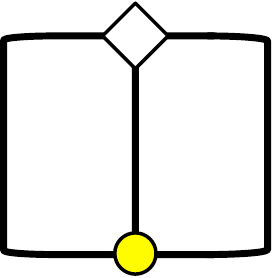}}} \\
    (a) $\langle \Psi_{\mps,0}^{(N,M)}|\hat{P}^{S}_{M,M}|\Psi_{\mps,1}^{(N,M)}\rangle=0$ & (b) $\sum_{l_{k-1}n_k r_{k}}B^{n_k}_{l_{k-1}r_{k}}[k]A^{n_k}_{l_{k-1}r_{k}}[k]=0$ \\
    \end{tabular}
\caption{Orthogonality constraint to be imposed for optimizing excited states with SP-MPS based on Eq. \eqref{orthoB}.}\label{fig:spmpsex}
\end{figure}

One  important feature of the combination of MPS with spin projection
is that compared with other spin projected methods based on Slater determinants,
it is more straightforward to compute excited states, due to the simplicity
in imposing orthogonality constraints using MPS. Specifically, to compute
the first excited state using SP-MPS, aside from using the state-averaged algorithm,
one can directly target the state by imposing
the constraint $\langle \Psi_{\spmps,0}^{(N,S,M)}|\Psi_{\spmps,1}^{(N,S,M)}\rangle=0$
between the excited state $|\Psi_{\spmps,1}^{(N,S,M)}\rangle$ to be optimized
and the ground state $|\Psi_{\spmps,0}^{(N,S,M)}\rangle$, which is assumed optimized
by the method introduced in the previous section. This simply requires
that at each local optimization step, a vector $B^{n_k}_{l_{k-1}r_{k}}[k]$ is constructed in the following way,
\begin{eqnarray}
0&=&\langle \Psi_{\spmps,0}^{(N,S,M)}|\Psi_{\spmps,1}^{(N,S,M)}\rangle=
\sum_{l_{k-1}n_k r_{k}}B^{n_k}_{l_{k-1}r_{k}}[k]A^{n_k}_{l_{k-1}r_{k}}[k],\nonumber\\
B[k]&\triangleq&
\langle \Psi_{\spmps,0}^{(N,S,M)}|\frac{\partial\Psi_{\spmps,1}^{(N,S,M)}}{\partial A[k]}\rangle
=\langle \Psi_{\mps,0}^{(N,M)}|\hat{P}^{S}_{M,M}|
\frac{\partial\Psi_{\mps,1}^{(N,M)}}{\partial A[k]}\rangle,\label{orthoB}
\end{eqnarray}
such that the local optimization based on Eq. \eqref{geproblem} for
$A[k]$ is subject to the constraint \eqref{orthoB}.
The graphical representation for such a condition is shown in Figure \ref{fig:spmpsex}.
It is seen that all the environmental tensors connected to $A[k]$ (in yellow
 in Figure \ref{fig:spmpsex}(a)) can be contracted into a three-way tensor $B[k]$
(open diamond in Figure \ref{fig:spmpsex}(b)). This constraint allows to tackle excited states
within the same symmetry as the ground state, and the generalization to multiple constraints is straightforward, viz., only a set of $B[k]$ vectors \eqref{orthoB} corresponding to each constraint needs to be constructed, then a projector $Q=1-VV^T$
to implement the set of orthogonality constraints can be defined within the set of orthonormal basis vectors $V$, which can be obtained from the QR decomposition of the set of $B[k]$ vectors, which are in general not orthonormal.

Finally, it should be mentioned that to reduce the cost of DMRG optimizations based on SP-MPS
for both ground and excited states, instead of solving the optimization problem \eqref{geproblem},
perturbation theory can be used to approximately decompose the original optimization problem
into a variational step plus a perturbation correction step.
The first step can be carried out with a small bond dimension $D_0$, while the latter
step can be performed with a large bond dimension $D_1$ but using a much simpler zeroth order Hamiltonian
to reduce the computational cost. Such an idea has been used in the MPSPT\cite{sharma2014communication}.
Although not fully explored in this paper, we present a possible generalization based on SP-MPS
in Appendix \ref{appendix1}.

\subsection{Properties}\label{theory:properties}
While we have shown that the SP-MPS can be easily used in every situation where the standard
SA-MPS is currently used, we emphasize that the SP-MPS is not intended to be a replacement for SA-MPS.
Rather, these two classes of MPS have completely different sets of merits and demerits, making
them quite complementary in their applicability. In particular,
the unique feature of SP-MPS is its closer connection with mean-field states,
which could be exploited, for example, in the future development of DMET\cite{knizia2012density,knizia2013density} for open-shell systems. To this end, we discuss here the evaluation of two essential ingredients in DMET using SP-MPS,
namely, the one-body reduced density matrix (1RDM) $\langle a_{p_\sigma}^\dagger a_{q_\tau}\rangle$ and the energy component for a fragment $X$ of the whole molecule defined as a sum of expectation values for $\hat{H}_p$,
viz., $e_X=\sum_{p\in X}\langle \hat{H}_p\rangle$.

For spin-independent (spin-free) operators $\hat{O}_{sf}$, due to the commutation relation
$[\mathcal{P}^S_{M,M},\hat{O}_{sf}]=0$, similarly to the energy functional \eqref{efunctional} the expectation value for $\hat{O}_{sf}$
can be simplified as
\begin{eqnarray}
\langle \hat{O}_{sf}\rangle
=
\frac{\langle\Psi_{\spmps}^{(N,S,M)}|\hat{O}_{sf}|\Psi_{\spmps}^{(N,S,M)}\rangle}
{\langle\Psi_{\spmps}^{(N,S,M)}|\Psi_{\spmps}^{(N,S,M)}\rangle}
=
\frac{\langle\Psi_{\mps}^{(N,M)}
|\hat{O}_{sf}\hat{P}^{S}_{M,M}|\Psi_{\mps}^{(N,M)}\rangle}
{\langle\Psi_{\mps}^{(N,M)}|\hat{P}^{S}_{M,M}|\Psi_{\mps}^{(N,M)}\rangle}.
\label{PropSpinFree}
\end{eqnarray}
Operators belonging to this case include $E_{pq}=\sum_{\sigma}a_{p\sigma}^\dagger a_{q\sigma}$, whose expectation value gives rise to
the spin-free 1RDM, and spin-spin correlation functions $\vec{S}_{X}\cdot\vec{S}_{Y}$.
For the energy component, if both the $\alpha$ and $\beta$ orbitals
for the same spatial orbital are selected in the same
group $X$, then $\sum_{p\in X}\hat{H}_p$ is also spin-free, such that the total energy \eqref{efunctional} can be rewritten as
\begin{eqnarray}
E[|\Psi_{\spmps}^{(N,S,M)}\rangle]=\sum_{p=1}^{K}e_p,\quad e_p=\sum_{\sigma}\frac{\langle\Psi_{\mps}^{(N,M)}|\hat{H}_{p_\sigma} \hat{P}^{S}_{M,M}|\Psi_{\mps}^{(N,M)}\rangle}{\langle\Psi_{\mps}^{(N,M)}|\hat{P}^{S}_{M,M}
|\Psi_{\mps}^{(N,M)}\rangle}.
\end{eqnarray}
This expression also fits into the parallelization scheme using the sum of MPOs representation,
meaning that each $e_p$ can be evaluated independently and in parallel.

The evaluation of expectation values of spin-dependent operators is in general more complicated than
for spin-free operators. One of the most important spin-dependent properties is the
spin density matrix for nonsinglet states
$\langle \Psi_{\spmps}^{(N,S,M)}|T_{pq}(1,0)|\Psi_{\spmps}^{(N,S,M)}\rangle$.
Another important property is the spin-orbit coupling matrix element $\langle \Psi_{\spmps,I}^{(N,S,M)}|H_{SOC}|\Psi_{\spmps,J}^{(N,S',M')}\rangle$ between two spin states $I$ and $J$,
which at the one-electron level only requires the transition density matrices of form
$\langle \Psi_{\spmps,I}^{(N,S,M)}|T_{pq}(1,\mu)|\Psi_{\spmps,J}^{(N,S',M')}\rangle$ ($\mu=1,0,-1$).
More specifically, assuming that only the high-spin reference MPS
$|\Psi_{\mps}^{(N,M=S)}\rangle$ is used for each SP-MPS, then only  transition density matrices
of the form $\langle \Psi_{\spmps,I}^{(N,S,M)}|T_{pq}(1,\mu)|\Psi_{\spmps,J}^{
(N,S'=S-\mu,M'=M-\mu)}\rangle$ need  be computed
by virtue of the Wigner-Eckart theorem\cite{li2013combining}.
These properties can be obtained in a straightforward approach using double integrations\cite{scuseria2011projected}
to discretize both the bra and ket spin projectors,
or by using a single integration based on \eqref{PropSpinFree},
in conjunction with higher-order spin-free density matrices\cite{luzanov1985calculating,gould1990spin}.
However, in both approaches, the computational scaling is higher
than that for evaluating the spin-free 1RDM \eqref{PropSpinFree}. Fortunately,
by using the transformation properties of $T_{pq}(1,\mu)$
under the action of spin rotations $\hat{R}(\Omega)$, the following expressions
for the spin-dependent one-body (transition) density matrices can be derived (for details, see Appendix \ref{appendix2}),
\begin{eqnarray}
\langle\Psi_{\spmps,I}^{(N,S,M=S)}|T_{pq}(1,1)|\Psi_{\spmps,J}^{(N,S'=S-1,M'=S-1)}\rangle&=&
\frac{2S-1}{2S+1}
\langle\Psi_{\mps,I}^{(N,M=S)}|T_{pq}(1,1)\hat{P}^{S-1}_{S-1,S-1}|\Psi_{\mps,J}^{(N,M'=S-1)}\rangle,\;\; S\ge 1,\label{SpinTpqA}\\
\langle\Psi_{\spmps,I}^{(N,S,M=S)}|T_{pq}(1,0)|\Psi_{\spmps,J}^{(N,S'=S,M'=S)}\rangle&=&
-\frac{\sqrt{S}}{S+1}
\langle\Psi_{\mps,I}^{(N,M=S)}|T_{pq}(1,1)\hat{P}^{S}_{S-1,S}|\Psi_{\mps,J}^{(N,M'=S)}\rangle\nonumber\\
&&+
\frac{S}{S+1}
\langle\Psi_{\mps,I}^{(N,M=S)}|T_{pq}(1,0)\hat{P}^{S}_{S,S}|\Psi_{\mps,J}^{(N,M'=S)}\rangle,\;\; S\ge 1/2,\label{SpinTpqB}\\
\langle\Psi_{\spmps,I}^{(N,S,M=S)}|T_{pq}(1,-1)|\Psi_{\spmps,J}^{(N,S'=S+1,M'=S+1)}\rangle&=&
\frac{1}{\sqrt{(S+1)(2S+1)}}
\langle\Psi_{\mps,I}^{(N,M=S)}|T_{pq}(1,1)\hat{P}^{S+1}_{S-1,S+1}|\Psi_{\mps,J}^{(N,M'=S+1)}\rangle\nonumber\\
&&-\frac{1}{\sqrt{S+1}}
\langle\Psi_{\mps,I}^{(N,M=S)}|T_{pq}(1,0)\hat{P}^{S+1}_{S,S+1}|\Psi_{\mps,J}^{(N,M'=S+1)}\rangle\nonumber\\
&&+
\langle\Psi_{\mps,I}^{(N,M=S)}|T_{pq}(1,-1)\hat{P}^{S+1}_{S+1,S+1}|\Psi_{\mps,J}^{(N,M'=S+1)}\rangle,\; S\ge 0,\label{SpinTpqC}
\end{eqnarray}
which are sufficient for the state interaction treatment of spin-orbit coupling\cite{li2013combining,sayfutyarova2016state}
as well as the computation of spin density matrices (with Eq. \eqref{SpinTpqB})
for nonsinglet states. These formulae show that rather than changing the computational scaling,
the evaluation of spin-dependent properties within SP-MPS only changes the prefactor
by a small factor compared with the evaluation of spin-free properties.

\section{Numerical examples}\label{results}
\subsection{Two-dimensional Hubbard model}
\begin{figure}
    \begin{tabular}{ccc}
    {\resizebox{0.34\textwidth}{!}{\includegraphics{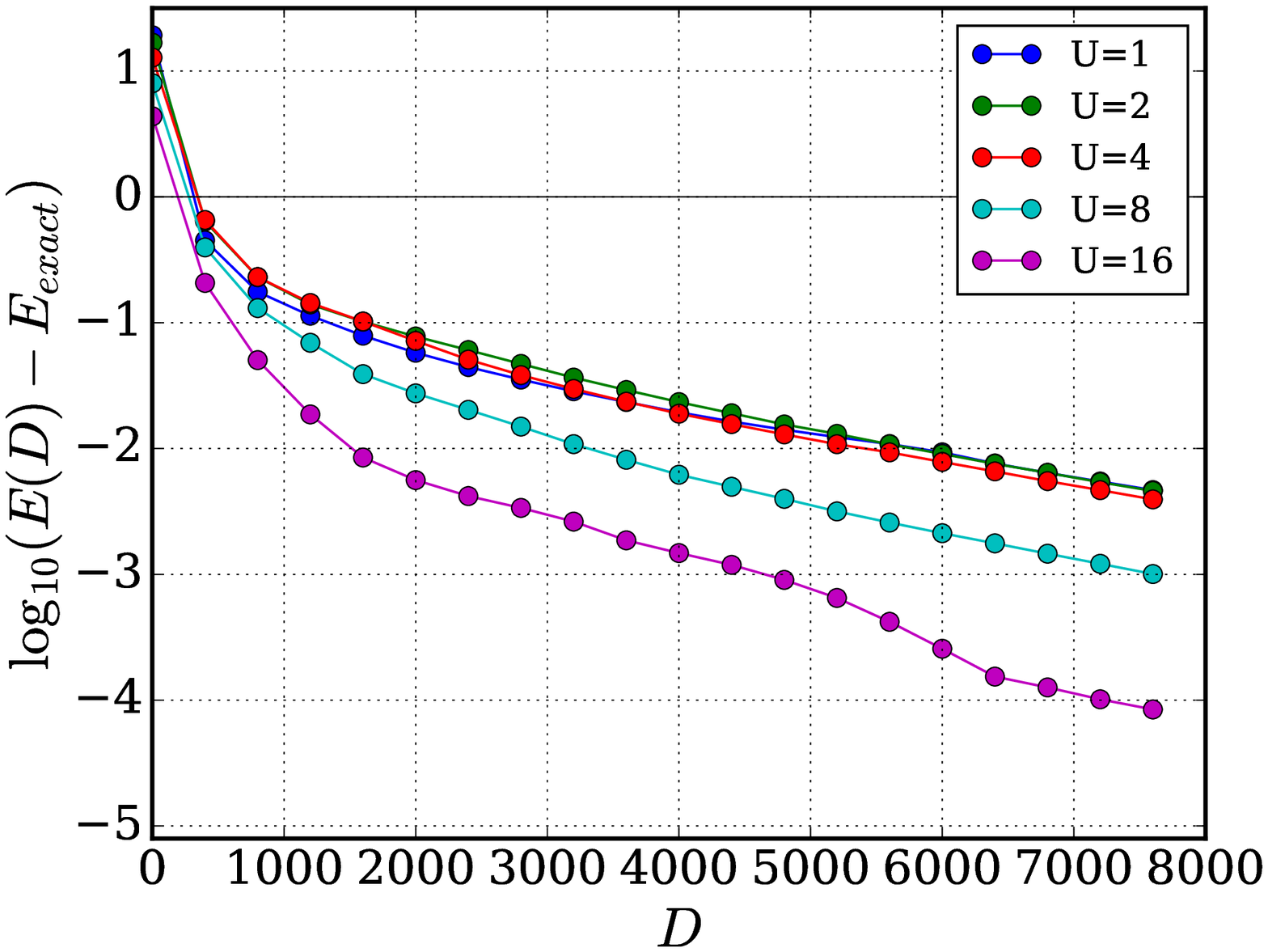}}} &
    {\resizebox{0.34\textwidth}{!}{\includegraphics{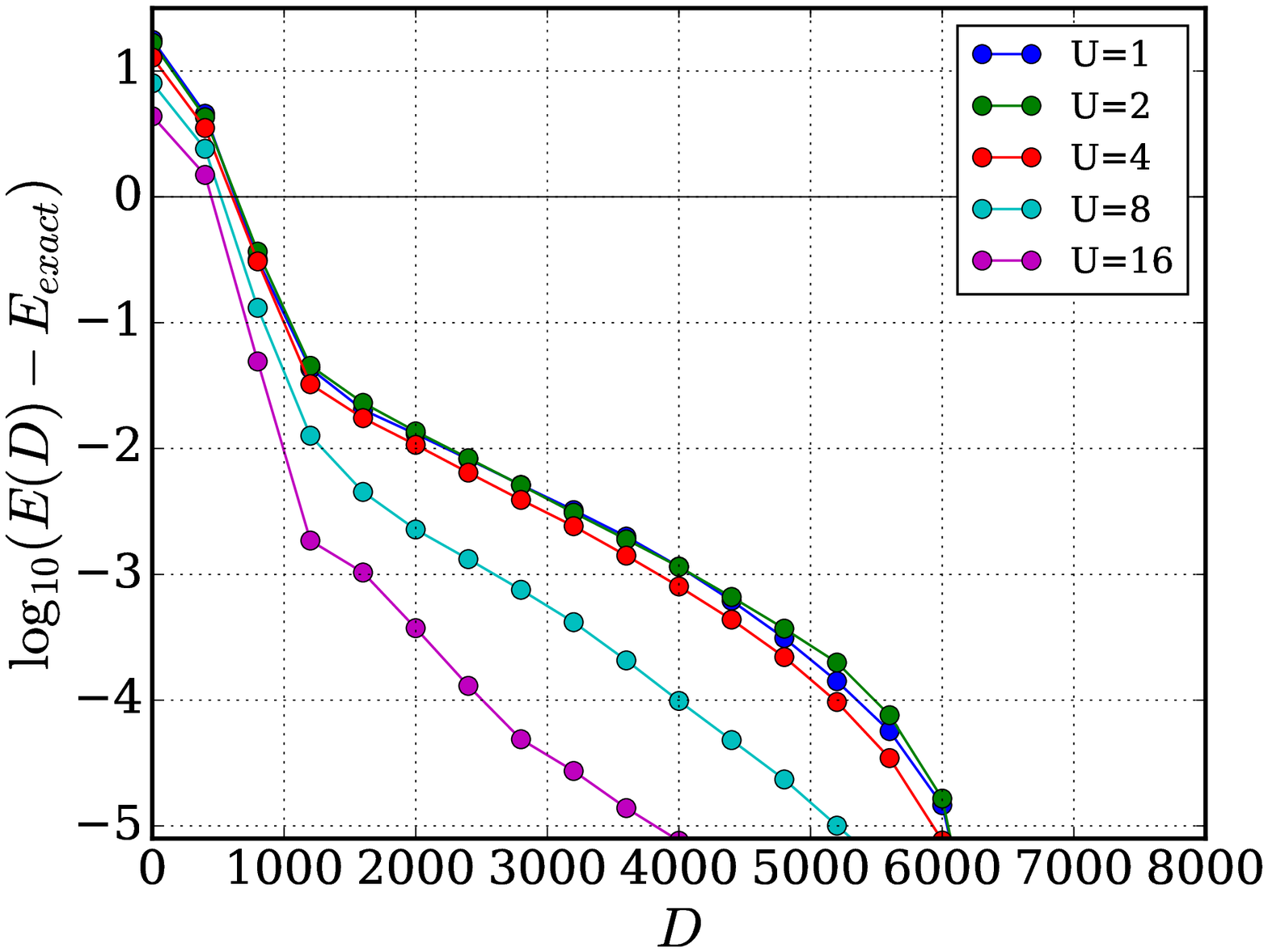}}} &
    {\resizebox{0.34\textwidth}{!}{\includegraphics{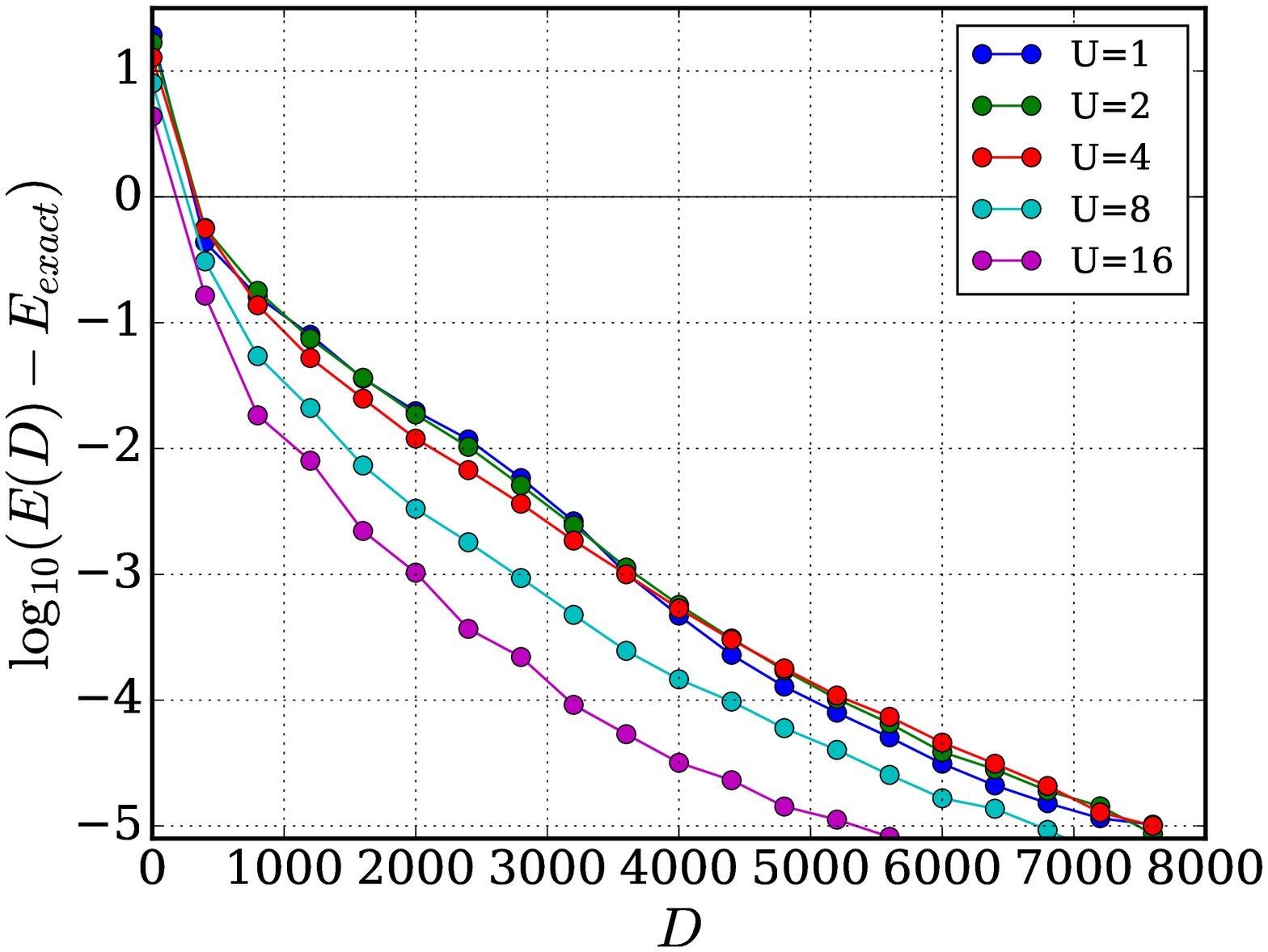}}} \\
    (a) MPS without spin-adaptation & (b) SA-MPS & (c) SP-MPS
    \end{tabular}
\caption{Energy convergence as a function of the bond dimension $D$
for the two-dimensional Hubbard model on a 4$\times$4 cluster at half-filling with
different values of $U$ for three different kinds of MPS.}\label{fig:hubbard2Denergy}
\end{figure}

\begin{figure}
    \begin{tabular}{ccc}
    {\resizebox{0.34\textwidth}{!}{\includegraphics{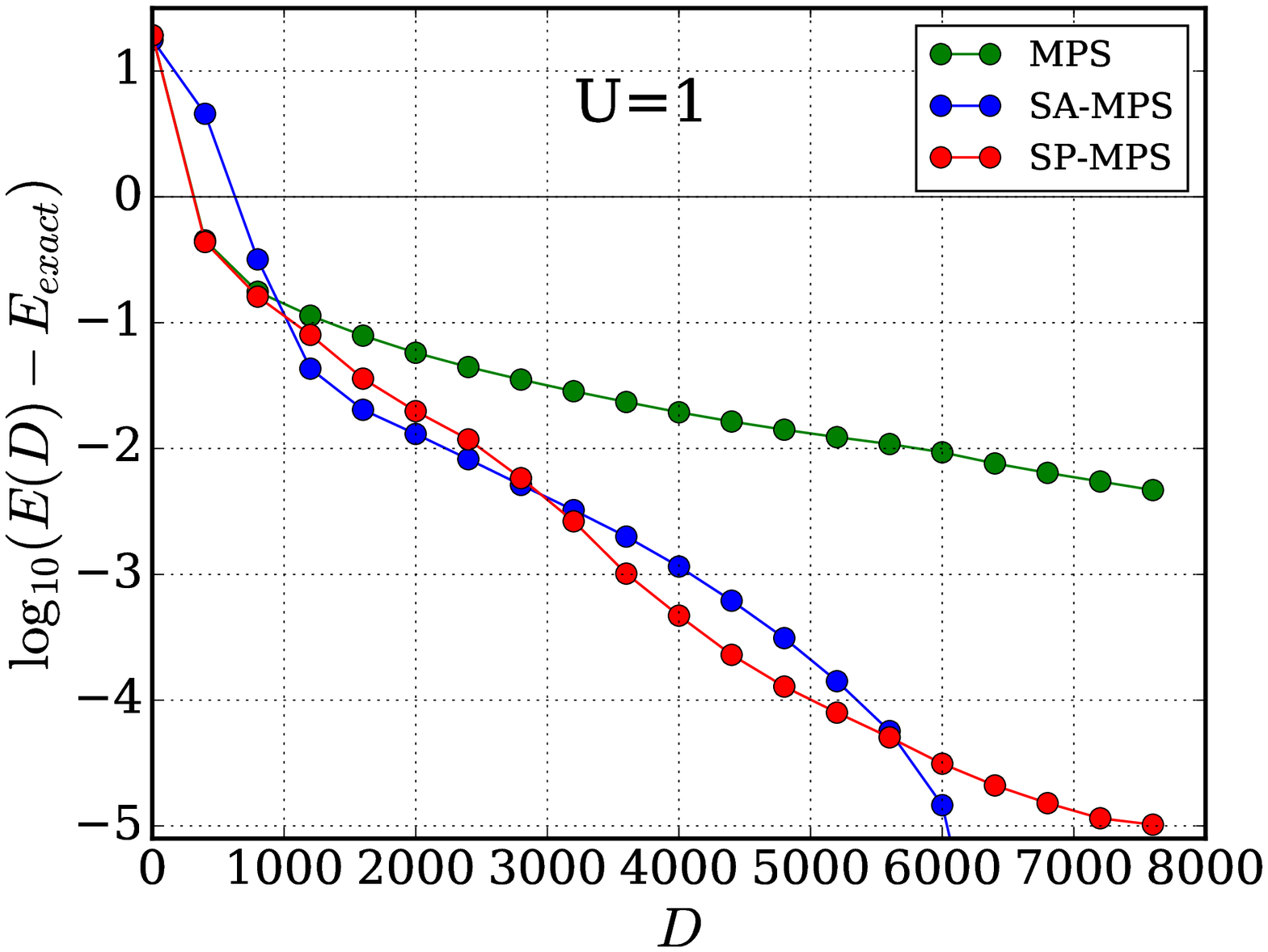}}} &
    {\resizebox{0.34\textwidth}{!}{\includegraphics{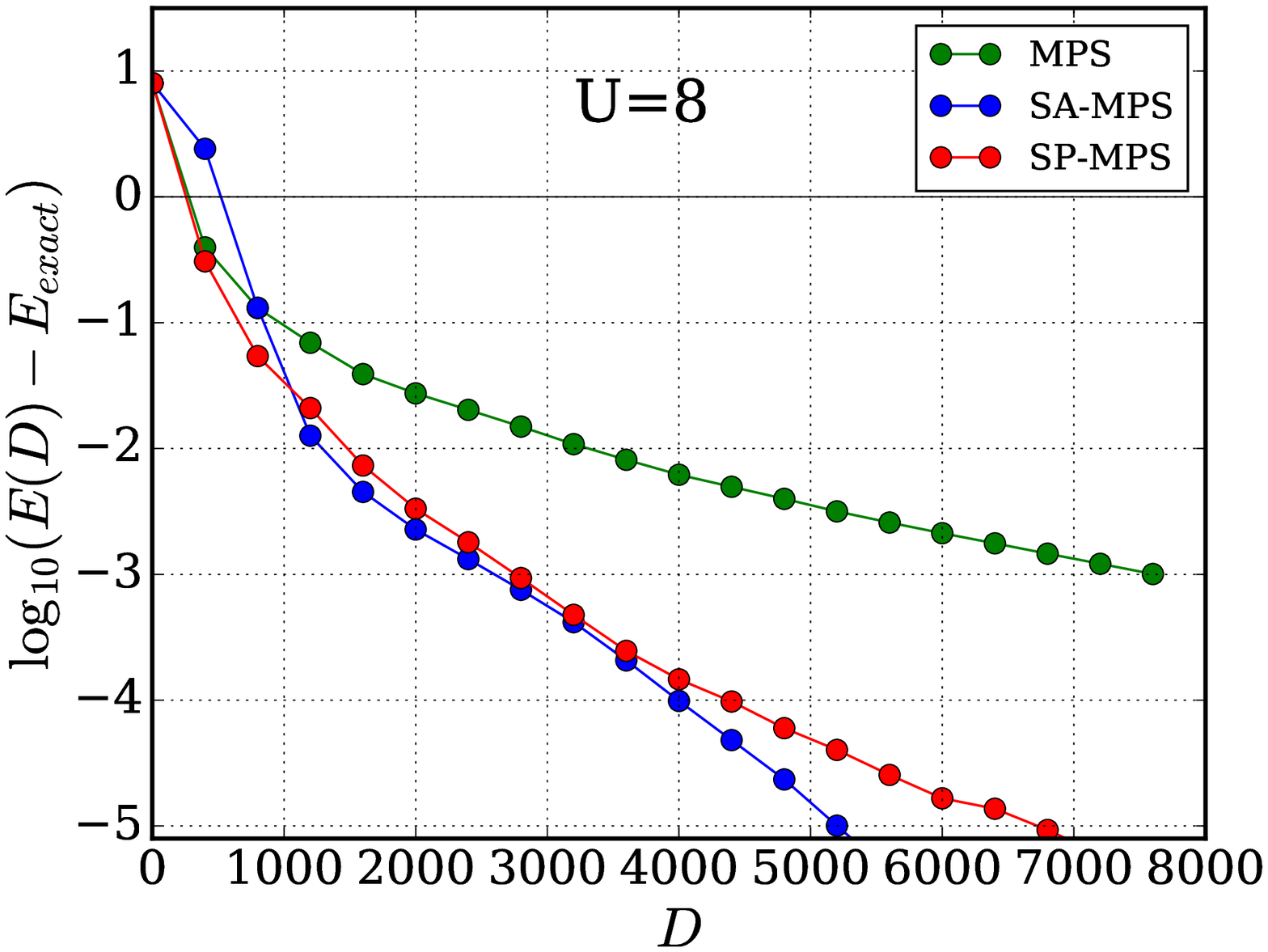}}} &
    {\resizebox{0.34\textwidth}{!}{\includegraphics{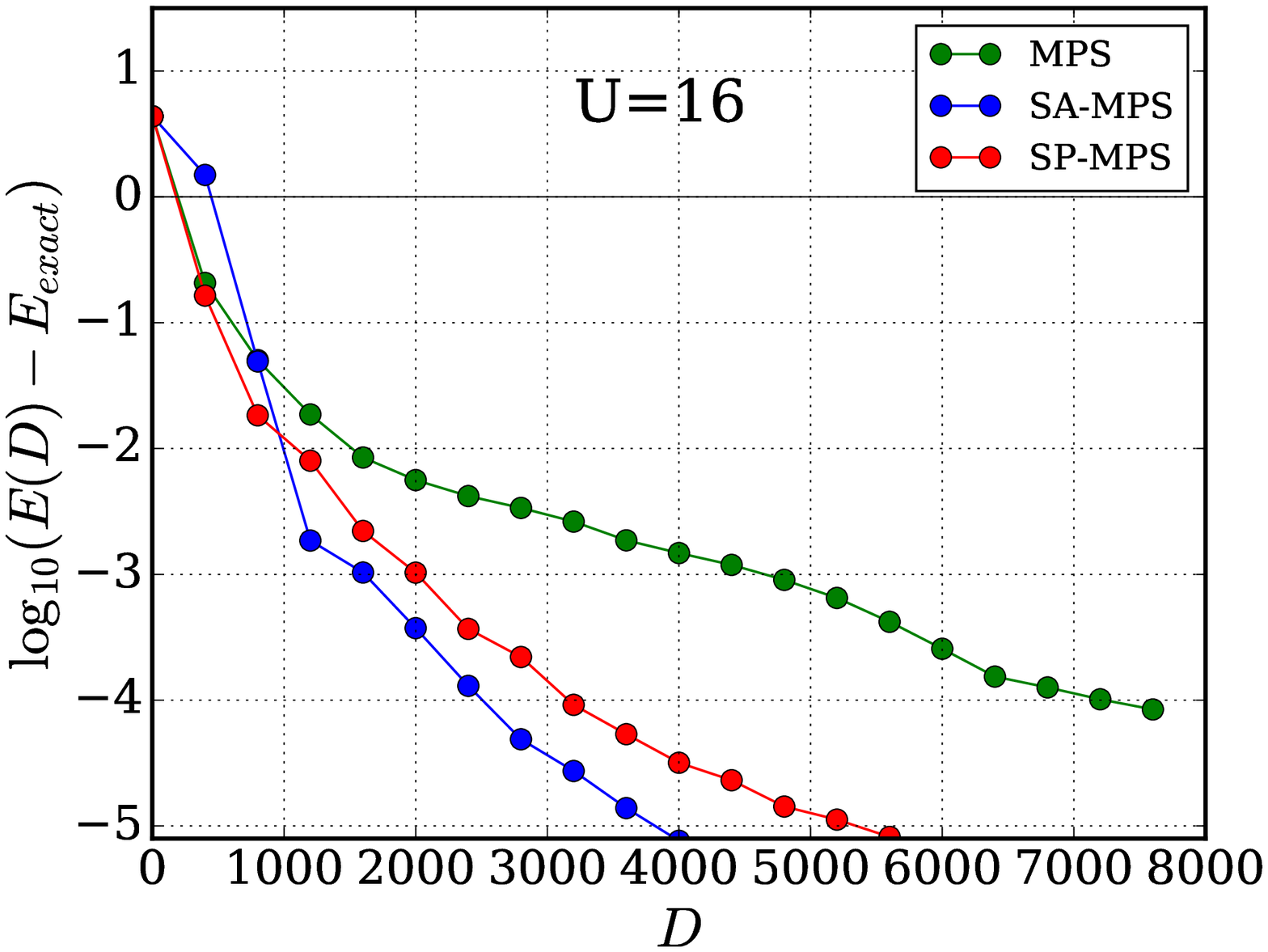}}} \\
    (a) $U=1$ & (b) $U=8$ & (c) $U=16$
    \end{tabular}
\caption{Comparison of the convergence of three different kinds of MPS
for three typical values of $U$.}\label{fig:hubbard2DdifferentU}
\end{figure}

As a proof-of-principle calculation, we consider the ground state of
the two-dimensional Hubbard model on a 4$\times$4 cluster with the Hamiltonian
$\hat{H}_{\mathrm{Hubbard}}\triangleq-t\sum_{\langle ij\rangle}\sum_{\sigma}(a_{i\sigma}^\dagger a_{j\sigma}+H.c.)
+U\sum_{i}n_{i\alpha}n_{i\beta}$, where $\langle ij\rangle$ represents
the sum over nearest neighbor pairs, for various values of $U$ ($U=1,2,4,8,16$, $t=1$) and at half-filling.
This model can be solved by exact diagonalization\cite{fano1990hole}.
Here we use it  to compare the performance of
MPS without spin-adaptation (denoted MPS for brevity), spin-adapted MPS (SA-MPS), and spin-projected
MPS (SP-MPS) for the singlet ground state.
For SP-MPS, to focus on the representational properties of SP-MPS,
we eliminate the angular integration error in \eqref{pmpo} by using
$N_{g}=5$ in the numerical quadrature, which is exact
in this model according to Eq. \eqref{exactQuadrature}.
The convergence of the ground state energies
using these three kinds of MPS in the site basis (ordered in a zigzag ordering for rows) as the bond dimension $D$
is increased is shown in Figure \ref{fig:hubbard2Denergy}.
It is clear that without spin adaptation, the
convergence is painfully slow, see Figure \ref{fig:hubbard2Denergy}(a).
For instance, it is not possible to converge to 10$^{-5}$ ($t$) even for
$D$ greater than 7000. As shown in Figure \ref{fig:hubbard2Denergy}(b),
 spin adaptation significantly improves the convergence
with respect to the bond dimension due to the use of reduced
renormalized states. It can be seen from Figure \ref{fig:hubbard2Denergy}(c) that the SP-MPS also
accelerates the convergence as a result of the spin projection.
To obtain a better comparison of SA-MPS and SP-MPS, the convergence for $U$=1, 8, and 16 is compared
in Figure \ref{fig:hubbard2DdifferentU}. In general, we find
 that SP-MPS achieves an accuracy of 10$^{-5}$ in the energy
with a bond dimension 1.3-1.4 times that needed with SA-MPS.
For small $U$(=1), the SP-MPS yields  lower energies than SA-MPS for
bond dimensions in the range 4000-5000, whereas for large $U$,
while both SA-MPS and SP-MPS converge faster than in the corresponding
$U=1$ case and significantly better than MPS without spin adaptation,
the SA-MPS generally tends to perform better than SP-MPS.

\begin{figure}
    \begin{tabular}{ccc}
    {\resizebox{0.34\textwidth}{!}{\includegraphics{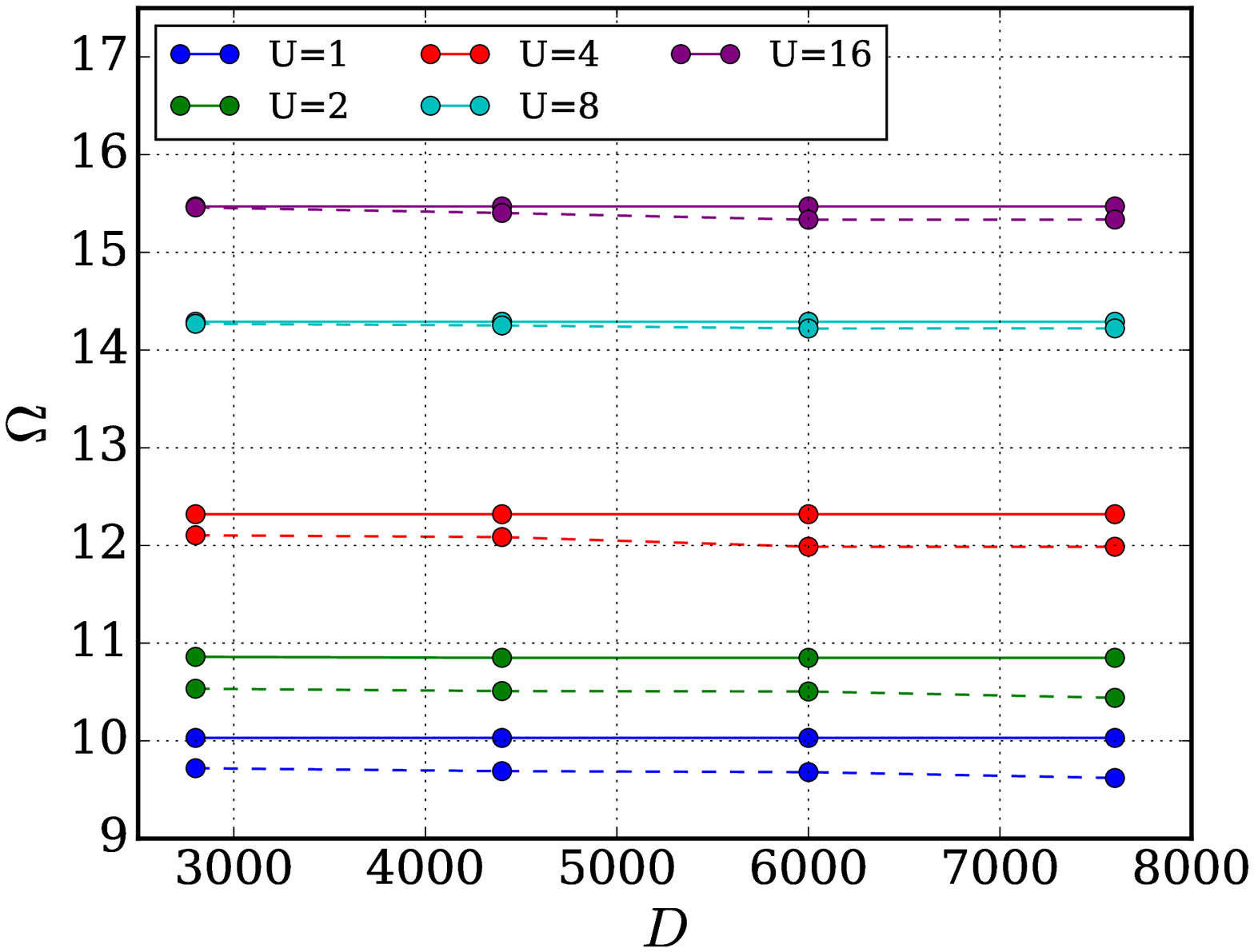}}} &
    {\resizebox{0.34\textwidth}{!}{\includegraphics{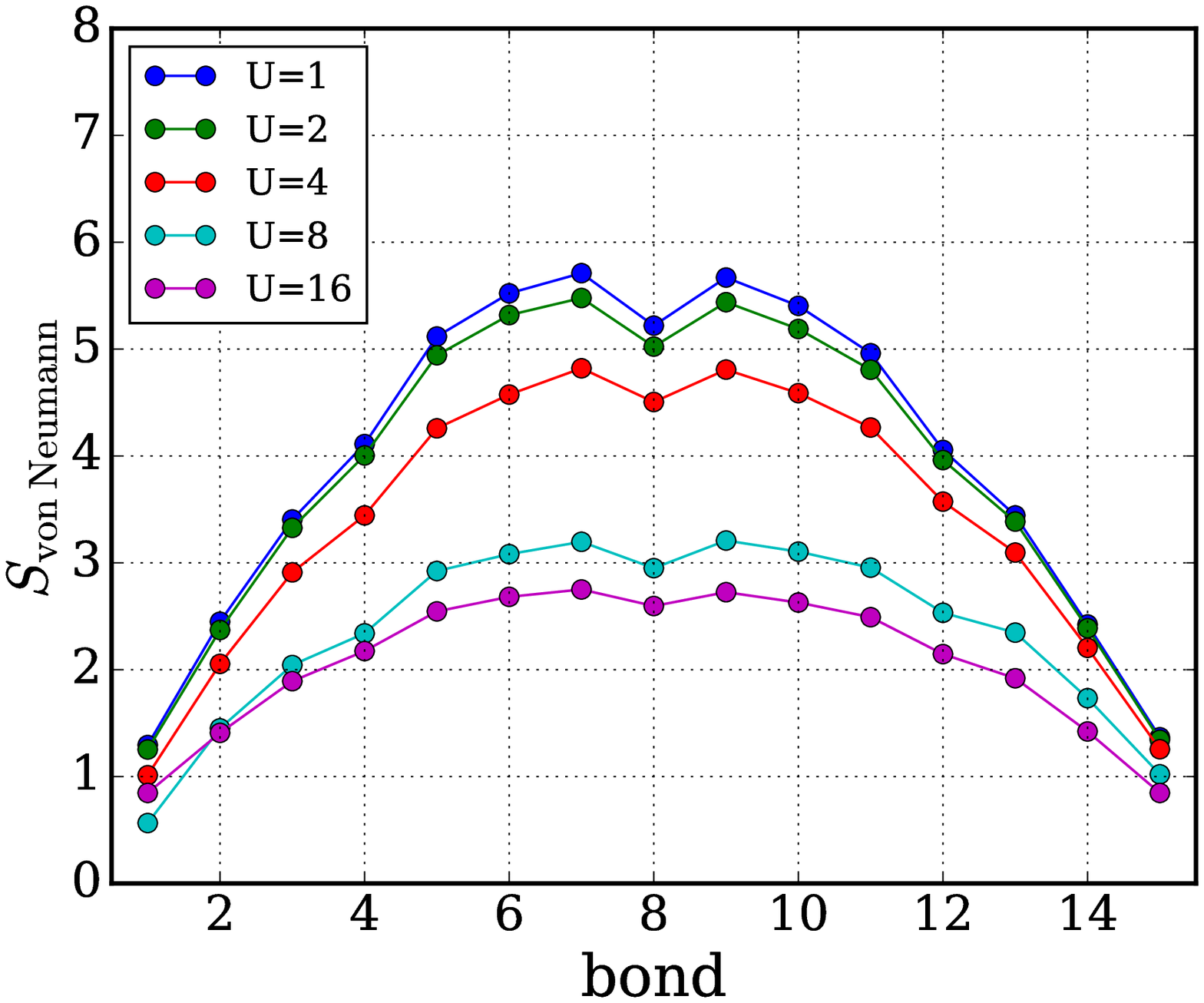}}} &
    {\resizebox{0.34\textwidth}{!}{\includegraphics{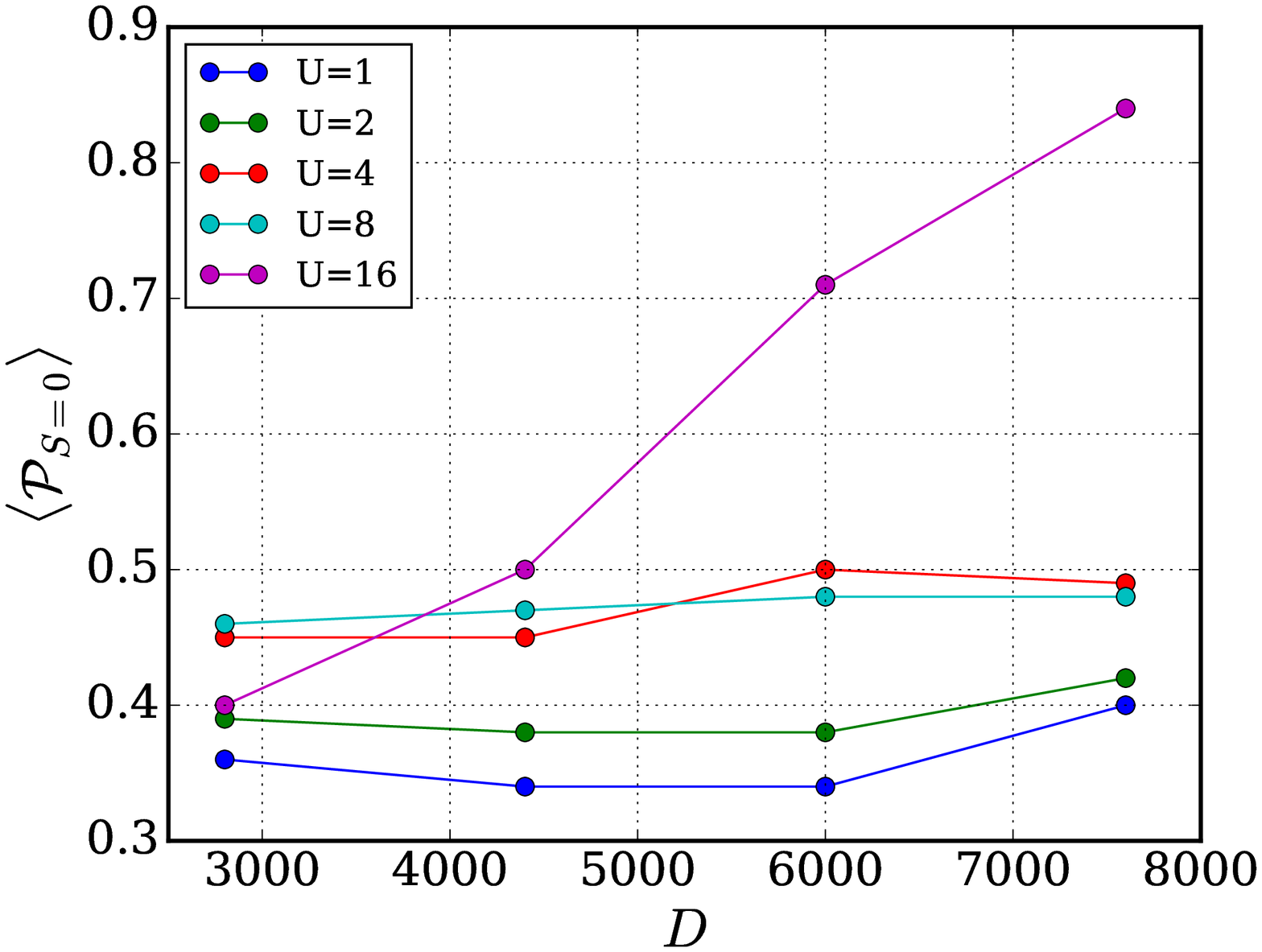}}} \\
    (a) seniority number $\Omega$ & (b) entropy $S_{\mathrm{von\;Neumann}}$ & (c) singlet component $\langle\mathcal{P}_{S=0}\rangle$
    \end{tabular}
\caption{Representative properties of SP-MPS and its
underlying MPS: (a) seniority numbers for
SP-MPS (solid) and underlying MPS (dashed) with
several bond dimensions, (b) von Neumann entropy $S_{\mathrm{von\;Neumann}}$
of the underlying MPS with $D=7600$ at each bond,
(c) singlet components of the underlying MPS for
different values of $U$ and bond dimensions $D$.}\label{fig:hubbard2Dproperty}
\end{figure}

To understand such differences, we can compute various properties
of the converged SP-MPS wavefunctions.
Figure \ref{fig:hubbard2Dproperty}(a) plots
the expectation value of the seniority number operator $\langle\hat{\Omega}\rangle$
for SP-MPS (solid lines) and its underlying MPS (dashed lines). The operator $\hat{\Omega}$
is used to measure the number of singly occupied (open-shell) orbitals, which is
also a sum of local operators,
\begin{eqnarray}
\hat{\Omega}=\sum_k\hat{\Omega}_k,\quad
\protect[\hat{\Omega}_k] &=& \left[\begin{array}{cccc}
0 & 0 & 0 & 0 \\
0 & 1 & 0 & 0 \\
0 & 0 & 1 & 0 \\
0 & 0 & 0 & 0 \\
\end{array}\right],
\end{eqnarray}
and hence can be written as an MPO with bond dimension 2.
Figure \ref{fig:hubbard2Dproperty}(b) displays the
von Neumann entropy defined by
$S_{\mathrm{von\;Neumann}}=-\tr(\rho\ln\rho)=-\sum_{i}\lambda_i\ln\lambda_i$ of
the underlying MPS with bond dimension 7600 at each bond, where
$\rho$ is the reduced density matrix of the system or environment, and $\lambda_i$
is its associated eigenvalue. In Figure \ref{fig:hubbard2Dproperty}(c),
the weight of the singlet components in the underlying MPS $\langle \mathcal{P}_{S=0}\rangle
\triangleq\langle\Psi^{(N,M=0)}_{\mps}|\mathcal{P}^{S=0}_{M=0,M=0}
|\langle\Psi^{(N,M=0)}_{\mps}\rangle$,
which is also the overlap between SP-MPS, that is, the denominator of
the energy functional \eqref{efunctional}, is compared for different values of $U$.
According to Figure \ref{fig:hubbard2Dproperty}(a),
the number of open-shell orbitals
decreases as $U$ decreases due to the enhanced
hopping to other sites, and the underlying
state in the site basis becomes more entangled for small $U$,
as demonstrated by the increased $S_{\mathrm{von\;Neumann}}$
shown in Figure \ref{fig:hubbard2Dproperty}(b).
[NB: We emphasize that the definition of entanglement
depends on the one-particle basis used to define the partitioning
when computing $S_{\mathrm{von\;Neumann}}$. Thus,
while the large $U$ case is more entangled in the
momentum or mean-field basis, the small $U$ case
is more entangled with the site basis employed here.]
In such a situation, the underlying MPS tends to break spin symmetry,
as demonstrated by the small values of the singlet
component (0.3-0.4) in Figure \ref{fig:hubbard2Dproperty}(c) at
small $U$, in order to describe the increased
entanglement in the ground state wavefunctions and
to recover more correlation energy.
Therefore, we can expect the SP-MPS
to perform better than SA-MPS in highly entangled situations,
and to become less superior in less entangled systems.

\subsection{Iron-sulfur clusters}
Having established the basic convergence properties and representational power of SP-MPS,
we consider two practical applications of SP-MPS to iron-sulfur
clusters. These have been discussed as typical of challenging
strongly correlated systems, because they have competing low-energy states corresponding to
different kinds of spin and charge fluctuations\cite{sharma2014communication}.
The first application is to \ce{[Fe2S2(SCH3)4]^{2-}}, where
two different initial guesses and two different active spaces are constructed
to illustrate the ability of SP-MPS to describe the correct
spin states. The second is to a larger cluster with four iron atoms \ce{[Fe4S4(SCH3)4]^{2-}}.
Here we examine the possibility of using different broken symmetry initial guesses in conjunction
with SP-MPS with very small bond dimensions, to construct a map of the physically relevant
low energy states in the Hilbert space.

\begin{figure}
    \begin{tabular}{c}
    {\resizebox{0.5\textwidth}{!}{\includegraphics{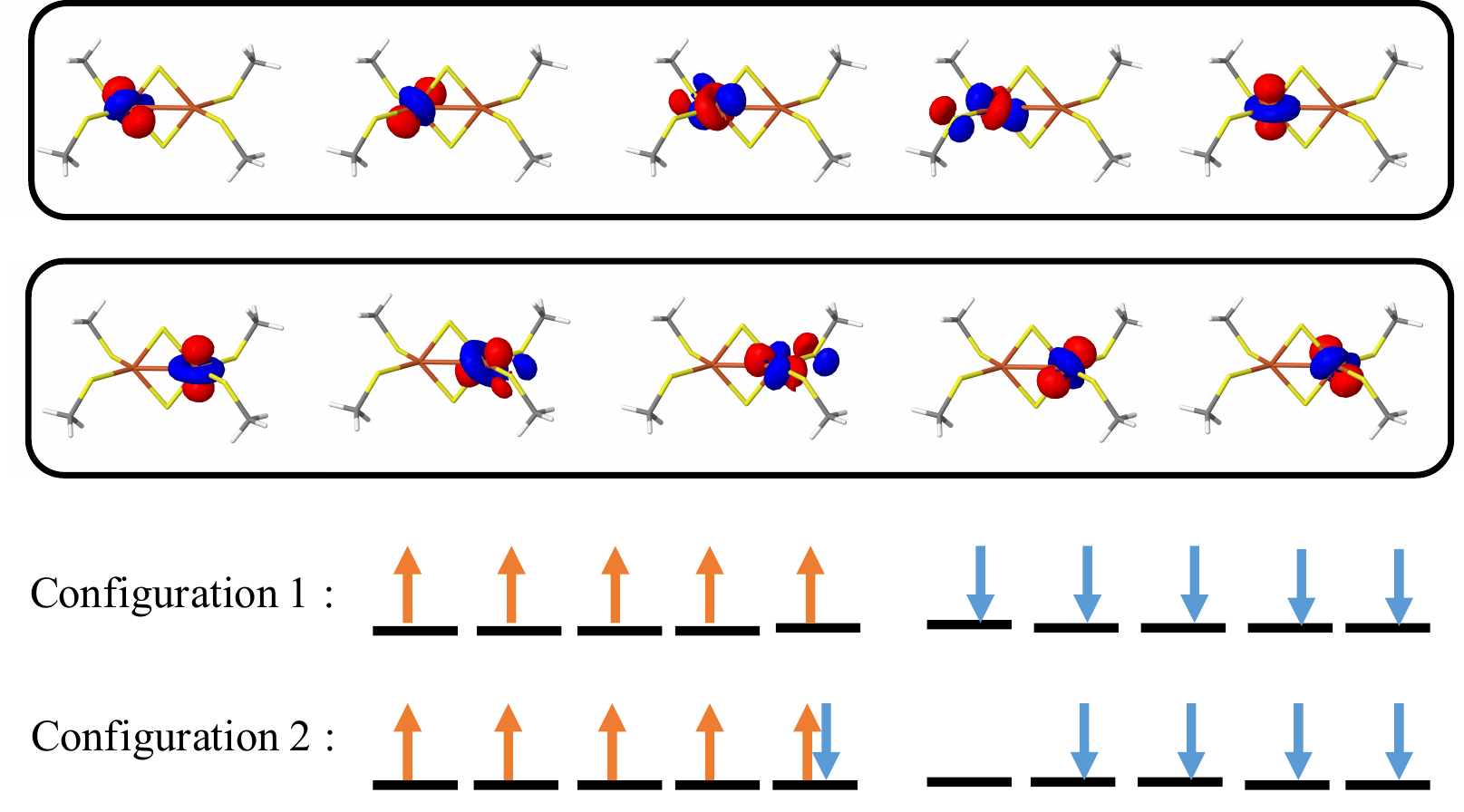}}} \\
    (a) Active space and initial configurations \\
    {\resizebox{0.5\textwidth}{!}{\includegraphics{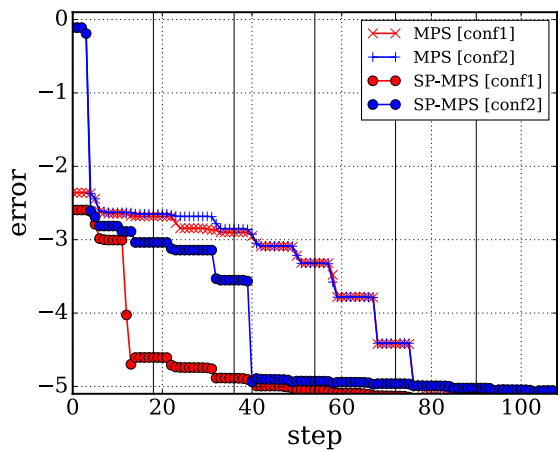}}} \\
    (b) Energy convergence
    \end{tabular}
\caption{Calculations for the complex \ce{[Fe2S2(SCH3)4]^{2-}} with CAS(10e,10o).
(a) Active space orbitals and initial configurations. (b)
Absolute errors of energies (in Hartrees) on a logarithmic scale for MPS and SP-MPS with $D=20$ as a function
of the local optimization iteration step, starting from two different initial configurations. Each region separated by
a solid vertical line corresponds to a full sweep.
The absolute errors of MPS and SP-MPS are relative to
the exact energies for $S=5$ and $S=0$, respectively.}
\label{fig:fe2s2Small}
\end{figure}

First, a small active space, CAS(10e,10o) with only  3$d_{\mathrm{Fe}}$ orbitals
constructed from localized DFT (density functional theory) orbitals
using the BP86 functional\cite{becke1988density,perdew1986density},
the TZP-DKH basis\cite{jorge2009contracted},
and the sf-X2C (spin-free exact two-component) Hamiltonian\cite{liu2010ideas,li2012spin}
to include scalar relativistic effects, is considered
with two different initial configurations, viz., Fe(III)-Fe(III)
and Fe(II)-Fe(IV) shown in Figure \ref{fig:fe2s2Small}(a).
The bond dimension $D=20$ is used
in all calculations, and the absolute errors on a logarithmic
scale are shown in Figure \ref{fig:fe2s2Small}(b).
While both the MPS without spin projection
and SP-MPS for the singlet state converge to an accuracy of 10$^{-5}$ Hartrees,
the final converged states are actually qualitatively different.
This is because while the exact energy within the active space
of the singlet state is -27.887643 Hartrees, the state with $S=5$
representing the ferromagnetically coupled iron centers is of lower energy (-27.890357 Hartrees).
Consequently, without the spin projection, the normal MPS calculation without spin adaptation
converges to the lowest energy (high spin) state. Including sulfur 3$p_{\mathrm{S}}$ orbitals
is important to recover the correct spin-state ordering, as the superexchange
effect  stabilizes the singlet state\cite{sharma2014low}.
Despite such deficiencies of the small active space, it is interesting
to see that while the energy of the second initial configuration
is higher as expected, after several sweeps the SP-MPS (blue line) starts to
relax to the same state as when starting from the first initial configuration (red line),
showing that even with this small bond dimension it is possible to recover from
the poor initial guess.

\begin{figure}
    \begin{tabular}{c}
    {\resizebox{0.5\textwidth}{!}{\includegraphics{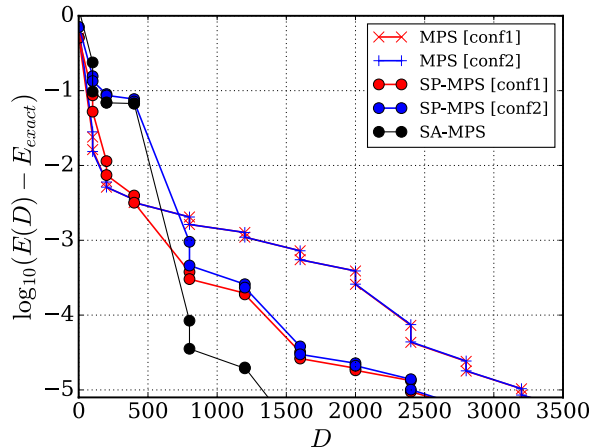}}} \\
    \end{tabular}
\caption{Errors of energies (in Hartrees) as a function of bond dimension $D$
for the complex \ce{[Fe2S2(SCH3)4]^{2-}} with CAS(30e,20o)
starting from two different initial configurations
shown in Figure \ref{fig:fe2s2Small}(a).}\label{fig:fe2s2Large}
\end{figure}

Including sulfur 3$p_{\mathrm{S}}$ orbitals gives rise to an enlarged active space CAS(30e,20o),
and the energy convergence is shown in Figure \ref{fig:fe2s2Large} as
a function of the bond dimension $D$. In this case, both MPS and SP-MPS
converge to the same singlet ground state, while the latter converges
faster with $D$. For comparison, the convergence of spin-adapted MPS (SA-MPS)
is also depicted in the same figure (black lines). We see that
for this case the SA-MPS converges to an accuracy of 10$^{-5}$ Hartrees much more quickly,
mainly because the ground state is not very highly entangled.
However, if the accuracy required is  10$^{-3}$ Hartrees, then
it may be advantageous to use SP-MPS with different initial guesses
to sample the space spanned by low-lying states, especially for
larger iron-sulfur clusters with many competing local minima in the
energy landscape of MPS with a fixed small bond dimension. Here, the connection of SP-MPS
to the underlying broken symmetry determinants is crucial as it makes it easy to setup and enumerate
all the possible starting low-energy broken symmetry configurations.

\begin{figure}
    \begin{tabular}{c}
    {\resizebox{0.8\textwidth}{!}{\includegraphics{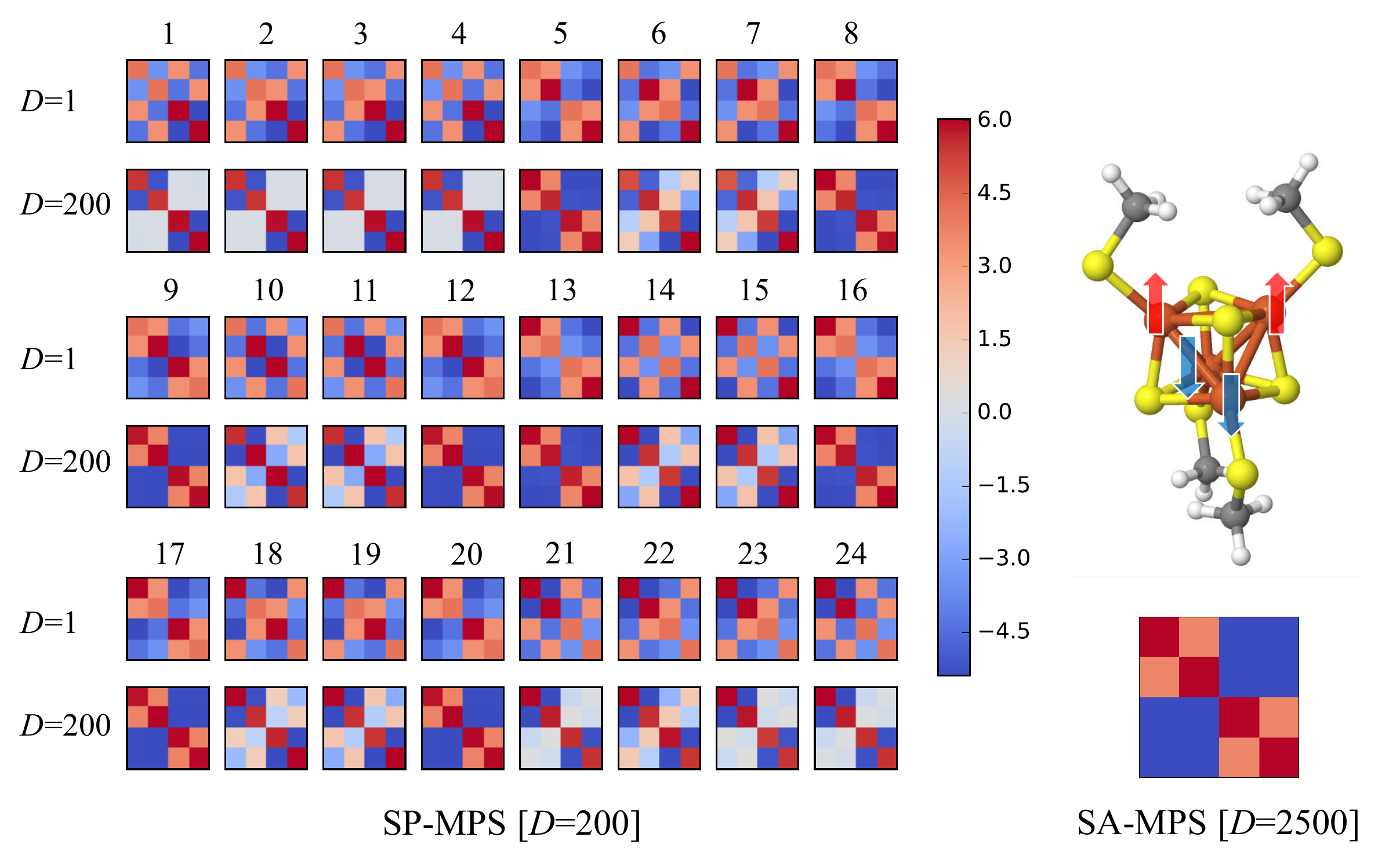}}} \\
    (a) Spin-spin correlation functions among four irons \\
    {\resizebox{0.8\textwidth}{!}{\includegraphics{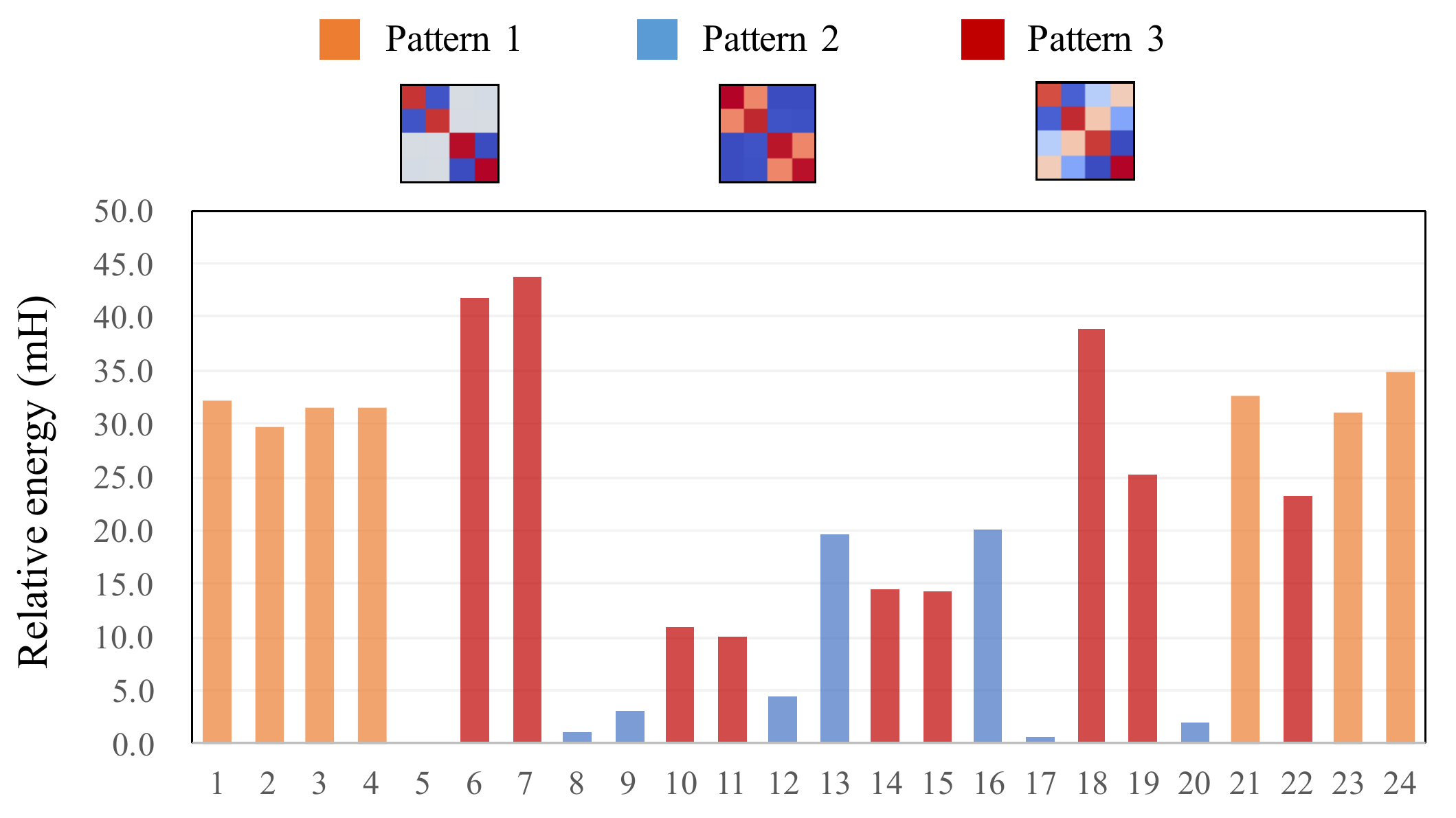}}} \\
    (b) Relative energies of the 24 SP-MPS states
    \end{tabular}
\caption{Calculations for the complex \ce{[Fe4S4(SCH3)4]^{2-}} with CAS(54e,36o).
(a) Converged spin-spin correlation functions between the four irons for 24 SP-MPS with $D=200$ (even rows)
starting from 24 different initial configurations (odd rows).
(b) Relative energies of the 24 converged SP-MPS states.}\label{fig:Fe4S4}
\end{figure}

We next consider the complex \ce{[Fe4S4(SCH3)4]^{2-}} shown in Figure \ref{fig:Fe4S4}(a)
 using an active space CAS(54e,36o) with all 3$d_{\mathrm{Fe}}$ orbitals and
3$p_{\mathrm{S}}$ orbitals. For this complex, 24(=4!) different physically meaningful
initial guesses can be constructed by distributing four different kinds of
iron oxidation and spin states, viz., spin-up/down Fe(II) and spin-up/down Fe(III), in
four different positions. SP-MPS calculations for singlet states with a small bond dimension
$D=200$ are carried out starting from these broken symmetry initial guesses for $|\Psi^{(N,M)}_{\mps}\rangle$.
After convergence, the spin-spin correlation functions
$\langle\vec{S}_{\mathrm{Fe}_{i}}\cdot\vec{S}_{\mathrm{Fe}_{j}}\rangle$ ($i,j\in\{1,2,3,4\}$)
between the four irons are computed by Eq. \eqref{PropSpinFree},
and the results are depicted in Figure \ref{fig:Fe4S4}(a),
where the odd rows contain the spin-spin correlation patterns for the SP-MPS state with initial
product state configurations,
and the even rows contain the corresponding converged SP-MPS results with $D=200$. Clearly,
there are three distinct patterns in the final results, and in all cases
the charges on the irons delocalize. In Figure \ref{fig:Fe4S4}(b),
the energies relative to the lowest SP-MPS energy (the 5th state)
for all SP-MPS states are compared for the three patterns.
We observe that the blue bars are associated with lower energies.
Thus, the corresponding spin-spin correlation pattern has a higher
chance to be the true pattern for the ground state.
In fact, this is indeed the case as demonstrated in an earlier
study\cite{sharma2014low} as well as in a state-averaged
SA-MPS calculation that we have performed for the lowest two states with $D=2500$, with the results
shown in the right panel of Figure \ref{fig:Fe4S4}(a).
Both calculations demonstrate that the ground state spin-spin correlation
pattern shown in Figure \ref{fig:Fe4S4}(a)
is the same as the pattern with the lowest energy discovered using the SP-MPS,
where two pairs of parallel spins are anti-ferromagetically coupled
to form a global singlet state. We thus see that SP-MPS
with small bond dimension, starting from different
physically motivated broken symmetry initial guesses, can help in
identifying the low energy electronic structure when there are many competing
low-lying states. Extending this approach to study even larger iron-sulfur clusters
will be pursued in the future.

\section{Conclusions and outlook}\label{conclusion}
\begin{figure}
    \begin{tabular}{cc}
    {\resizebox{!}{0.2\textheight}{\includegraphics{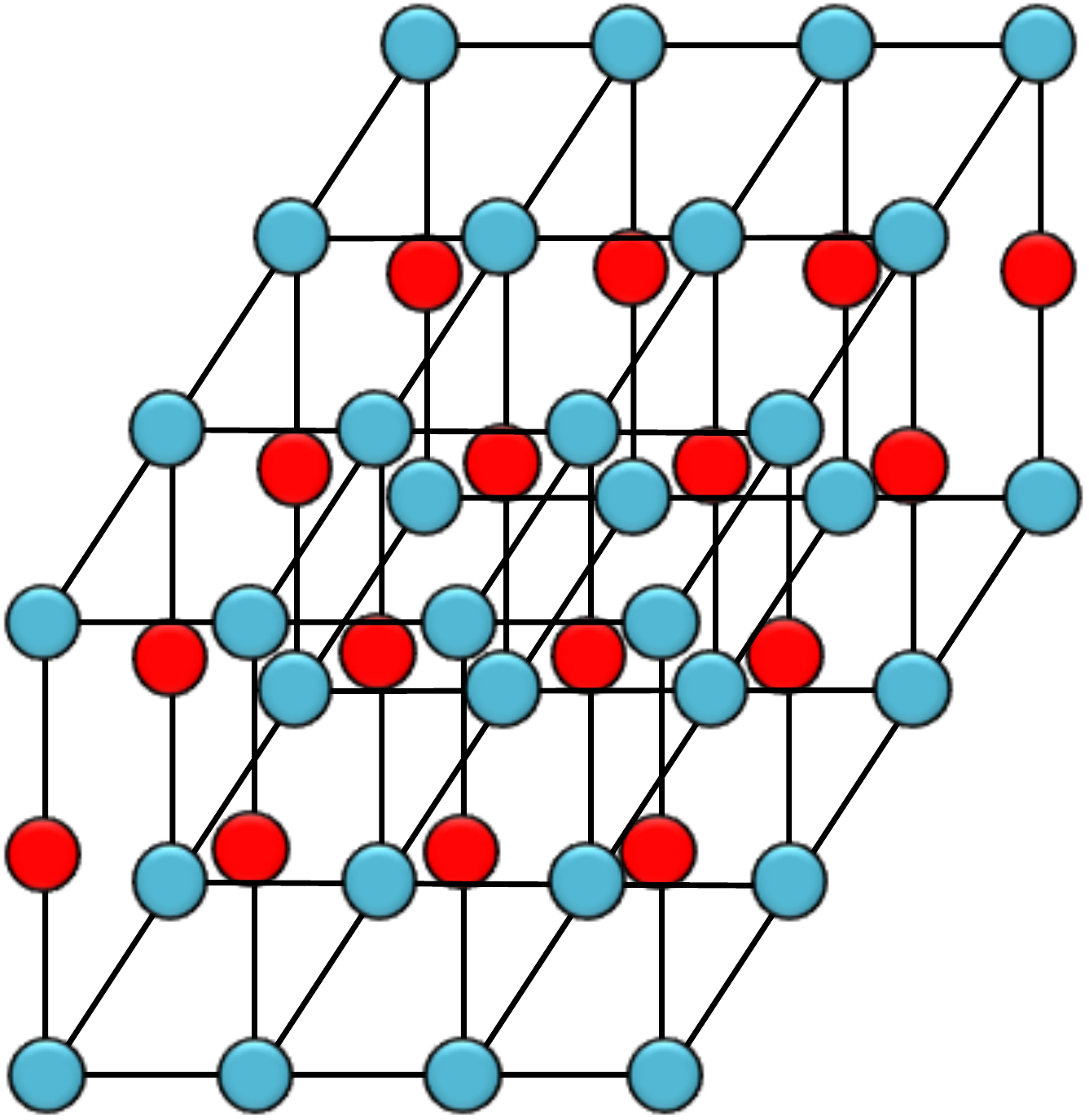}}} &
    {\resizebox{!}{0.13\textheight}{\includegraphics{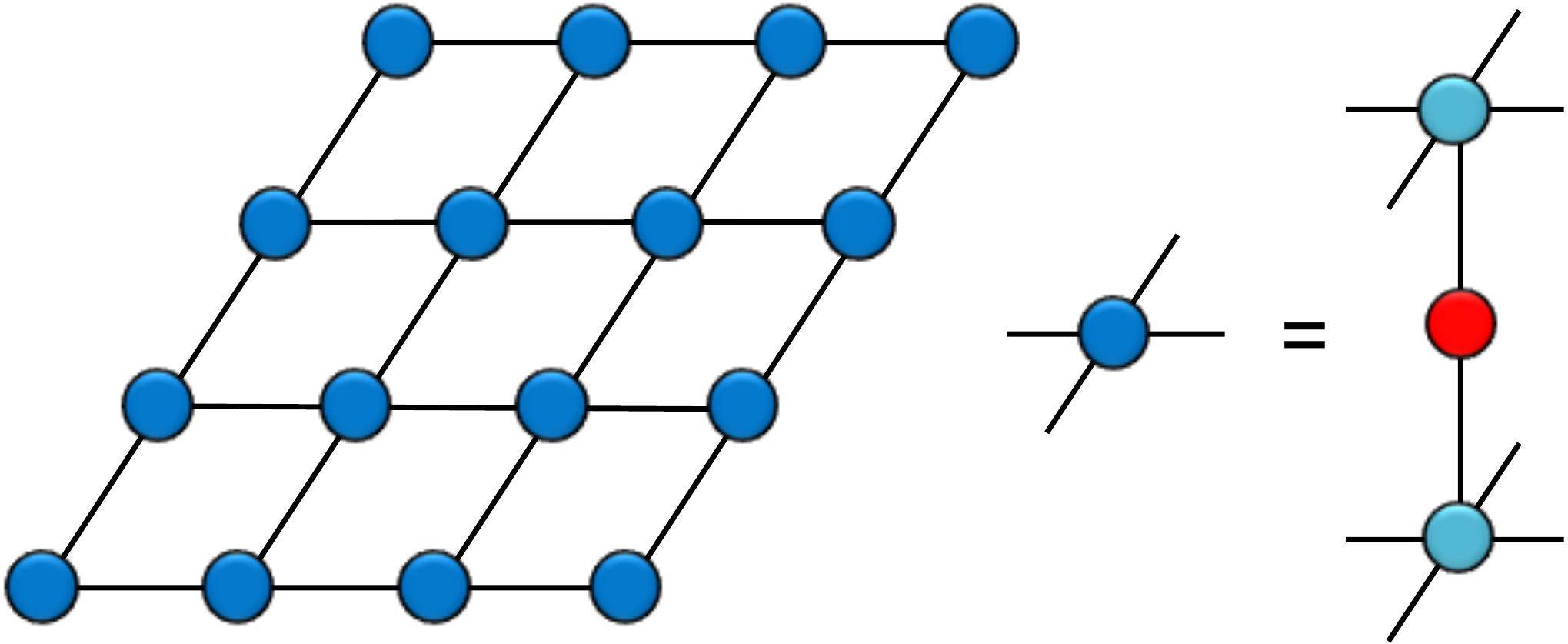}}} \\
    (a) $\langle\Psi_{\mathrm{PEPS}}|e^{-\ii\beta_g \hat{S}_y}|\Psi_{\mathrm{PEPS}}\rangle$ &
    (b) Contracted $\langle\Psi_{\mathrm{PEPS}}|e^{-\ii\beta_g \hat{S}_y}|\Psi_{\mathrm{PEPS}}\rangle$
    \end{tabular}
\caption{Overlap between two PEPS with projectors \eqref{pmpo},
$\langle\Psi_{\mathrm{PEPS}}|e^{-\ii\beta_g \hat{S}_y}|\Psi_{\mathrm{PEPS}}\rangle$.}\label{fig:peps}
\end{figure}

In this work, we have developed a versatile tool for strongly correlated systems by
combining the idea of spin projection in quantum chemistry with
the simplest TNS, the MPS, which has an underlying one dimensional connectivity.
Such an approach could  be even more advantageous within two possible generalizations.
The first generalization is to extend to non-Abelian point
group symmetries, which would otherwise involve
the use of the generalized 6$j$ and 9$j$ symbols
for the non-Abelian point groups if the symmetry
adaptation is carried out in a similar fashion
as SA-MPS. However, using symmetry projection,
only the representation matrix $D^{\Gamma}(\hat{R})$ is needed,
where $\Gamma$ represents the target irreducible representation (irrep).
In view of the fact that for most of the non-Abelian point groups,
the dimension of the degenerate irrep is two (actually
the largest dimension is 5 for the irrep $H_g$ of the $I_h$ group\cite{altmann1994point}),
the bond dimension to represent each symmetry operation is small
so long as orbitals belonging to the same irrep are placed in adjacent sites.
The second generalization is to construct spin eigenfunctions for higher-dimensional
tensor network states such as the PEPS\cite{verstraete2004renormalization}.
This is also quite straightforward due to the simple structure of the ``projector'' $\hat{P}^{S}_{M,M}$ \eqref{pmpo}
that is a sum of products of local spin rotations.
For instance, the overlap between two PEPS with projectors \eqref{pmpo},
$\langle\Psi_{\mathrm{PEPS}}|e^{-\ii\beta_g \hat{S}_y}|\Psi_{\mathrm{PEPS}}\rangle$
shown in Figure \ref{fig:peps}, reveals that the introduction
of spin projectors adds no complications to the contraction of PEPS.

Regarding SP-MPS themselves, we have shown that they possess several distinct features that are not shared by the spin-adapted MPS with non-Abelian symmetry, such as a simple formulation and implementation
which only requires the use of Abelian symmetries for the underlying MPS, and avoids
the use of the singlet embedding scheme. 
Perhaps the most important feature is the close connection to traditional ``broken symmetry'' determinants.
This gives the ability to seed SP-MPS from initial ``broken symmetry'' determinants,
providing a route to connect chemical intuition about broken symmetry configurations to realistic calculations
that properly incorporate fluctuations and correlations.
This further opens up the possibility to
fully map out the low energy landscape of competing states in finite chemical systems,
in particular the polymetallic transition metal compounds, similar to what is already
done in condensed phase problems~\cite{zheng2016stripe}.
In such applications, the computed energies of SP-MPS can be improved by using the SP-MPS based perturbation theory, see Appendix \ref{appendix1}.
Further, a combination of SP-MPS and SA-MPS is also possible by
using the optimized SP-MPS with small bond dimension to initialize spin-adapted DMRG calculations with larger bond dimensions, to improve the computed energies efficiently.
The connection to broken symmetry mean-field states may also help in the future development of quantum embedding methods
for open-shell systems. Specifically, the SP-MPS can be used as an impurity
solver for the embedded system in DMET, and such a combination
may help overcome the difficulties in using MPS to describe very high-dimensional
entanglement, as can be found in large transition metal clusters. These directions are being explored in our laboratory.

\newpage
\section*{Appendix 1: SP-MPS perturbation theory}\label{appendix1}
We here describe how to incorporate SP-MPS into the MPSPT framework\cite{sharma2014communication}, and in particular, how to deal with the complications arising from the spin projectors. In the MPSPT, once a partition of $\hat{H}=\hat{H}_0+\hat{V}$ is given, the first order wavefunction represented in
the MPS form is to be obtained by minimizing the Hylleraas functional\cite{sharma2014communication} (assuming
real MPS for simplicity),
\begin{eqnarray}
L[\Psi_1] = \langle \Psi_1|\hat{H}_0-E_0|\Psi_1\rangle + 2\langle \Psi_1 |\hat{V}|\Psi_0\rangle,\quad
\mathrm{subject\;to}\quad
\langle\Psi_0|\Psi_1\rangle = 0.\label{Hylleraas}
\end{eqnarray}
In the SP-MPS case, both $|\Psi_0\rangle$ and $|\Psi_1\rangle$ are to be represented
by the ansatz \eqref{SPMPS}, and the only slight modification is that
the reference $|\Psi_0\rangle$ is explicitly normalized via,
\begin{eqnarray}
|\Psi_0\rangle=\mathcal{N}\mathcal{P}^S_{M,M}|\Psi_{\mps,0}^{(N,M)}\rangle,\quad
\mathcal{N}=1/\sqrt{\langle\Psi_{\mps,0}^{(N,M)}|
\hat{P}^S_{M,M}|\Psi_{\mps,0}^{(N,M)}\rangle}.
\end{eqnarray}
The orthogonality constraint in Eq. \eqref{Hylleraas} can be implemented as in the excited-state calculations discussed above. The key to avoid complications due to double integrations for $L[\Psi_1]$ is to choose
a good spin-free $\hat{H}_0$. Then the perturbation $\hat{V}=\hat{H}-\hat{H}_0$ will also be spin-free, such that
in both terms of $L[\Psi_1]$ \eqref{Hylleraas}, the commutator $[\mathcal{P}^S_{M,M},\hat{O}_{sf}]=0$
can be used to bring $L[\Psi_1]$ into a form that only involves a single
integration in $\hat{P}^{S}_{M,M}$ similar to that in the energy functional \eqref{efunctional}.
Then, the stationary condition for Eq. \eqref{Hylleraas} will lead to a simple linear equation to be solved in each local optimization problem, i.e.,
\begin{eqnarray}
\frac{\partial L[\Psi_1]}{\partial A[k]}&=&
2(H_{0,\mathrm{eff}}A[k]+V_{\mathrm{eff}})=0,\nonumber\\
H_{0,\mathrm{eff}}A[k]&\triangleq &\langle \frac{\partial\Psi_{\mps,1}^{(N,M)}}{\partial A[k]}|\hat{H}_0\hat{P}^{S}_{M,M}-E_0\hat{P}^S_{M,M}|\Psi_{\mps,1}^{(N,M)}\rangle,\nonumber\\
V_{\mathrm{eff}}&\triangleq &\langle\frac{\partial\Psi_{\mps,1}^{(N,M)}}{\partial A[k]} |\hat{V}\hat{P}^{S}_{M,M}|\Psi_{\mps,0}^{(N,M)}\rangle.\label{linear}
\end{eqnarray}
subject to the orthogonality constraint similar to Eq. \eqref{orthoB}. Here, all MPS
in Eq. \eqref{linear} refer to the underlying MPS such that the equation can be solved
similarly to in MPSPT.

For spin-free $\hat{H}_0$, we propose to use a simple analog of the Epstein-Nesbet partition in the space of determinants,
\begin{eqnarray}
\hat{H}_d=\sum_{p}h_{pp}\sum_{\sigma}n_{p_\sigma} + \frac{1}{2}
\sum_{pq}
[pp|qq]
\sum_{\sigma}n_{p_\sigma}\sum_{\tau}n_{q_\tau}
-
\frac{1}{2}
\sum_{pq}
[pq|qp]
\sum_{\sigma}n_{p_\sigma} n_{q_\sigma},
\end{eqnarray}
where $h_{pq}$ and $[pq|rs]$ are the one- and two-electron integrals over spatial orbitals. However, while the first two terms are singlet operators, which can be reexpressed by using $E_{pq}$, the last term containing
$\sum_{\sigma}n_{p_\sigma} n_{q_\sigma}$ is not a pure singlet operator. To show this, it is rewritten
into a combination of spin tensor operators,
\begin{eqnarray}
\sum_{\sigma}n_{p_\sigma} n_{q_\sigma}&=&S_{pp}(0,0)S_{qq}(0,0)+T_{pp}(1,0)T_{qq}(1,0),\\
S_{pq}(0,0) &=& \frac{1}{\sqrt{2}}(a_{p_\alpha}^+a_{q_\alpha} +a_{p_\beta}^+a_{q_\beta})\;=\;\frac{1}{\sqrt{2}}E_{pq},\label{Sai00}\\
T_{pq}(1,0) &=&\frac{1}{\sqrt{2}}(a_{p_\alpha}^+a_{q_\alpha} -a_{p_\beta}^+a_{q_\beta}),\label{Tai10}\\
T_{pq}(1,1) &=&-a_{p_\alpha}^+a_{q_\beta},\label{Tai11}\\
T_{pq}(1,-1)&=& a_{p_\beta}^+a_{q_\alpha},\label{Tai1m1}
\end{eqnarray}
where $S_{pq}$ and $T_{pq}$ are singlet and triplet operators, respectively.
Extracting the singlet component of $T_{pp}(1,0)T_{qq}(1,0)$, viz.,
$[T_{pp}(1)\times T_{qq}(1)]^0_0 C^{00}_{10,10}$ with $C^{S_3M_3}_{S_1M_1,S_2M_2}$ being the Clebsch-Gordan coefficient, gives the following form for the singlet component of $\hat{H}_d$,
\begin{eqnarray}
\hat{H}_d^{\mathrm{singlet}}&=&
\sum_{p}\epsilon_{p}E_{pp}+\sum_{pq}J_{pq}E_{pp}E_{qq}+\sum_{pq}K_{pq}E_{pq}E_{qp},\nonumber\\
\epsilon_{p} &\triangleq& h_{pp} - \frac{1}{6}\sum_{q}[pq|qp] - \frac{1}{3}[pp|pp],\nonumber \\
J_{pq} &\triangleq& \frac{1}{2}[pp|qq] - \frac{1}{6}[pq|qp],\nonumber \\
K_{pq} &\triangleq& \frac{1}{6}[pq|qp],
\end{eqnarray}
which can be rewritten as a sum of $2K$ MPO by the same splitting as in Eq. \eqref{sumH} for $\hat{H}$, each of which is of bond dimension $5$. Compared with the bond
dimension $O(K^2)$ for $\hat{H}$, it is seen that $\hat{H}_d^{\mathrm{singlet}}$ is a significant simplification. For $\hat{H}_0$ used in the Hylleraas function \eqref{Hylleraas}, taking into account the fact that $|\Psi_0\rangle$ is not
an eigenfunction of $\hat{H}_d^{\mathrm{singlet}}$, the following form
could be chosen,
\begin{eqnarray}
\hat{H}_0=PE_0P+Q\hat{H}_d^{\mathrm{singlet}}Q,\quad P=|\Psi_0\rangle\langle\Psi_0|,\quad Q=1-P,
\end{eqnarray}
where $E_0$ could be chosen as the DMRG energy for $|\Psi_0\rangle$ for simplicity.
The performance of such perturbation theory will be studied in future.

\section*{Appendix 2: Derivations for Eqs. \eqref{SpinTpqA}, \eqref{SpinTpqB}, and \eqref{SpinTpqC}}\label{appendix2}
To derive a general expression for $\langle\Psi_{\spmps,I}^{(N,S,M)}|T_{pq}(1,\mu)
|\Psi_{\spmps,J}^{(N,S',M')}\rangle$, we first examine the action of $\mathcal{P}^{S}_{M,M}$
on $T_{pq}(1,\mu)$,
\begin{eqnarray}
\mathcal{P}^S_{M,M}T_{pq}(1,\mu)
&=&
\frac{2S+1}{8\pi^2}\int d\Omega
D^{S*}_{M,M}(\Omega)
\hat{R}(\Omega)T_{pq}(1,\mu)\nonumber\\
&=&
\frac{2S+1}{8\pi^2}\int d\Omega
D^{S*}_{M,M}(\Omega)
\left(\hat{R}(\Omega)T_{pq}(1,\mu)\hat{R}(\Omega)^{-1}\right)\hat{R}(\Omega)\nonumber\\
&=&
\frac{2S+1}{8\pi^2}\int d\Omega
D^{S*}_{M,M}(\Omega)
\left(\sum_{\nu} T_{pq}(1,\nu)D_{\nu,\mu}^{1}(\Omega)\right)\hat{R}(\Omega)\nonumber\\
&=&
\sum_{\nu}T_{pq}(1,\nu)
\left(\frac{2S+1}{8\pi^2}\int d\Omega
[D^{S*}_{M,M}(\Omega)
D_{\nu,\mu}^{1}(\Omega)]\hat{R}(\Omega)\right).\label{PT}
\end{eqnarray}
The term in the bracket can be recast into
a linear combination of projectors, by using
the following Clebsch-Gordan series\cite{varshalovich1988quantum},
\begin{eqnarray}
D^{S*}_{M,M}(\Omega)D_{\nu\mu}^{1}(\Omega)&=&
(-1)^{\nu-\mu}D^{S*}_{M,M}(\Omega)D_{-\nu,-\mu}^{1*}(\Omega)\nonumber\\
&=&
(-1)^{\nu-\mu}
\sum_{S'}
C_{SM,1(-\nu)}^{S'(M-\nu)} D^{S'*}_{M-\nu,M-\mu}(\Omega)
C_{SM,1(-\mu)}^{S'(M-\mu)},\label{DD2}
\end{eqnarray}
where $C_{S_1M_1,S_2M_2}^{S_3M_3}$ represents the
(real) Clebsch-Gordan coefficients. With the decomposition \eqref{DD2}, Eq. \eqref{PT}
becomes,
\begin{eqnarray}
\mathcal{P}^S_{M,M}T_{pq}(1,\mu)
&=&
\sum_{\nu}
\sum_{S'}
(-1)^{\nu-\mu}
C_{SM,1(-\nu)}^{S'(M-\nu)}
C_{SM,1(-\mu)}^{S'(M-\mu)}
T_{pq}(1,\nu)
\mathcal{P}^{S'}_{M-\nu,M-\mu}.
\end{eqnarray}
Using this result, the matrix elements $\langle\Psi_{\spmps,I}^{(N,S,M)}|T_{pq}(1,\mu)
|\Psi_{\spmps,J}^{(N,S',M')}\rangle$ are derived as
\begin{eqnarray}
&&\langle\Psi_{\spmps,I}^{(N,S,M)}|T_{pq}(1,\mu)|\Psi_{\spmps,J}^{(N,S',M')}\rangle\nonumber\\
&=&
\langle\Psi_{\mps,I}^{(N,M)}|\mathcal{P}^{S}_{M,M}
T_{pq}(1,\mu)\mathcal{P}^{S'}_{M',M'}|\Psi_{\mps,J}^{(N,M')}\rangle\nonumber\\
&=&
\sum_{\nu}
\sum_{S''}
(-1)^{\nu-\mu}
C_{SM,1(-\nu)}^{S''(M-\nu)}
C_{SM,1(-\mu)}^{S''(M-\mu)}
\langle\Psi_{\mps,I}^{(N,M)}|T_{pq}(1,\nu)
\mathcal{P}^{S''}_{M-\nu,M-\mu}
\mathcal{P}^{S'}_{M',M'}|\Psi_{\mps,J}^{(N,M')}\rangle\nonumber\\
&=&
\delta_{M-\mu,M'}
C_{SM,1(-\mu)}^{S'M'}
\sum_{\nu}
(-1)^{\nu-\mu}
C_{SM,1(-\nu)}^{S'(M-\nu)}
\langle\Psi_{\mps,I}^{(N,M)}|T_{pq}(1,\nu)
\mathcal{P}^{S'}_{M-\nu,M'}|\Psi_{\mps,J}^{(N,M')}\rangle\nonumber\\
&=&
\delta_{M-\mu,M'}
C_{SM,1(-\mu)}^{S'M'}
\sum_{\nu}
(-1)^{\nu-\mu}
C_{SM,1(-\nu)}^{S'(M-\nu)}
\langle\Psi_{\mps,I}^{(N,M)}|T_{pq}(1,\nu)
\hat{P}^{S'}_{M-\nu,M'}|\Psi_{\mps,J}^{(N,M')}\rangle,\label{FinalSpinDep}
\end{eqnarray}
where the projector property
$\mathcal{P}^{S}_{M_1,M_2}\mathcal{P}^{S'}_{M_1',M_2'}=
\delta_{SS'}\delta_{M_2M_1'}
\mathcal{P}^{S}_{M_1,M_2'}$ has been used.
Note that the nonvanishing condition for the Clebsch-Gordan
coefficient imposes the triangular condition
for angular momentum coupling in Eq. \eqref{FinalSpinDep}, $|S-1|\le S'\le S+1$,
otherwise the matrix elements are zero.
If $M$ and $M'$ are further limited to the high spin case,
Eq. \eqref{FinalSpinDep}  simplifies to
\begin{eqnarray}
&&\langle\Psi_{\spmps,I}^{(N,S,M=S)}|T_{pq}(1,\mu)|\Psi_{\spmps,J}^{(N,S'=S-\mu,M'=S-\mu)}\rangle\nonumber\\
&=&
C_{SS,1(-\mu)}^{(S-\mu)(S-\mu)}
\sum_{\nu}
(-1)^{\nu-\mu}
C_{SS,1(-\nu)}^{(S-\mu)(S-\nu)}
\langle\Psi_{\mps,I}^{(N,M=S)}|T_{pq}(1,\nu)
\hat{P}^{S-\mu}_{S-\nu,S-\mu}|\Psi_{\mps,J}^{(N,M'=S-\mu)}\rangle,
\end{eqnarray}
which gives rise to Eqs. \eqref{SpinTpqA}-\eqref{SpinTpqC} by substituting in the value of $\mu=1,0,-1$, respectively.

\begin{acknowledgement}
This work was supported by the NSF through award SI2-SSI:1657286, with additional support from award 1650436.
\end{acknowledgement}


\bibliography{refSPMPS}

\providecommand{\latin}[1]{#1}
\makeatletter
\providecommand{\doi}
  {\begingroup\let\do\@makeother\dospecials
  \catcode`\{=1 \catcode`\}=2\doi@aux}
\providecommand{\doi@aux}[1]{\endgroup\texttt{#1}}
\makeatother
\providecommand*\mcitethebibliography{\thebibliography}
\csname @ifundefined\endcsname{endmcitethebibliography}
  {\let\endmcitethebibliography\endthebibliography}{}
\begin{mcitethebibliography}{89}
\providecommand*\natexlab[1]{#1}
\providecommand*\mciteSetBstSublistMode[1]{}
\providecommand*\mciteSetBstMaxWidthForm[2]{}
\providecommand*\mciteBstWouldAddEndPuncttrue
  {\def\EndOfBibitem{\unskip.}}
\providecommand*\mciteBstWouldAddEndPunctfalse
  {\let\EndOfBibitem\relax}
\providecommand*\mciteSetBstMidEndSepPunct[3]{}
\providecommand*\mciteSetBstSublistLabelBeginEnd[3]{}
\providecommand*\EndOfBibitem{}
\mciteSetBstSublistMode{f}
\mciteSetBstMaxWidthForm{subitem}{(\alph{mcitesubitemcount})}
\mciteSetBstSublistLabelBeginEnd
  {\mcitemaxwidthsubitemform\space}
  {\relax}
  {\relax}

\bibitem[White(1992)]{white1992density}
White,~S.~R. Density matrix formulation for quantum renormalization groups.
  \emph{Phys. Rev. Lett.} \textbf{1992}, \emph{69}, 2863\relax
\mciteBstWouldAddEndPuncttrue
\mciteSetBstMidEndSepPunct{\mcitedefaultmidpunct}
{\mcitedefaultendpunct}{\mcitedefaultseppunct}\relax
\EndOfBibitem
\bibitem[White(1993)]{white1993density}
White,~S.~R. Density-matrix algorithms for quantum renormalization groups.
  \emph{Phys. Rev. B} \textbf{1993}, \emph{48}, 10345\relax
\mciteBstWouldAddEndPuncttrue
\mciteSetBstMidEndSepPunct{\mcitedefaultmidpunct}
{\mcitedefaultendpunct}{\mcitedefaultseppunct}\relax
\EndOfBibitem
\bibitem[White and Martin(1999)White, and Martin]{white1999ab}
White,~S.~R.; Martin,~R.~L. Ab initio quantum chemistry using the density
  matrix renormalization group. \emph{J. Chem. Phys.} \textbf{1999},
  \emph{110}, 4127--4130\relax
\mciteBstWouldAddEndPuncttrue
\mciteSetBstMidEndSepPunct{\mcitedefaultmidpunct}
{\mcitedefaultendpunct}{\mcitedefaultseppunct}\relax
\EndOfBibitem
\bibitem[Daul \latin{et~al.}(2000)Daul, Ciofini, Daul, and White]{daul2000full}
Daul,~S.; Ciofini,~I.; Daul,~C.; White,~S.~R. Full-CI quantum chemistry using
  the density matrix renormalization group. \emph{Int. J. Quantum Chem.}
  \textbf{2000}, \emph{79}, 331--342\relax
\mciteBstWouldAddEndPuncttrue
\mciteSetBstMidEndSepPunct{\mcitedefaultmidpunct}
{\mcitedefaultendpunct}{\mcitedefaultseppunct}\relax
\EndOfBibitem
\bibitem[Mitrushenkov \latin{et~al.}(2001)Mitrushenkov, Fano, Ortolani,
  Linguerri, and Palmieri]{mitrushenkov2001quantum}
Mitrushenkov,~A.~O.; Fano,~G.; Ortolani,~F.; Linguerri,~R.; Palmieri,~P.
  Quantum chemistry using the density matrix renormalization group. \emph{J.
  Chem. Phys.} \textbf{2001}, \emph{115}, 6815--6821\relax
\mciteBstWouldAddEndPuncttrue
\mciteSetBstMidEndSepPunct{\mcitedefaultmidpunct}
{\mcitedefaultendpunct}{\mcitedefaultseppunct}\relax
\EndOfBibitem
\bibitem[Chan and Head-Gordon(2002)Chan, and Head-Gordon]{chan2002highly}
Chan,~G. K.-L.; Head-Gordon,~M. Highly correlated calculations with a
  polynomial cost algorithm: A study of the density matrix renormalization
  group. \emph{J. Chem. Phys.} \textbf{2002}, \emph{116}, 4462--4476\relax
\mciteBstWouldAddEndPuncttrue
\mciteSetBstMidEndSepPunct{\mcitedefaultmidpunct}
{\mcitedefaultendpunct}{\mcitedefaultseppunct}\relax
\EndOfBibitem
\bibitem[Chan and Head-Gordon(2003)Chan, and Head-Gordon]{chan2003exact}
Chan,~G. K.-L.; Head-Gordon,~M. Exact solution (within a triple-zeta, double
  polarization basis set) of the electronic Schr{\"o}dinger equation for water.
  \emph{J. Chem. Phys.} \textbf{2003}, \emph{118}, 8551--8554\relax
\mciteBstWouldAddEndPuncttrue
\mciteSetBstMidEndSepPunct{\mcitedefaultmidpunct}
{\mcitedefaultendpunct}{\mcitedefaultseppunct}\relax
\EndOfBibitem
\bibitem[Legeza \latin{et~al.}(2003)Legeza, R{\"o}der, and
  Hess]{legeza2003controlling}
Legeza,~{\"O}.; R{\"o}der,~J.; Hess,~B. Controlling the accuracy of the
  density-matrix renormalization-group method: The dynamical block state
  selection approach. \emph{Phys. Rev. B} \textbf{2003}, \emph{67},
  125114\relax
\mciteBstWouldAddEndPuncttrue
\mciteSetBstMidEndSepPunct{\mcitedefaultmidpunct}
{\mcitedefaultendpunct}{\mcitedefaultseppunct}\relax
\EndOfBibitem
\bibitem[Legeza and S{\'o}lyom(2003)Legeza, and
  S{\'o}lyom]{legeza2003optimizing}
Legeza,~{\"O}.; S{\'o}lyom,~J. Optimizing the density-matrix renormalization
  group method using quantum information entropy. \emph{Phys. Rev. B}
  \textbf{2003}, \emph{68}, 195116\relax
\mciteBstWouldAddEndPuncttrue
\mciteSetBstMidEndSepPunct{\mcitedefaultmidpunct}
{\mcitedefaultendpunct}{\mcitedefaultseppunct}\relax
\EndOfBibitem
\bibitem[Chan(2004)]{chan2004algorithm}
Chan,~G. K.-L. An algorithm for large scale density matrix renormalization
  group calculations. \emph{J. Chem. Phys.} \textbf{2004}, \emph{120},
  3172--3178\relax
\mciteBstWouldAddEndPuncttrue
\mciteSetBstMidEndSepPunct{\mcitedefaultmidpunct}
{\mcitedefaultendpunct}{\mcitedefaultseppunct}\relax
\EndOfBibitem
\bibitem[Mitrushenkov \latin{et~al.}(2003)Mitrushenkov, Linguerri, Palmieri,
  and Fano]{mitrushenkov2003quantum}
Mitrushenkov,~A.; Linguerri,~R.; Palmieri,~P.; Fano,~G. Quantum chemistry using
  the density matrix renormalization group II. \emph{J. Chem. Phys.}
  \textbf{2003}, \emph{119}, 4148--4158\relax
\mciteBstWouldAddEndPuncttrue
\mciteSetBstMidEndSepPunct{\mcitedefaultmidpunct}
{\mcitedefaultendpunct}{\mcitedefaultseppunct}\relax
\EndOfBibitem
\bibitem[Chan \latin{et~al.}(2004)Chan, K{\'a}llay, and Gauss]{chan2004state}
Chan,~G. K.-L.; K{\'a}llay,~M.; Gauss,~J. State-of-the-art density matrix
  renormalization group and coupled cluster theory studies of the nitrogen
  binding curve. \emph{J. Chem. Phys.} \textbf{2004}, \emph{121},
  6110--6116\relax
\mciteBstWouldAddEndPuncttrue
\mciteSetBstMidEndSepPunct{\mcitedefaultmidpunct}
{\mcitedefaultendpunct}{\mcitedefaultseppunct}\relax
\EndOfBibitem
\bibitem[Chan and Van~Voorhis(2005)Chan, and Van~Voorhis]{chan2005density}
Chan,~G. K.-L.; Van~Voorhis,~T. Density-matrix renormalization-group algorithms
  with nonorthogonal orbitals and non-Hermitian operators, and applications to
  polyenes. \emph{J. Chem. Phys.} \textbf{2005}, \emph{122}, 204101\relax
\mciteBstWouldAddEndPuncttrue
\mciteSetBstMidEndSepPunct{\mcitedefaultmidpunct}
{\mcitedefaultendpunct}{\mcitedefaultseppunct}\relax
\EndOfBibitem
\bibitem[Moritz and Reiher(2006)Moritz, and Reiher]{moritz2006construction}
Moritz,~G.; Reiher,~M. Construction of environment states in quantum-chemical
  density-matrix renormalization group calculations. \emph{J. Chem. Phys.}
  \textbf{2006}, \emph{124}, 034103\relax
\mciteBstWouldAddEndPuncttrue
\mciteSetBstMidEndSepPunct{\mcitedefaultmidpunct}
{\mcitedefaultendpunct}{\mcitedefaultseppunct}\relax
\EndOfBibitem
\bibitem[Hachmann \latin{et~al.}(2006)Hachmann, Cardoen, and
  Chan]{hachmann2006multireference}
Hachmann,~J.; Cardoen,~W.; Chan,~G. K.-L. Multireference correlation in long
  molecules with the quadratic scaling density matrix renormalization group.
  \emph{J. Chem. Phys.} \textbf{2006}, \emph{125}, 144101\relax
\mciteBstWouldAddEndPuncttrue
\mciteSetBstMidEndSepPunct{\mcitedefaultmidpunct}
{\mcitedefaultendpunct}{\mcitedefaultseppunct}\relax
\EndOfBibitem
\bibitem[Marti \latin{et~al.}(2008)Marti, Ond{\'\i}k, Moritz, and
  Reiher]{marti2008density}
Marti,~K.~H.; Ond{\'\i}k,~I.~M.; Moritz,~G.; Reiher,~M. Density matrix
  renormalization group calculations on relative energies of transition metal
  complexes and clusters. \emph{J. Chem. Phys.} \textbf{2008}, \emph{128},
  014104\relax
\mciteBstWouldAddEndPuncttrue
\mciteSetBstMidEndSepPunct{\mcitedefaultmidpunct}
{\mcitedefaultendpunct}{\mcitedefaultseppunct}\relax
\EndOfBibitem
\bibitem[Ghosh \latin{et~al.}(2008)Ghosh, Hachmann, Yanai, and
  Chan]{ghosh2008orbital}
Ghosh,~D.; Hachmann,~J.; Yanai,~T.; Chan,~G. K.-L. Orbital optimization in the
  density matrix renormalization group, with applications to polyenes and
  $\beta$-carotene. \emph{J. Chem. Phys.} \textbf{2008}, \emph{128},
  144117\relax
\mciteBstWouldAddEndPuncttrue
\mciteSetBstMidEndSepPunct{\mcitedefaultmidpunct}
{\mcitedefaultendpunct}{\mcitedefaultseppunct}\relax
\EndOfBibitem
\bibitem[Chan(2008)]{chan2008density}
Chan,~G. K.-L. Density matrix renormalisation group Lagrangians. \emph{Phys.
  Chem. Chem. Phys.} \textbf{2008}, \emph{10}, 3454--3459\relax
\mciteBstWouldAddEndPuncttrue
\mciteSetBstMidEndSepPunct{\mcitedefaultmidpunct}
{\mcitedefaultendpunct}{\mcitedefaultseppunct}\relax
\EndOfBibitem
\bibitem[Zgid and Nooijen(2008)Zgid, and Nooijen]{zgid2008obtaining}
Zgid,~D.; Nooijen,~M. Obtaining the two-body density matrix in the density
  matrix renormalization group method. \emph{J. Chem. Phys.} \textbf{2008},
  \emph{128}, 144115\relax
\mciteBstWouldAddEndPuncttrue
\mciteSetBstMidEndSepPunct{\mcitedefaultmidpunct}
{\mcitedefaultendpunct}{\mcitedefaultseppunct}\relax
\EndOfBibitem
\bibitem[Marti and Reiher(2010)Marti, and Reiher]{marti2010density}
Marti,~K.~H.; Reiher,~M. The density matrix renormalization group algorithm in
  quantum chemistry. \emph{Z. Phys. Chem.} \textbf{2010}, \emph{224},
  583--599\relax
\mciteBstWouldAddEndPuncttrue
\mciteSetBstMidEndSepPunct{\mcitedefaultmidpunct}
{\mcitedefaultendpunct}{\mcitedefaultseppunct}\relax
\EndOfBibitem
\bibitem[Luo \latin{et~al.}(2010)Luo, Qin, and Xiang]{luo2010optimizing}
Luo,~H.-G.; Qin,~M.-P.; Xiang,~T. Optimizing Hartree-Fock orbitals by the
  density-matrix renormalization group. \emph{Phys. Rev. B} \textbf{2010},
  \emph{81}, 235129\relax
\mciteBstWouldAddEndPuncttrue
\mciteSetBstMidEndSepPunct{\mcitedefaultmidpunct}
{\mcitedefaultendpunct}{\mcitedefaultseppunct}\relax
\EndOfBibitem
\bibitem[Marti and Reiher(2011)Marti, and Reiher]{marti2011new}
Marti,~K.~H.; Reiher,~M. New electron correlation theories for transition metal
  chemistry. \emph{Phys. Chem. Chem. Phys.} \textbf{2011}, \emph{13},
  6750--6759\relax
\mciteBstWouldAddEndPuncttrue
\mciteSetBstMidEndSepPunct{\mcitedefaultmidpunct}
{\mcitedefaultendpunct}{\mcitedefaultseppunct}\relax
\EndOfBibitem
\bibitem[Kurashige and Yanai(2011)Kurashige, and Yanai]{kurashige2011second}
Kurashige,~Y.; Yanai,~T. Second-order perturbation theory with a density matrix
  renormalization group self-consistent field reference function: Theory and
  application to the study of chromium dimer. \emph{J. Chem. Phys.}
  \textbf{2011}, \emph{135}, 094104\relax
\mciteBstWouldAddEndPuncttrue
\mciteSetBstMidEndSepPunct{\mcitedefaultmidpunct}
{\mcitedefaultendpunct}{\mcitedefaultseppunct}\relax
\EndOfBibitem
\bibitem[Sharma and Chan(2012)Sharma, and Chan]{sharma2012spin}
Sharma,~S.; Chan,~G. K.-L. Spin-adapted density matrix renormalization group
  algorithms for quantum chemistry. \emph{J. Chem. Phys.} \textbf{2012},
  \emph{136}, 124121\relax
\mciteBstWouldAddEndPuncttrue
\mciteSetBstMidEndSepPunct{\mcitedefaultmidpunct}
{\mcitedefaultendpunct}{\mcitedefaultseppunct}\relax
\EndOfBibitem
\bibitem[Chan(2012)]{chan2012low}
Chan,~G.~K. Low entanglement wavefunctions. \emph{Wiley Interdiscip. Rev.:
  Comput. Mol. Sci.} \textbf{2012}, \emph{2}, 907--920\relax
\mciteBstWouldAddEndPuncttrue
\mciteSetBstMidEndSepPunct{\mcitedefaultmidpunct}
{\mcitedefaultendpunct}{\mcitedefaultseppunct}\relax
\EndOfBibitem
\bibitem[Wouters \latin{et~al.}(2012)Wouters, Limacher, Van~Neck, and
  Ayers]{wouters2012longitudinal}
Wouters,~S.; Limacher,~P.~A.; Van~Neck,~D.; Ayers,~P.~W. Longitudinal static
  optical properties of hydrogen chains: Finite field extrapolations of matrix
  product state calculations. \emph{J. Chem. Phys.} \textbf{2012}, \emph{136},
  134110\relax
\mciteBstWouldAddEndPuncttrue
\mciteSetBstMidEndSepPunct{\mcitedefaultmidpunct}
{\mcitedefaultendpunct}{\mcitedefaultseppunct}\relax
\EndOfBibitem
\bibitem[Mizukami \latin{et~al.}(2012)Mizukami, Kurashige, and
  Yanai]{mizukami2012more}
Mizukami,~W.; Kurashige,~Y.; Yanai,~T. More $\pi$ electrons make a difference:
  Emergence of many radicals on graphene nanoribbons studied by ab initio DMRG
  theory. \emph{J. Chem. Theory Comput.} \textbf{2012}, \emph{9},
  401--407\relax
\mciteBstWouldAddEndPuncttrue
\mciteSetBstMidEndSepPunct{\mcitedefaultmidpunct}
{\mcitedefaultendpunct}{\mcitedefaultseppunct}\relax
\EndOfBibitem
\bibitem[Kurashige \latin{et~al.}(2013)Kurashige, Chan, and
  Yanai]{kurashige2013entangled}
Kurashige,~Y.; Chan,~G. K.-L.; Yanai,~T. Entangled quantum electronic
  wavefunctions of the Mn4CaO5 cluster in photosystem II. \emph{Nat. Chem.}
  \textbf{2013}, \emph{5}, 660--666\relax
\mciteBstWouldAddEndPuncttrue
\mciteSetBstMidEndSepPunct{\mcitedefaultmidpunct}
{\mcitedefaultendpunct}{\mcitedefaultseppunct}\relax
\EndOfBibitem
\bibitem[Sharma \latin{et~al.}(2014)Sharma, Sivalingam, Neese, and
  Chan]{sharma2014low}
Sharma,~S.; Sivalingam,~K.; Neese,~F.; Chan,~G. K.-L. Low-energy spectrum of
  iron--sulfur clusters directly from many-particle quantum mechanics.
  \emph{Nat. Chem.} \textbf{2014}, \emph{6}, 927--933\relax
\mciteBstWouldAddEndPuncttrue
\mciteSetBstMidEndSepPunct{\mcitedefaultmidpunct}
{\mcitedefaultendpunct}{\mcitedefaultseppunct}\relax
\EndOfBibitem
\bibitem[Wouters and Van~Neck(2014)Wouters, and Van~Neck]{wouters2014density}
Wouters,~S.; Van~Neck,~D. The density matrix renormalization group for ab
  initio quantum chemistry. \emph{Eur. Phys. J. D} \textbf{2014}, \emph{68},
  1--20\relax
\mciteBstWouldAddEndPuncttrue
\mciteSetBstMidEndSepPunct{\mcitedefaultmidpunct}
{\mcitedefaultendpunct}{\mcitedefaultseppunct}\relax
\EndOfBibitem
\bibitem[Wouters \latin{et~al.}(2014)Wouters, Poelmans, Ayers, and
  Van~Neck]{wouters2014chemps2}
Wouters,~S.; Poelmans,~W.; Ayers,~P.~W.; Van~Neck,~D. CheMPS2: A free
  open-source spin-adapted implementation of the density matrix renormalization
  group for ab initio quantum chemistry. \emph{Comput. Phys. Commun.}
  \textbf{2014}, \emph{185}, 1501--1514\relax
\mciteBstWouldAddEndPuncttrue
\mciteSetBstMidEndSepPunct{\mcitedefaultmidpunct}
{\mcitedefaultendpunct}{\mcitedefaultseppunct}\relax
\EndOfBibitem
\bibitem[Fertitta \latin{et~al.}(2014)Fertitta, Paulus, Barcza, and
  Legeza]{fertitta2014investigation}
Fertitta,~E.; Paulus,~B.; Barcza,~G.; Legeza,~{\"O}. Investigation of
  metal--insulator-like transition through the ab initio density matrix
  renormalization group approach. \emph{Phys. Rev. B} \textbf{2014}, \emph{90},
  245129\relax
\mciteBstWouldAddEndPuncttrue
\mciteSetBstMidEndSepPunct{\mcitedefaultmidpunct}
{\mcitedefaultendpunct}{\mcitedefaultseppunct}\relax
\EndOfBibitem
\bibitem[Knecht \latin{et~al.}(2014)Knecht, Legeza, and
  Reiher]{knecht2014communication}
Knecht,~S.; Legeza,~{\"O}.; Reiher,~M. Communication: Four-component density
  matrix renormalization group. \emph{J. Chem. Phys.} \textbf{2014},
  \emph{140}, 041101\relax
\mciteBstWouldAddEndPuncttrue
\mciteSetBstMidEndSepPunct{\mcitedefaultmidpunct}
{\mcitedefaultendpunct}{\mcitedefaultseppunct}\relax
\EndOfBibitem
\bibitem[Szalay \latin{et~al.}(2015)Szalay, Pfeffer, Murg, Barcza, Verstraete,
  Schneider, and Legeza]{szalay2015tensor}
Szalay,~S.; Pfeffer,~M.; Murg,~V.; Barcza,~G.; Verstraete,~F.; Schneider,~R.;
  Legeza,~{\"O}. Tensor product methods and entanglement optimization for ab
  initio quantum chemistry. \emph{Int. J. Quantum Chem.} \textbf{2015},
  \emph{115}, 1342--1391\relax
\mciteBstWouldAddEndPuncttrue
\mciteSetBstMidEndSepPunct{\mcitedefaultmidpunct}
{\mcitedefaultendpunct}{\mcitedefaultseppunct}\relax
\EndOfBibitem
\bibitem[Yanai \latin{et~al.}(2015)Yanai, Kurashige, Mizukami, Chalupsk{\`y},
  Lan, and Saitow]{yanai2015density}
Yanai,~T.; Kurashige,~Y.; Mizukami,~W.; Chalupsk{\`y},~J.; Lan,~T.~N.;
  Saitow,~M. Density matrix renormalization group for ab initio Calculations
  and associated dynamic correlation methods: A review of theory and
  applications. \emph{Int. J. Quantum Chem.} \textbf{2015}, \emph{115},
  283--299\relax
\mciteBstWouldAddEndPuncttrue
\mciteSetBstMidEndSepPunct{\mcitedefaultmidpunct}
{\mcitedefaultendpunct}{\mcitedefaultseppunct}\relax
\EndOfBibitem
\bibitem[Olivares-Amaya \latin{et~al.}(2015)Olivares-Amaya, Hu, Nakatani,
  Sharma, Yang, and Chan]{olivares2015ab}
Olivares-Amaya,~R.; Hu,~W.; Nakatani,~N.; Sharma,~S.; Yang,~J.; Chan,~G. K.-L.
  The ab-initio density matrix renormalization group in practice. \emph{J.
  Chem. Phys.} \textbf{2015}, \emph{142}, 034102\relax
\mciteBstWouldAddEndPuncttrue
\mciteSetBstMidEndSepPunct{\mcitedefaultmidpunct}
{\mcitedefaultendpunct}{\mcitedefaultseppunct}\relax
\EndOfBibitem
\bibitem[{\"O}stlund and Rommer(1995){\"O}stlund, and
  Rommer]{ostlund1995thermodynamic}
{\"O}stlund,~S.; Rommer,~S. Thermodynamic limit of density matrix
  renormalization. \emph{Phys. Rev. Lett.} \textbf{1995}, \emph{75}, 3537\relax
\mciteBstWouldAddEndPuncttrue
\mciteSetBstMidEndSepPunct{\mcitedefaultmidpunct}
{\mcitedefaultendpunct}{\mcitedefaultseppunct}\relax
\EndOfBibitem
\bibitem[Rommer and {\"O}stlund(1997)Rommer, and {\"O}stlund]{rommer1997class}
Rommer,~S.; {\"O}stlund,~S. Class of ansatz wave functions for one-dimensional
  spin systems and their relation to the density matrix renormalization group.
  \emph{Phys. Rev. B} \textbf{1997}, \emph{55}, 2164\relax
\mciteBstWouldAddEndPuncttrue
\mciteSetBstMidEndSepPunct{\mcitedefaultmidpunct}
{\mcitedefaultendpunct}{\mcitedefaultseppunct}\relax
\EndOfBibitem
\bibitem[Wouters \latin{et~al.}(2013)Wouters, Nakatani, Van~Neck, and
  Chan]{wouters2013thouless}
Wouters,~S.; Nakatani,~N.; Van~Neck,~D.; Chan,~G. K.-L. Thouless theorem for
  matrix product states and subsequent post density matrix renormalization
  group methods. \emph{Phys. Rev. B} \textbf{2013}, \emph{88}, 075122\relax
\mciteBstWouldAddEndPuncttrue
\mciteSetBstMidEndSepPunct{\mcitedefaultmidpunct}
{\mcitedefaultendpunct}{\mcitedefaultseppunct}\relax
\EndOfBibitem
\bibitem[Dorando \latin{et~al.}(2009)Dorando, Hachmann, and
  Chan]{dorando2009analytic}
Dorando,~J.~J.; Hachmann,~J.; Chan,~G. K.-L. Analytic response theory for the
  density matrix renormalization group. \emph{J. Chem. Phys.} \textbf{2009},
  \emph{130}, 184111\relax
\mciteBstWouldAddEndPuncttrue
\mciteSetBstMidEndSepPunct{\mcitedefaultmidpunct}
{\mcitedefaultendpunct}{\mcitedefaultseppunct}\relax
\EndOfBibitem
\bibitem[Kinder \latin{et~al.}(2011)Kinder, Ralph, and
  Chan]{kinder2011analytic}
Kinder,~J.~M.; Ralph,~C.~C.; Chan,~G.~K. Analytic time evolution, random phase
  approximation, and Green functions for matrix product states. \emph{arXiv
  preprint arXiv:1103.2155} \textbf{2011}, \relax
\mciteBstWouldAddEndPunctfalse
\mciteSetBstMidEndSepPunct{\mcitedefaultmidpunct}
{}{\mcitedefaultseppunct}\relax
\EndOfBibitem
\bibitem[Nakatani \latin{et~al.}(2014)Nakatani, Wouters, Van~Neck, and
  Chan]{nakatani2014linear}
Nakatani,~N.; Wouters,~S.; Van~Neck,~D.; Chan,~G. K.-L. Linear response theory
  for the density matrix renormalization group: Efficient algorithms for
  strongly correlated excited states. \emph{J. Chem. Phys.} \textbf{2014},
  \emph{140}, 024108\relax
\mciteBstWouldAddEndPuncttrue
\mciteSetBstMidEndSepPunct{\mcitedefaultmidpunct}
{\mcitedefaultendpunct}{\mcitedefaultseppunct}\relax
\EndOfBibitem
\bibitem[Haegeman \latin{et~al.}(2011)Haegeman, Cirac, Osborne, Pi{\v{z}}orn,
  Verschelde, and Verstraete]{haegeman2011time}
Haegeman,~J.; Cirac,~J.~I.; Osborne,~T.~J.; Pi{\v{z}}orn,~I.; Verschelde,~H.;
  Verstraete,~F. Time-dependent variational principle for quantum lattices.
  \emph{Phys. Rev. Lett.} \textbf{2011}, \emph{107}, 070601\relax
\mciteBstWouldAddEndPuncttrue
\mciteSetBstMidEndSepPunct{\mcitedefaultmidpunct}
{\mcitedefaultendpunct}{\mcitedefaultseppunct}\relax
\EndOfBibitem
\bibitem[Sharma and Chan(2014)Sharma, and Chan]{sharma2014communication}
Sharma,~S.; Chan,~G. K.-L. Communication: A flexible multi-reference
  perturbation theory by minimizing the Hylleraas functional with matrix
  product states. \emph{J. Chem. Phys.} \textbf{2014}, \emph{141}, 111101\relax
\mciteBstWouldAddEndPuncttrue
\mciteSetBstMidEndSepPunct{\mcitedefaultmidpunct}
{\mcitedefaultendpunct}{\mcitedefaultseppunct}\relax
\EndOfBibitem
\bibitem[Li and Chan(2016)Li, and Chan]{li2016hilbert}
Li,~Z.; Chan,~G. K.-L. Hilbert space renormalization for the many-electron
  problem. \emph{J. Chem. Phys.} \textbf{2016}, \emph{144}, 084103\relax
\mciteBstWouldAddEndPuncttrue
\mciteSetBstMidEndSepPunct{\mcitedefaultmidpunct}
{\mcitedefaultendpunct}{\mcitedefaultseppunct}\relax
\EndOfBibitem
\bibitem[Verstraete \latin{et~al.}(2004)Verstraete, Garcia-Ripoll, and
  Cirac]{verstraete2004matrix}
Verstraete,~F.; Garcia-Ripoll,~J.~J.; Cirac,~J.~I. Matrix product density
  operators: simulation of finite-temperature and dissipative systems.
  \emph{Phys. Rev. Lett.} \textbf{2004}, \emph{93}, 207204\relax
\mciteBstWouldAddEndPuncttrue
\mciteSetBstMidEndSepPunct{\mcitedefaultmidpunct}
{\mcitedefaultendpunct}{\mcitedefaultseppunct}\relax
\EndOfBibitem
\bibitem[McCulloch(2007)]{mcculloch2007density}
McCulloch,~I.~P. From density-matrix renormalization group to matrix product
  states. \emph{J. Stat. Mech: Theory Exp.} \textbf{2007}, \emph{2007},
  P10014\relax
\mciteBstWouldAddEndPuncttrue
\mciteSetBstMidEndSepPunct{\mcitedefaultmidpunct}
{\mcitedefaultendpunct}{\mcitedefaultseppunct}\relax
\EndOfBibitem
\bibitem[Verstraete \latin{et~al.}(2008)Verstraete, Murg, and
  Cirac]{verstraete2008matrix}
Verstraete,~F.; Murg,~V.; Cirac,~J.~I. Matrix product states, projected
  entangled pair states, and variational renormalization group methods for
  quantum spin systems. \emph{Adv. Phys.} \textbf{2008}, \emph{57},
  143--224\relax
\mciteBstWouldAddEndPuncttrue
\mciteSetBstMidEndSepPunct{\mcitedefaultmidpunct}
{\mcitedefaultendpunct}{\mcitedefaultseppunct}\relax
\EndOfBibitem
\bibitem[Pirvu \latin{et~al.}(2010)Pirvu, Murg, Cirac, and
  Verstraete]{pirvu2010matrix}
Pirvu,~B.; Murg,~V.; Cirac,~J.~I.; Verstraete,~F. Matrix product operator
  representations. \emph{New J. Phys.} \textbf{2010}, \emph{12}, 025012\relax
\mciteBstWouldAddEndPuncttrue
\mciteSetBstMidEndSepPunct{\mcitedefaultmidpunct}
{\mcitedefaultendpunct}{\mcitedefaultseppunct}\relax
\EndOfBibitem
\bibitem[Chan \latin{et~al.}(2016)Chan, Keselman, Nakatani, Li, and
  White]{chan2016mpo}
Chan,~G. K.-L.; Keselman,~A.; Nakatani,~N.; Li,~Z.; White,~S.~R. Matrix product
  operators, matrix product states, and ab initio density matrix
  renormalization group algorithms. \emph{J. Chem. Phys.} \textbf{2016},
  \emph{145}, 014102\relax
\mciteBstWouldAddEndPuncttrue
\mciteSetBstMidEndSepPunct{\mcitedefaultmidpunct}
{\mcitedefaultendpunct}{\mcitedefaultseppunct}\relax
\EndOfBibitem
\bibitem[Sierra and Nishino(1997)Sierra, and Nishino]{sierra1997density}
Sierra,~G.; Nishino,~T. The density matrix renormalization group method applied
  to interaction round a face Hamiltonians. \emph{Nucl. Phys. B} \textbf{1997},
  \emph{495}, 505--532\relax
\mciteBstWouldAddEndPuncttrue
\mciteSetBstMidEndSepPunct{\mcitedefaultmidpunct}
{\mcitedefaultendpunct}{\mcitedefaultseppunct}\relax
\EndOfBibitem
\bibitem[McCulloch and Gul{\'a}csi(2000)McCulloch, and
  Gul{\'a}csi]{mcculloch2000density}
McCulloch,~I.~P.; Gul{\'a}csi,~M. Density matrix renormalisation group method
  and symmetries of the Hamiltonian. \emph{Aust. J. Phys.} \textbf{2000},
  \emph{53}, 597--612\relax
\mciteBstWouldAddEndPuncttrue
\mciteSetBstMidEndSepPunct{\mcitedefaultmidpunct}
{\mcitedefaultendpunct}{\mcitedefaultseppunct}\relax
\EndOfBibitem
\bibitem[McCulloch and Gul{\'a}csi(2001)McCulloch, and
  Gul{\'a}csi]{mcculloch2001total}
McCulloch,~I.~P.; Gul{\'a}csi,~M. Total spin in the density matrix
  renormalization group algorithm. \emph{Philos. Mag. Lett.} \textbf{2001},
  \emph{81}, 447--453\relax
\mciteBstWouldAddEndPuncttrue
\mciteSetBstMidEndSepPunct{\mcitedefaultmidpunct}
{\mcitedefaultendpunct}{\mcitedefaultseppunct}\relax
\EndOfBibitem
\bibitem[McCulloch and Gul{\'a}csi(2002)McCulloch, and
  Gul{\'a}csi]{mcculloch2002non}
McCulloch,~I.~P.; Gul{\'a}csi,~M. The non-Abelian density matrix
  renormalization group algorithm. \emph{EPL. Europhys. Lett.} \textbf{2002},
  \emph{57}, 852\relax
\mciteBstWouldAddEndPuncttrue
\mciteSetBstMidEndSepPunct{\mcitedefaultmidpunct}
{\mcitedefaultendpunct}{\mcitedefaultseppunct}\relax
\EndOfBibitem
\bibitem[Keller and Reiher(2016)Keller, and Reiher]{keller2016spin}
Keller,~S.; Reiher,~M. Spin-adapted matrix product states and operators.
  \emph{J. Chem. Phys.} \textbf{2016}, \emph{144}, 134101\relax
\mciteBstWouldAddEndPuncttrue
\mciteSetBstMidEndSepPunct{\mcitedefaultmidpunct}
{\mcitedefaultendpunct}{\mcitedefaultseppunct}\relax
\EndOfBibitem
\bibitem[Zgid and Nooijen(2008)Zgid, and Nooijen]{zgid2008spin}
Zgid,~D.; Nooijen,~M. On the spin and symmetry adaptation of the density matrix
  renormalization group method. \emph{J. Chem. Phys.} \textbf{2008},
  \emph{128}, 014107\relax
\mciteBstWouldAddEndPuncttrue
\mciteSetBstMidEndSepPunct{\mcitedefaultmidpunct}
{\mcitedefaultendpunct}{\mcitedefaultseppunct}\relax
\EndOfBibitem
\bibitem[Tatsuaki(2000)]{tatsuaki2000interaction}
Tatsuaki,~W. Interaction-round-a-face density-matrix renormalization-group
  method applied to rotational-invariant quantum spin chains. \emph{Phys. Rev.
  E} \textbf{2000}, \emph{61}, 3199\relax
\mciteBstWouldAddEndPuncttrue
\mciteSetBstMidEndSepPunct{\mcitedefaultmidpunct}
{\mcitedefaultendpunct}{\mcitedefaultseppunct}\relax
\EndOfBibitem
\bibitem[Sayfutyarova and Chan(2016)Sayfutyarova, and
  Chan]{sayfutyarova2016state}
Sayfutyarova,~E.~R.; Chan,~G. K.-L. A state interaction spin-orbit coupling
  density matrix renormalization group method. \emph{J. Chem. Phys.}
  \textbf{2016}, \emph{144}, 234301\relax
\mciteBstWouldAddEndPuncttrue
\mciteSetBstMidEndSepPunct{\mcitedefaultmidpunct}
{\mcitedefaultendpunct}{\mcitedefaultseppunct}\relax
\EndOfBibitem
\bibitem[Noodleman \latin{et~al.}(1988)Noodleman, Case, and
  Aizman]{noodleman1988broken}
Noodleman,~L.; Case,~D.~A.; Aizman,~A. Broken symmetry analysis of spin
  coupling in iron-sulfur clusters. \emph{J. Am. Chem. Soc.} \textbf{1988},
  \emph{110}, 1001--1005\relax
\mciteBstWouldAddEndPuncttrue
\mciteSetBstMidEndSepPunct{\mcitedefaultmidpunct}
{\mcitedefaultendpunct}{\mcitedefaultseppunct}\relax
\EndOfBibitem
\bibitem[L{\"o}wdin(1955)]{lowdin1955quantum}
L{\"o}wdin,~P.-O. Quantum theory of many-particle systems. III. Extension of
  the Hartree-Fock scheme to include degenerate systems and correlation
  effects. \emph{Phys. Rev.} \textbf{1955}, \emph{97}, 1509\relax
\mciteBstWouldAddEndPuncttrue
\mciteSetBstMidEndSepPunct{\mcitedefaultmidpunct}
{\mcitedefaultendpunct}{\mcitedefaultseppunct}\relax
\EndOfBibitem
\bibitem[Scuseria \latin{et~al.}(2011)Scuseria, Jim{\'e}nez-Hoyos, Henderson,
  Samanta, and Ellis]{scuseria2011projected}
Scuseria,~G.~E.; Jim{\'e}nez-Hoyos,~C.~A.; Henderson,~T.~M.; Samanta,~K.;
  Ellis,~J.~K. Projected quasiparticle theory for molecular electronic
  structure. \emph{J. Chem. Phys.} \textbf{2011}, \emph{135}, 124108\relax
\mciteBstWouldAddEndPuncttrue
\mciteSetBstMidEndSepPunct{\mcitedefaultmidpunct}
{\mcitedefaultendpunct}{\mcitedefaultseppunct}\relax
\EndOfBibitem
\bibitem[Jimenez-Hoyos \latin{et~al.}(2012)Jimenez-Hoyos, Henderson,
  Tsuchimochi, and Scuseria]{jimenez2012projected}
Jimenez-Hoyos,~C.~A.; Henderson,~T.~M.; Tsuchimochi,~T.; Scuseria,~G.~E.
  Projected hartree--fock theory. \emph{J. Chem. Phys.} \textbf{2012},
  \emph{136}, 164109\relax
\mciteBstWouldAddEndPuncttrue
\mciteSetBstMidEndSepPunct{\mcitedefaultmidpunct}
{\mcitedefaultendpunct}{\mcitedefaultseppunct}\relax
\EndOfBibitem
\bibitem[Jim{\'e}nez-Hoyos \latin{et~al.}(2013)Jim{\'e}nez-Hoyos,
  Rodr{\'\i}guez-Guzm{\'a}n, and Scuseria]{jimenez2013multi}
Jim{\'e}nez-Hoyos,~C.~A.; Rodr{\'\i}guez-Guzm{\'a}n,~R.; Scuseria,~G.~E.
  Multi-component symmetry-projected approach for molecular ground state
  correlations. \emph{J. Chem. Phys.} \textbf{2013}, \emph{139}, 204102\relax
\mciteBstWouldAddEndPuncttrue
\mciteSetBstMidEndSepPunct{\mcitedefaultmidpunct}
{\mcitedefaultendpunct}{\mcitedefaultseppunct}\relax
\EndOfBibitem
\bibitem[Jim{\'e}nez-Hoyos \latin{et~al.}(2013)Jim{\'e}nez-Hoyos,
  Rodr{\'\i}guez-Guzm{\'a}n, and Scuseria]{jimenez2013excited}
Jim{\'e}nez-Hoyos,~C.~A.; Rodr{\'\i}guez-Guzm{\'a}n,~R.; Scuseria,~G.~E.
  Excited electronic states from a variational approach based on
  symmetry-projected Hartree--Fock configurations. \emph{J. Chem. Phys.}
  \textbf{2013}, \emph{139}, 224110\relax
\mciteBstWouldAddEndPuncttrue
\mciteSetBstMidEndSepPunct{\mcitedefaultmidpunct}
{\mcitedefaultendpunct}{\mcitedefaultseppunct}\relax
\EndOfBibitem
\bibitem[Tsuchimochi and Ten-no(2016)Tsuchimochi, and
  Ten-no]{tsuchimochi2016communication}
Tsuchimochi,~T.; Ten-no,~S. Communication: Configuration interaction combined
  with spin-projection for strongly correlated molecular electronic structures.
  \emph{J. Chem. Phys.} \textbf{2016}, \emph{144}, 011101\relax
\mciteBstWouldAddEndPuncttrue
\mciteSetBstMidEndSepPunct{\mcitedefaultmidpunct}
{\mcitedefaultendpunct}{\mcitedefaultseppunct}\relax
\EndOfBibitem
\bibitem[Tsuchimochi and Ten-no(2016)Tsuchimochi, and
  Ten-no]{tsuchimochi2016black}
Tsuchimochi,~T.; Ten-no,~S. Black-Box Description of Electron Correlation with
  the Spin-Extended Configuration Interaction Model: Implementation and
  Assessment. \emph{J. Chem. Theory Comput.} \textbf{2016}, \emph{12},
  1741--1759\relax
\mciteBstWouldAddEndPuncttrue
\mciteSetBstMidEndSepPunct{\mcitedefaultmidpunct}
{\mcitedefaultendpunct}{\mcitedefaultseppunct}\relax
\EndOfBibitem
\bibitem[Knizia and Chan(2012)Knizia, and Chan]{knizia2012density}
Knizia,~G.; Chan,~G. K.-L. Density matrix embedding: A simple alternative to
  dynamical mean-field theory. \emph{Phys. Rev. Lett.} \textbf{2012},
  \emph{109}, 186404\relax
\mciteBstWouldAddEndPuncttrue
\mciteSetBstMidEndSepPunct{\mcitedefaultmidpunct}
{\mcitedefaultendpunct}{\mcitedefaultseppunct}\relax
\EndOfBibitem
\bibitem[Knizia and Chan(2013)Knizia, and Chan]{knizia2013density}
Knizia,~G.; Chan,~G. K.-L. Density matrix embedding: A strong-coupling quantum
  embedding theory. \emph{J. Chem. Theory Comput.} \textbf{2013}, \emph{9},
  1428--1432\relax
\mciteBstWouldAddEndPuncttrue
\mciteSetBstMidEndSepPunct{\mcitedefaultmidpunct}
{\mcitedefaultendpunct}{\mcitedefaultseppunct}\relax
\EndOfBibitem
\bibitem[Verstraete and Cirac(2004)Verstraete, and
  Cirac]{verstraete2004renormalization}
Verstraete,~F.; Cirac,~J.~I. Renormalization algorithms for quantum-many body
  systems in two and higher dimensions. \emph{arXiv preprint cond-mat/0407066}
  \textbf{2004}, \relax
\mciteBstWouldAddEndPunctfalse
\mciteSetBstMidEndSepPunct{\mcitedefaultmidpunct}
{}{\mcitedefaultseppunct}\relax
\EndOfBibitem
\bibitem[Schollw{\"o}ck(2011)]{schollwock2011density}
Schollw{\"o}ck,~U. The density-matrix renormalization group in the age of
  matrix product states. \emph{Ann. Phys.} \textbf{2011}, \emph{326},
  96--192\relax
\mciteBstWouldAddEndPuncttrue
\mciteSetBstMidEndSepPunct{\mcitedefaultmidpunct}
{\mcitedefaultendpunct}{\mcitedefaultseppunct}\relax
\EndOfBibitem
\bibitem[Keller \latin{et~al.}(2015)Keller, Dolfi, Troyer, and
  Reiher]{keller2015efficient}
Keller,~S.; Dolfi,~M.; Troyer,~M.; Reiher,~M. An efficient matrix product
  operator representation of the quantum chemical Hamiltonian. \emph{J. Chem.
  Phys.} \textbf{2015}, \emph{143}, 244118\relax
\mciteBstWouldAddEndPuncttrue
\mciteSetBstMidEndSepPunct{\mcitedefaultmidpunct}
{\mcitedefaultendpunct}{\mcitedefaultseppunct}\relax
\EndOfBibitem
\bibitem[Jordan and Wigner(1928)Jordan, and Wigner]{jordan1928pauli}
Jordan,~P.; Wigner,~E.~P. About the Pauli exclusion principle. \emph{Z. Phys.}
  \textbf{1928}, \emph{47}, 14--75\relax
\mciteBstWouldAddEndPuncttrue
\mciteSetBstMidEndSepPunct{\mcitedefaultmidpunct}
{\mcitedefaultendpunct}{\mcitedefaultseppunct}\relax
\EndOfBibitem
\bibitem[Percus and Rotenberg(1962)Percus, and Rotenberg]{percus1962exact}
Percus,~J.; Rotenberg,~A. Exact eigenfunctions of angular momentum by
  rotational projection. \emph{J. Math. Phys.} \textbf{1962}, \emph{3},
  928--932\relax
\mciteBstWouldAddEndPuncttrue
\mciteSetBstMidEndSepPunct{\mcitedefaultmidpunct}
{\mcitedefaultendpunct}{\mcitedefaultseppunct}\relax
\EndOfBibitem
\bibitem[Press \latin{et~al.}(2007)Press, Teukolsky, Vetterling, and
  Flannery]{press2007numerical}
Press,~W.~H.; Teukolsky,~S.~A.; Vetterling,~W.~T.; Flannery,~B.~P.
  \emph{Numerical recipes 3rd edition: The art of scientific computing};
  Cambridge university press, 2007\relax
\mciteBstWouldAddEndPuncttrue
\mciteSetBstMidEndSepPunct{\mcitedefaultmidpunct}
{\mcitedefaultendpunct}{\mcitedefaultseppunct}\relax
\EndOfBibitem
\bibitem[Verstraete \latin{et~al.}(2004)Verstraete, Porras, and
  Cirac]{verstraete2004density}
Verstraete,~F.; Porras,~D.; Cirac,~J.~I. Density matrix renormalization group
  and periodic boundary conditions: a quantum information perspective.
  \emph{Phys. Rev. Lett.} \textbf{2004}, \emph{93}, 227205\relax
\mciteBstWouldAddEndPuncttrue
\mciteSetBstMidEndSepPunct{\mcitedefaultmidpunct}
{\mcitedefaultendpunct}{\mcitedefaultseppunct}\relax
\EndOfBibitem
\bibitem[Pippan \latin{et~al.}(2010)Pippan, White, and
  Evertz]{pippan2010efficient}
Pippan,~P.; White,~S.~R.; Evertz,~H.~G. Efficient matrix-product state method
  for periodic boundary conditions. \emph{Phys. Rev. B} \textbf{2010},
  \emph{81}, 081103\relax
\mciteBstWouldAddEndPuncttrue
\mciteSetBstMidEndSepPunct{\mcitedefaultmidpunct}
{\mcitedefaultendpunct}{\mcitedefaultseppunct}\relax
\EndOfBibitem
\bibitem[Li \latin{et~al.}(2013)Li, Suo, Zhang, Xiao, and Liu]{li2013combining}
Li,~Z.; Suo,~B.; Zhang,~Y.; Xiao,~Y.; Liu,~W. Combining spin-adapted open-shell
  TD-DFT with spin--orbit coupling. \emph{Mol. Phys.} \textbf{2013},
  \emph{111}, 3741--3755\relax
\mciteBstWouldAddEndPuncttrue
\mciteSetBstMidEndSepPunct{\mcitedefaultmidpunct}
{\mcitedefaultendpunct}{\mcitedefaultseppunct}\relax
\EndOfBibitem
\bibitem[Luzanov(1985)]{luzanov1985calculating}
Luzanov,~A. Calculating spin density in the quantum-chemical unitary formalism.
  \emph{Theor. Exp. Chem.} \textbf{1985}, \emph{21}, 329--331\relax
\mciteBstWouldAddEndPuncttrue
\mciteSetBstMidEndSepPunct{\mcitedefaultmidpunct}
{\mcitedefaultendpunct}{\mcitedefaultseppunct}\relax
\EndOfBibitem
\bibitem[Gould and Paldus(1990)Gould, and Paldus]{gould1990spin}
Gould,~M.; Paldus,~J. Spin-dependent unitary group approach. I. General
  formalism. \emph{J. Chem. Phys.} \textbf{1990}, \emph{92}, 7394--7401\relax
\mciteBstWouldAddEndPuncttrue
\mciteSetBstMidEndSepPunct{\mcitedefaultmidpunct}
{\mcitedefaultendpunct}{\mcitedefaultseppunct}\relax
\EndOfBibitem
\bibitem[Fano \latin{et~al.}(1990)Fano, Ortolani, and Parola]{fano1990hole}
Fano,~G.; Ortolani,~F.; Parola,~A. Hole-hole effective interaction in the
  two-dimensional Hubbard model. \emph{Phys. Rev. B} \textbf{1990}, \emph{42},
  6877\relax
\mciteBstWouldAddEndPuncttrue
\mciteSetBstMidEndSepPunct{\mcitedefaultmidpunct}
{\mcitedefaultendpunct}{\mcitedefaultseppunct}\relax
\EndOfBibitem
\bibitem[Becke(1988)]{becke1988density}
Becke,~A.~D. Density-functional exchange-energy approximation with correct
  asymptotic behavior. \emph{Phys. Rev. A} \textbf{1988}, \emph{38}, 3098\relax
\mciteBstWouldAddEndPuncttrue
\mciteSetBstMidEndSepPunct{\mcitedefaultmidpunct}
{\mcitedefaultendpunct}{\mcitedefaultseppunct}\relax
\EndOfBibitem
\bibitem[Perdew(1986)]{perdew1986density}
Perdew,~J.~P. Density-functional approximation for the correlation energy of
  the inhomogeneous electron gas. \emph{Phys. Rev. B} \textbf{1986}, \emph{33},
  8822\relax
\mciteBstWouldAddEndPuncttrue
\mciteSetBstMidEndSepPunct{\mcitedefaultmidpunct}
{\mcitedefaultendpunct}{\mcitedefaultseppunct}\relax
\EndOfBibitem
\bibitem[Jorge \latin{et~al.}(2009)Jorge, Canal, Camiletti, and
  Machado]{jorge2009contracted}
Jorge,~F.; Canal,~N.~A.; Camiletti,~G.; Machado,~S. Contracted Gaussian basis
  sets for Douglas-Kroll-Hess calculations: Estimating scalar relativistic
  effects of some atomic and molecular properties. \emph{J. Chem. Phys.}
  \textbf{2009}, \emph{130}, 064108--064108\relax
\mciteBstWouldAddEndPuncttrue
\mciteSetBstMidEndSepPunct{\mcitedefaultmidpunct}
{\mcitedefaultendpunct}{\mcitedefaultseppunct}\relax
\EndOfBibitem
\bibitem[Liu(2010)]{liu2010ideas}
Liu,~W. Ideas of relativistic quantum chemistry. \emph{Mol. Phys.}
  \textbf{2010}, \emph{108}, 1679--1706\relax
\mciteBstWouldAddEndPuncttrue
\mciteSetBstMidEndSepPunct{\mcitedefaultmidpunct}
{\mcitedefaultendpunct}{\mcitedefaultseppunct}\relax
\EndOfBibitem
\bibitem[Li \latin{et~al.}(2012)Li, Xiao, and Liu]{li2012spin}
Li,~Z.; Xiao,~Y.; Liu,~W. On the spin separation of algebraic two-component
  relativistic Hamiltonians. \emph{J. Chem. Phys.} \textbf{2012}, \emph{137},
  154114\relax
\mciteBstWouldAddEndPuncttrue
\mciteSetBstMidEndSepPunct{\mcitedefaultmidpunct}
{\mcitedefaultendpunct}{\mcitedefaultseppunct}\relax
\EndOfBibitem
\bibitem[Altmann and Herzig(1994)Altmann, and Herzig]{altmann1994point}
Altmann,~S.; Herzig,~P. \emph{Point-group theory tables}; Oxford, 1994\relax
\mciteBstWouldAddEndPuncttrue
\mciteSetBstMidEndSepPunct{\mcitedefaultmidpunct}
{\mcitedefaultendpunct}{\mcitedefaultseppunct}\relax
\EndOfBibitem
\bibitem[Zheng \latin{et~al.}(2016)Zheng, Chung, Corboz, Ehlers, Qin, Noack,
  Shi, White, Zhang, and Chan]{zheng2016stripe}
Zheng,~B.-X.; Chung,~C.-M.; Corboz,~P.; Ehlers,~G.; Qin,~M.-P.; Noack,~R.~M.;
  Shi,~H.; White,~S.~R.; Zhang,~S.; Chan,~G. K.-L. Stripe order in the
  underdoped region of the two-dimensional Hubbard model. \emph{arXiv preprint
  arXiv:1701.00054} \textbf{2016}, \relax
\mciteBstWouldAddEndPunctfalse
\mciteSetBstMidEndSepPunct{\mcitedefaultmidpunct}
{}{\mcitedefaultseppunct}\relax
\EndOfBibitem
\bibitem[Varshalovich \latin{et~al.}(1988)Varshalovich, Moskalev, and
  Khersonskii]{varshalovich1988quantum}
Varshalovich,~D.~A.; Moskalev,~A.; Khersonskii,~V. \emph{Quantum theory of
  angular momentum}; World Scientific, 1988; p~84\relax
\mciteBstWouldAddEndPuncttrue
\mciteSetBstMidEndSepPunct{\mcitedefaultmidpunct}
{\mcitedefaultendpunct}{\mcitedefaultseppunct}\relax
\EndOfBibitem
\end{mcitethebibliography}

\clearpage
\newpage
For Table of Contents Only
\begin{figure}
\includegraphics[width=0.9\textwidth]{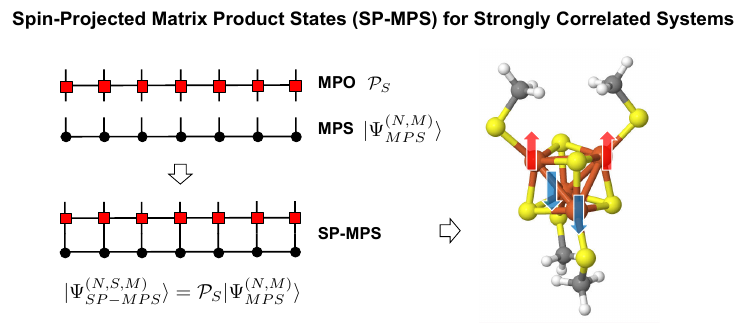}
\end{figure}

\end{document}